\documentclass[article,nojss,shortnames]{jss}

\let\maxwidth\textwidth

\usepackage[final]{changes}

\usepackage{thumbpdf}
\usepackage{amsfonts,amstext,amsmath,amssymb,amsthm}
\usepackage{accents}
\usepackage{xcolor}
\usepackage{rotating}
\usepackage{verbatim}
\usepackage[graphicx]{realboxes}
\usepackage{xspace}
\usepackage{lscape}

\usepackage{accents}

\theoremstyle{example}

\newcommand{\NAMI}{nonparanormal adjusted marginal inference\xspace}

\usepackage{booktabs}
\usepackage{nicefrac}

\newcommand{\rY}{Y}
\newcommand{\rX}{\mX}

\newcommand{\ry}{y}
\newcommand{\rx}{\xvec}

\newcommand{\samY}{\mathcal{Y}}
\newcommand{\samX}{\mathcal{X}}

\newcommand{\h}{h}

\newcommand{\basisy}{\avec}

\newcommand{\parm}{\varthetavec}
\newcommand{\eparm}{\vartheta}

\newcommand{\shiftparm}{\betavec}

\renewcommand{\Prob}{\mathbb{P}}
\newcommand{\Ex}{\mathbb{E}}
\newcommand{\Var}{\mathbb{V}}
\newcommand{\RR}{\mathbb{R}}
\newcommand{\NN}{\mathbb{N}}

\usepackage{dsfont}

\def \expit {\text{logit}^{-1}}
\def \cloglog {\text{cloglog}}

 \DeclareMathOperator{\logit}{logit}

 \DeclareMathOperator{\Cov}{Cov}

 \DeclareMathOperator{\diag}{diag}

 \DeclareMathOperator{\ND}{N}

\def \avec {\text{\boldmath$a$}}    \def \mA {\text{\boldmath$A$}}
    \def \mB {\text{\boldmath$B$}}
    
    \def \mD {\text{\boldmath$D$}}

\def \hvec {\text{\boldmath$h$}}    \def \mH {\text{\boldmath$H$}}
    \def \mI {\text{\boldmath$I$}}

\def \xvec {\text{\boldmath$x$}}    \def \mX {\text{\boldmath$X$}}

 \def \calH {\mathcal H}

\def \betavec         {\text{\boldmath$\beta$}}

\def \varthetavec     {\text{\boldmath$\vartheta$}}

\def \lambdavec       {\text{\boldmath$\lambda$}}

\def \mLambda  {\mathbf{\Lambda}}

\def \mSigma   {\mathbf{\Sigma}}

\def \mOmega   {\mathbf{\Omega}}

\def \nullvec {\mathbf{0}}

\newtheorem{lem}{Lemma}

\newcommand{\App}{Appendix} 
\newcommand{\AppFigure}{Figure~}
\newcommand{\AppFigures}{Figures~}
\newcommand{\AppTable}{Table~}
\newcommand{\Figure}{Figure}
\newcommand{\Table}{Table}

\usepackage[utf8]{inputenc}
\usepackage{enumitem}

\usepackage{afterpage}

\makeatletter
\renewcommand*\env@matrix[1][\arraystretch]{%
	\edef\arraystretch{#1}%
	\hskip -\arraycolsep
	\let\@ifnextchar\new@ifnextchar
	\array{*\c@MaxMatrixCols c}}
\makeatother

\usepackage{colortbl}

\usepackage{makecell, cellspace, caption}
\usepackage{subcaption}

\makeatletter
\def\maxwidth{ %
  \ifdim\Gin@nat@width>\linewidth
    \linewidth
  \else
    \Gin@nat@width
  \fi
}
\makeatother

\definecolor{fgcolor}{rgb}{0.345, 0.345, 0.345}
\newcommand{\hlnum}[1]{\textcolor[rgb]{0.686,0.059,0.569}{#1}}%
\newcommand{\hlsng}[1]{\textcolor[rgb]{0.192,0.494,0.8}{#1}}%
\newcommand{\hlcom}[1]{\textcolor[rgb]{0.678,0.584,0.686}{\textit{#1}}}%
\newcommand{\hlopt}[1]{\textcolor[rgb]{0,0,0}{#1}}%
\newcommand{\hldef}[1]{\textcolor[rgb]{0.345,0.345,0.345}{#1}}%
\newcommand{\hlkwb}[1]{\textcolor[rgb]{0.69,0.353,0.396}{#1}}%
\newcommand{\hlkwc}[1]{\textcolor[rgb]{0.333,0.667,0.333}{#1}}%
\newcommand{\hlkwd}[1]{\textcolor[rgb]{0.737,0.353,0.396}{\textbf{#1}}}%

\usepackage{framed}
\makeatletter
\newenvironment{kframe}{%
 \def\at@end@of@kframe{}%
 \ifinner\ifhmode%
  \def\at@end@of@kframe{\end{minipage}}%
  \begin{minipage}{\columnwidth}%
 \fi\fi%
 \def\FrameCommand##1{\hskip\@totalleftmargin \hskip-\fboxsep
 \colorbox{shadecolor}{##1}\hskip-\fboxsep
     \hskip-\linewidth \hskip-\@totalleftmargin \hskip\columnwidth}%
 \MakeFramed {\advance\hsize-\width
   \@totalleftmargin\z@ \linewidth\hsize
   \@setminipage}}%
 {\par\unskip\endMakeFramed%
 \at@end@of@kframe}
\makeatother

\definecolor{shadecolor}{rgb}{.97, .97, .97}
\definecolor{messagecolor}{rgb}{0, 0, 0}
\definecolor{warningcolor}{rgb}{1, 0, 1}
\definecolor{errorcolor}{rgb}{1, 0, 0}
\newenvironment{knitrout}{}{} 

\usepackage{alltt}
\IfFileExists{upquote.sty}{\usepackage{upquote}}{}

\newcommand{\ARXIV}{1}

\author{Susanne Dandl \and Torsten Hothorn \\ Universit\"at Z\"urich}
\Plainauthor{Dandl \& Hothorn}

\title{Nonparanormal Adjusted Marginal Inference}
\Plaintitle{Nonparanormal Adjusted Marginal Inference}
\Shorttitle{Nonparanormal Adjusted Marginal Inference}

\Abstract{

Although treatment effects can be estimated from observed outcome
distributions obtained from proper randomization in clinical trials,
covariate adjustment is recommended to increase precision.  For important
treatment effects, such as odds or hazard ratios, conditioning on covariates
in binary logistic or proportional hazards models changes the
interpretation of the treatment effect and conditioning on different sets of
covariates renders the resulting effect estimates incomparable.

We propose a novel nonparanormal model formulation for adjusted marginal
inference.  This model for the joint distribution of outcome and covariates
directly features a marginally defined treatment effect parameter, such as a
marginal odds or hazard ratio.  Not only the marginal treatment effect of
interest can be estimated based on this model, it also provides an overall
coefficient of determination and covariate-specific measures of prognostic
strength.

For the special case of Cohen's standardized mean difference $d$, we
theoretically show that adjusting for an informative prognostic variable
improves the precision of the marginal, noncollapsible effect.  Empirical
results confirm this not only for Cohen's $d$ but also for odds and hazard
ratios in simulations and three applications.
A reference implementation is available in the \proglang{R}~add-on package
\href{https://doi.org/10.32614/CRAN.package.tram}{\pkg{tram}}.
 }

\Keywords{marginal effect, noncollapsibility, covariate adjustment, 
	  randomized trial, transformation model}
\Plainkeywords{marginal effect, noncollapsibility, covariate adjustment, randomised trial, transformation model}

\Address{
  Susanne Dandl \& Torsten Hothorn\\
  Institut f\"ur Epidemiologie, Biostatistik und Pr\"avention \\
  Universit\"at Z\"urich \\
  Hirschengraben 84, CH-8001 Z\"urich, Switzerland \\
  Email: \texttt{Susanne.Dandl@uzh.ch}
}

\begin{document}


\section{Introduction}

\deleted{Randomized clinical trials (RCTs) aim at the estimation of causally
interpretable treatment effects. Typically, treatment effect parameters are defined as
functions of marginal outcome distributions. In the simplest case, one compares the
distributions of the outcome observed in two groups of patients: those 
randomized to be treated by a standard therapy and those randomized to 
receive an innovative therapy. For ease of
communication and comparability between trials, treatment effects are 
expressed as differences in means, odds ratios, risk differences, hazard ratios,
restricted mean survival times or similar marginally defined measures.}

Increasing the precision of treatment effect estimates in randomized clinical trials (RCTs)
has kept
statisticians on their toes over the last century.  Two simple yet important
ideas are stratification and covariate adjustment, both relying on the
availability of prognostic information.  Prognostic variables are baseline
covariates that are associated with the outcome and
explain outcome heterogeneity.  By exploiting the information
in such prognostic variables, the precision of the treatment effect can be
increased, leading to narrower confidence intervals and, therefore, more
powerful inference about the true treatment effect.

Many classical contributions advocating for covariate adjustment in RCTs base their
arguments on differences in means in analysis of covariance (ANCOVA) 
models \citep[see][and the references therein]{senn_2024}.  The error term in
such models accounts for some forms of model misspecification (such as
missing or incorrectly transformed prognostic variables). Adding or removing
terms only affects the variance of the residual term but not the model parameter corresponding
to the average treatment effect. More flexible model formulations extending
ANCOVA models \citep[see, for example, 
the references in][]{siegfried_relevance_2023} gained novel interest in
the machine learning era \citep{Schuler_Walsh_Hall_2022}. Covariate
adjustment has been recommended for the analysis of RCTs by several authorities
(e.g., \href{https://www.ema.europa.eu/en/documents/scientific-guideline/guideline-adjustment-baseline-covariates-clinical-trials_en.pdf}{European Medicines Agency
	EMA/CHMP/295050/2013} and \href{https://www.fda.gov/regulatory-information/search-fda-guidance-documents/adjusting-covariates-randomized-clinical-trials-drugs-and-biological-products}{Food and Drug Administration FDA-2019-D-0934}).

For nonnormal models, such as binary logistic or Cox models,
lack of an explicit residual term introduces a dependency of the treatment effect parameter
on unadjusted outcome heterogeneity. Adding prognostic
variables \deleted{to reduce the error variance} leads to \deleted{stronger treatment effects,
i.e.,} higher effect magnitudes compared to the effect estimate of an
unadjusted model.  While this \replaced{increases}{leads to increased} power for null hypothesis
significance tests of the treatment effects, the \deleted{treatment} effect estimates
are not directly comparable between models with differing sets of prognostic
covariates \citep[see][among
others]{robinson_surprising_1991,martinussen_collapsibility_2013,daniel_making_2021}.
\deleted{In the literature,} This noncomparability issue is also known as noncollapsibility. 
A treatment effect is \deleted{called} noncollapsible if the marginal effect obtained from averaging over prognostic variables
in a conditional model \deleted{adjusting for variables} \replaced{differs from}{is not identical to} the effect obtained from a marginal model
\deleted{not adjusting for covariates} \citep{aalen_does_2015}. 

\replaced{Instead of}{As alternatives to the simple remedy of} reporting marginal and adjusted
\deleted{(``multivariable'')} treatment effect estimates side-by-side in the presence
of noncollapsibility, two conceptually different strategies were proposed.  
\added{Strategy I is to replace the noncollapsible model by a collapsible one} \citep[e.g., a Cox model by a Weibull accelerated failure time model, as done in][]{aalen_does_2015} \added{or to reformulate the treatment effect (e.g., the hazard ratio as restricted mean survival times or risk differences).}
\deleted{Strategy I is to replace the noncollapsible model by a collapsible one.
An example is to replace the Cox
proportional hazards model by a Weibull accelerated failure time model
\mbox{\citep{aalen_does_2015}} or to reformulate the treatment effect, for example as restricted
mean survival times or risk differences.}
\replaced{S}{In this context, s}everal generally applicable 
inference procedures have been suggested, most prominently G-computation or
standardization
\citep{daniel_making_2021,VanLancker_2024}.

Strategy II is to directly adjust marginal log-odds or log-hazard ratios for
prognostic covariates.  \cite{Tsiatis_Davidian_2008} suggested a framework for semiparametric locally
efficient adjustment based on the joint distribution of outcome, treatment,
and covariates. 
The approach was generalized to marginal binary logistic regression
models by \cite{Zhang_Tsiatis_2008} for inference on log-odds ratios, 
and to proportional hazards models by \cite{Lu_Tsiatis_2008} \replaced{for}{, enabling inference on} log-hazard ratios. 
\deleted{Recently, }\cite{Ye_Shao_Yi_2024} applied very similar
ideas to covariate adjustment for the comparison of marginal survivor curves\deleted{,
potentially also under proportional hazards}. 
The advantage of this strategy is the marginal interpretability on classical log-odds or
log-hazard ratio scales \citep{doi_redefining_2022}.

In the spirit of strategy II, we present a novel \replaced{\NAMI method}{nonparanormal model} featuring a 
marginal treatment effect parameter
along with an estimation procedure able to leverage prognostic information
for increasing the precision of corresponding \deleted{treatment effect} parameter
estimates. \replaced{This enables, for example,}{For example, the novel \NAMI method suggested here allows to estimate} the estimation of a marginal log-odds ratio whose standard error shrinks with increasing strength of available 
prognostic information. \replaced{This allows reporting}{Using \NAMI, we are able to report} interpretable,
comparable marginal treatment effect estimates with smaller standard errors
than unadjusted analysis, even when the parameter is
\deleted{classified as being} noncollapsible\deleted{ in conventional terminology \mbox{\citep{van_lancker_covariate_2024}}}.

On a more technical level, one can understand our contribution as a
nonparanormal alternative to the semiparametric approach for the joint
distribution of outcome, treatment, and covariates. Instead of leaving most
aspects of this joint distribution unspecified \citep[as done
by][]{Zhang_Tsiatis_2008},  we suggest to model the joint distribution of
outcome and covariates by a nonparanormal model
\citep{liu_etal_nonparanormal}. The marginal model
for the outcome features the treatment effect parameter of interest,
marginal covariate distributions are described in a model-free way, and
their joint distribution is characterized by a Gaussian copula. This \deleted{model gives
rise to a} novel formulation of the conditional distribution of outcome given
treatment and covariates \deleted{which} is, by design, collapsible. The approach is
based on marginal transformation models allowing broad definitions of
marginal treatment effects, such as Cohen's
standardized differences in means $d$, odds
and hazard ratios, or probabilistic indices\deleted{ and other noncollapsible treatment
effect parameters} \citep{Hothorn_Moest_Buehlmann_2017}. 
The nonparanormal model is fully parameterized \citep[exploiting connections
to multivariate transformation models proposed by][]{Klein_Hothorn_Barbanti_2020} and
thus standard maximum likelihood approaches in these models \citep{Hothorn_2024}
can be applied for parameter
estimation and the construction of confidence intervals or test procedures.
This does not only apply to the marginal treatment effect but also to
prognostic covariate effects. Unlike semiparametric approaches, where the
prognostic value of covariates is not directly quantified, our model
provides an overall coefficient of determination as well as
covariate-specific measures of prognostic strength.

\replaced{We introduce the general concept of \NAMI and its application to improved estimation of Cohen's $d$ for continuous, odds ratios for binary, and hazard ratios for survival outcomes in
Section~\ref{sec:methodology}. For Cohen's $d$, we derive an analytic standard error under covariate adjustment to explore the potential for sample size reductions theoretically. Sections~\ref{sec:simulation} and \ref{sec:application} evaluate \NAMI's ability to improve precision and assess prognostic strength in RCTs across diverse outcome types.}{We proceed by introducing the general concept for \NAMI and its application
to the improved estimation of Cohen's $d$ for continuous, odds ratios for
binary, and hazard ratios for survival outcomes in
Section~\ref{sec:methodology}.  For the special case of Cohen's $d$, we
present an analytic expression of the standard error under covariate
adjustment allowing to investigate the potential for sample size reductions
theoretically.  We evaluate the ability of \NAMI to increase precision in
marginal treatment effect parameter estimation and to assess prognostic
strength empirically for RCTs with continuous,
binary, and survival outcomes in Sections~\ref{sec:simulation} and \ref{sec:application}.}


\section{Nonparanormal adjusted marginal inference}
\label{sec:methodology}

We are interested in the effect of some binary treatment $W \sim B(1, \pi)$ 
on the distribution of an outcome $\rY \in \samY$
assuming an at least ordered sample space $\samY$.   The propensity score
$\pi = \Prob(W = 1)$ is constant and does not rely on covariates
in this randomized trial setting.  In addition to $\rY$ and $W$, $J - 1$
baseline covariates $\rX = (X_1, X_2, ..., X_{J-1})$ were observed, with $X_j \in \samX_j, j = 1,
..., J - 1$ assuming all covariate sample spaces $\samX_j$ are at least ordered. 

\subsection{Univariate marginal and conditional transformation models}

We denote the conditional cumulative distribution function of $\rY$ given $W
= 0$ (``control'') as $F_0(\ry) = \Prob(\rY \le \ry \mid W = 0)$ and the conditional
cumulative distribution function of $\rY$ given $W = 1$ (``treated'') as $F_1(\ry) =
\Prob(\rY \le \ry \mid W = 1)$.  The treatment effect $\tau := \tau(F_0,
F_1) \in \RR$ expresses the discrepancy between the two marginal (with
respect to covariates) distributions as 
an, ideally interpretable, scalar.  Because $F_0$ and $F_1$ do not rely on covariates $\rX$,
$F_0$ and $F_1$ are also called marginal models and $\tau$ reflects a
\textit{marginal} treatment effect. In randomized trials, it is possible to estimate
$\tau$ from $\rY$ and $W$ alone, ignoring covariates.

The two distribution functions $F_0$ and $F_1$ could be estimated
nonparametrically, e.g., as empirical distribution functions without
assuming a specific distributional form, however, the characterization
of the discrepancy between the two distributions in the form of an
\emph{interpretable} scalar treatment effect $\tau$ is somewhat challenging.  Alternatively, parametric
distributions for $F_0$ and $F_1$ allow specification of interpretable treatment
effects $\tau$, however, at the price of imposing strong 
assumptions on the outcome distribution.

Transformation models \citep[in the sense of][]{BoxCox_1964,Hothorn_Moest_Buehlmann_2017}
offer a compromise between the nonparametric and parametric worlds by
transforming outcomes via a monotone nondecreasing transformation function $h :
\samY \rightarrow \RR$ such that the \emph{transformed} outcome distribution is 
described by a simple cumulative distribution function $G: \RR \rightarrow
[0,1]$ with parameter-free log-concave absolute continuous density.  
This results in $F_0(\ry) = G\bigl(\h(\ry)\bigr)$ for the distribution under
control.  Throughout this manuscript, we assume that the treatment effect
$\tau$ is defined as a shift effect on the scale of the transformation,
i.e., $F_1(\ry) = G\bigl(\h(\ry) - \tau\bigr)$, resulting in the overall marginal
transformation model for the conditional distribution of the outcome $\rY$
given treatment group $W = 0$ or $W = 1$ as 
\begin{equation}
	F_w(\ry) = F(\ry \mid W = w)  = G\bigl(\h(\ry) - \tau w\bigr) = G\bigl(\h(y \mid w)\bigl).
	\label{eq:marginal}
\end{equation}
Different choices of $G$ and $\h$ to be discussed in Section~\ref{subsec:outcometypes}
give rise to numerous classical and novel treatment effects $\tau$ for
continuous, binary, ordered categorical, or survival outcomes $\rY$.

Leveraging the prognostic information about the outcome $\rY$ contained in
the covariates $\rX$ is possible in linear transformation models with
additional linear predictors. In such models, the conditional
cumulative distribution function of the outcome given treatment and
covariates is formulated as
\begin{eqnarray}
	F(\ry \mid W = w, \rX = \rx) = G\bigl(\h_{\rx}(\ry) - \tau_{\rx} w - \widetilde{\rx}^\top \shiftparm\bigr)
	\label{eq:ltm}
\end{eqnarray}
for some appropriate coding $\widetilde{\rx}$ of covariates $\rx$.
In the following, we call the treatment effect parameter $\tau_{\rx}$ the
\textit{conditional} effect to reflect that it is the effect when
conditioning on covariates $\rX$.  In general, the conditional effect
$\tau_{\rx}$ is not equal to the marginal effect $\tau$ in
model~(\ref{eq:marginal}), the same holds for the transformation functions $\h$
and $\h_\rx$.  If integrating the conditional model over the distribution of $\rX$ 
results in model~(\ref{eq:marginal}), the effect $\tau$ is called collapsible
\citep[see Chapter 6 in][for a more general definition
of collapsibility]{pearl_2009}. Section~\ref{subsec:outcometypes} 
shows that
Cohen's $d$ in the linear model, the log-odds ratio in a binary logistic regression
model, and the log-hazard ratio in the Cox model can be expressed as parameter
$\tau$ in model~(\ref{eq:marginal}). The corresponding conditional models are
noncollapsible, that is, $\tau \neq \tau_\rx$.

We \replaced{propose a Gaussian copula}{proceed by proposing a} model for the joint distribution of outcome given
treatment and covariates.\deleted{by means of a Gaussian copula.} The marginal outcome
distribution $F_w(\ry)$ stays intact, allowing estimation of the
marginal treatment effect $\tau$ in the presence of covariates. The model
\replaced{features}{gives rise to} a novel conditional distribution $F(\ry \mid W = w, \rX = \rx)$
\replaced{with collapsible}{featuring the} treatment effect $\tau$.
\deleted{ in a collapsible way.}

\subsection{Multivariate transformation models}
\label{subsec:multivar}

Transformation models for multivariate outcomes were introduced by \cite{Klein_Hothorn_Barbanti_2020}
extending univariate transformation models for a single $\rY$ to the
multivariate situation. We adapt this approach and treat the covariates $\rX$ as additional outcomes
in a joint model for $(\rX, \rY)$. We first 
parameterize the marginal covariate distributions as unconditional transformation models, i.e.,
\begin{equation}
	\Prob(X_j \le x_j) = \Phi\bigl(h_j(x_j)\bigr), j = 1, ..., J-1
	\label{eq:marginalx}
\end{equation}
with $h_j : \samY_j \rightarrow \RR$ serving as the nondecreasing transformation function for $X_j$
and $\Phi$ denoting the cumulative distribution function of $\ND(0, 1)$.
The outcome $\rY$ is also transformed to a latent normal scale 
via $h_J(\ry \mid w) := \Phi^{-1}\left[G\Bigl\{h\left(\ry \mid w\right)\Bigr\}\right]$
based on model~(\ref{eq:marginal}).
The standard normal distribution function $\Phi$ is attractive as it provides a direct link
to Gaussian copulas \citep{song_etal_gaussiancop_2009} and
nonparanormal models \citep{liu_etal_nonparanormal}.

The multivariate transformation function $\hvec: \samX_1 \times \samX_2 \times \cdots \times
\samX_{J - 1} \times \samY \rightarrow \RR^J$ defined as
$
\hvec(\rX, \rY \mid W = w) = \bigl(h_1(X_1), h_2(X_2), ..., h_{J-1}(X_{J-1}), h_J(Y \mid w)\bigr)^\top
$
formulates the joint cumulative distribution function of
covariates $\rX$ and outcome $\rY$ given treatment $W$ as
\begin{eqnarray}
	\Prob(\rX \le \rx, \rY \le \ry \mid W = w) =
	\Phi_{\mSigma} \bigl(\hvec(\rx, \ry \mid W = w)\bigr).
	\label{eq:joinedmultivar}
\end{eqnarray}
Here, $\Phi_{\mSigma}$ denotes the cumulative distribution function of $\ND_J(\nullvec,
\mSigma)$, the $J$-dimensional normal distribution with zero mean and $J
\times J$ correlation matrix $\mSigma$ ensuring identifiability of $\hvec$.

Based on \cite{Hothorn_2024}, we parameterize the correlation matrix $\mSigma$ in terms of
the inverse of its Cholesky factor, that is, using the factorization
$\mSigma = \mOmega^{-1} \mOmega^{-\top}$. The lower triangular $J \times J$ matrix $\mOmega
= (\omega_{j\jmath})$ 
has positive diagonals $\omega_{jj} > 0, j = 1, \dots, J$ and 
lower triangular elements $\omega_{j\jmath}, 1 \le \jmath < j \le J$.
We write $\mOmega(\lambdavec)$ with \emph{unconstrained} 
parameters $\lambdavec = (\lambda_{21}, \dots, \lambda_{J,J-1})^\top \in \RR^{\nicefrac{J (J - 1)}{2}}$ as the lower
triangular elements of a unit lower triangular matrix 
\begin{equation}
	\mLambda = \mLambda(\lambdavec) = \left(\lambda_{jj'}\right)_{1 \le j' < j \le J}
	\label{eq:mLambda}
\end{equation}
such that $\mOmega(\lambdavec) = \mLambda(\lambdavec) \diag(\mLambda^{-1}(\lambdavec) \mLambda^{-\top}(\lambdavec))^{1/2}$ in accordance with Section 2, Option 2
in \cite{Hothorn_2024}. The corresponding precision matrix is $\mSigma^{-1} = \mOmega^\top \mOmega$
and we refer to the correlations, on a latent transformed normal scale, 
between covariate $X_j$ and outcome $\rY$ in both treatment groups, as
the elements $\rho(\lambdavec)_{Jj}, j = 1, \dots, J - 1$, of the last row of
the correlation matrix $\mSigma$.
From the joint distribution of $\rX$ and $\rY$ given treatment $W$ in
model~(\ref{eq:joinedmultivar}), we can derive 
the conditional distribution of $\rY$ given $W$ and covariates $\rX
\subseteq \RR^{J-1}$ from an absolute continuous distribution as
\begin{eqnarray}
	\Prob(\rY \le \ry \mid W = w, \rX = \rx) = 
	\Phi\left(\sum\limits_{j = 1}^{J - 1} \omega_{Jj} \h_j(x_j) + \omega_{JJ}\h_J(\ry \mid w)\right).
	\label{eq:condmultivar}
\end{eqnarray}
The regression coefficients are obtained from the $J$th row of the 
precision matrix $\mSigma^{-1} = \mOmega^\top \mOmega$, which is identical
to the vector $\omega_{JJ} (\omega_{J1}, \dots, \omega_{JJ})^\top$.
It should be noted that the marginal model of $\rY$ given $W$ is identical
to model~(\ref{eq:marginal}), i.e., 
$F_w(\ry) = \Prob(Y \le \ry \mid W = w) = \Phi\bigl(h_J(y \mid w)\bigr) = \Phi\left[\Phi^{-1}\Bigl\{G\bigl(\h(y \mid w)\bigr)\Bigr\}\right] = G\bigl(\h(y) - \tau w\bigr)$ 
because the model is constrained to unit marginal variances in $\mSigma$. For
absolute continuous $(\rX, \rY)$, we can write the marginal model as
$\h_J(\rY \mid w) = \varepsilon \sim \ND(0, 1)$ and the conditional
distribution~(\ref{eq:condmultivar}) is equivalent to a normal linear regression
model for Gaussianized outcome and covariates. With
regression coefficients $-\omega_{JJ}^{-1} \omega_{Jj}$ for $\h_j(x_j), j =
1, \dots, J - 1$ and residual standard deviation $\omega_{JJ}^{-1}$
we have
$
\h_J(\rY \mid w) = \sum_{j = 1}^{J - 1} -\omega_{JJ}^{-1} \omega_{Jj} \h_j(x_j) + \omega_{JJ}^{-1}
\varepsilon.
$
Because $\Var\bigl(\h_j(X_j)\bigr) = 1$ for all $j = 1, \dots, J - 1$ by definition, we can
rank covariates with respect to their prognostic strengths $|\omega_{Jj}|$.
The gain obtained from adjusting for covariates can be measured by the
coefficient of determination 
$R^2 := 1 - \omega_{JJ}^{-2}$
defined by the
ratio of the residual variances $\omega_{JJ}^{-2}$ of the 
conditional model~(\ref{eq:condmultivar}) and the marginal model 
with residual standard deviation one. For noncontinuous variables, the same
arguments hold on a latent continuous scale.

\subsection{Specific model applications}
\label{subsec:outcometypes}
The choice of $G$ and $h$ depends on the outcome at hand and the desired interpretation of the treatment effect
$\tau$.
We follow \cite{Hothorn_Moest_Buehlmann_2017} and parameterize $h:\samY \rightarrow \RR$ as a linear combination of basis functions, i.e. 
\begin{equation}
	\h(y) = \basisy(y)^\top \parm.
	\label{eq:lincom}
\end{equation} 
Here, $\basisy : \samY \rightarrow \RR^{M}$  denotes a multidimensional basis function with $M \in \NN$  while $\parm$ corresponds to basis
coefficients.
In the following, different choices of $G$, $\basisy$ and
$\parm$ are discussed for continuous, binary and survival outcomes. These
choices give rise to various notions of  treatment effects $\tau$, for
example in terms of Cohen's $d$, probabilistic indices, log-odds ratios, or
log-hazard ratios.  It is also shown how popular models, like the (proportional
odds) logistic regression models or Cox proportional hazards models, can be
embedded in our methodology. Appropriate choices for the transformations $h_j, j = 1, ..., J-1$ for
covariates in $\rX$ 
as well as parameter estimation for model~(\ref{eq:joinedmultivar}) are discussed in Section~\ref{sec:covariates}.  
An attractive choice for $G$ in case of {continuous outcomes} is the standard normal distribution, i.e., $G = \Phi$.
\deleted{In the following,} We now discuss suitable choices of $h$ for
normal and non-normal outcomes. \deleted{when the outcome is normally distributed
and when this is not the case.}

\paragraph{Continuous outcome}

If the outcome $\rY$ is normally distributed, a linear transformation
function $\h(\ry) = \eparm_1 + \eparm_2 y$  can be chosen, resulting in a classical
\added{normal} linear model.  
The basis functions and parameters in model~(\ref{eq:lincom}) are then equal to $\basisy(y) = (1, y)^\top$  and $\parm = (\eparm_1, \eparm_2)^\top$.  
The parameterizations
$Y \mid W = 0 \sim \ND(-{\eparm_1} \eparm_2^{-1}, \eparm_2^{-2})$ and 
$Y \mid W = 1 \sim \ND\bigl(-{(\eparm_1 - \tau)} \eparm_2^{-1}, \eparm_2^{-2}\bigr)$
lead to the marginal model 
	$F_w(\ry) = \Phi\left(\eparm_1 + \eparm_2 \ry - \tau w \right)$.
The treatment effect $\tau$ is defined as the standardized difference in means -- known as Cohen's $d$ -- because 
$\E(\rY \mid W = 1) - \E(\rY \mid W = 0) = \tau \eparm_2^{-1}$.
Cohen's $d$ is a noncollapsible effect measure:  
If more of the variance of $Y$ is explained by adding a prognostic covariate
$X_j$ to the model, the conditional effect is larger than the
marginal one.  For normal outcomes, a closed form expression of the
standard error of Cohen's $d$ in an unadjusted analysis can be derived.

\begin{lem}
	In the model $Y \mid W = w \sim \ND(-(\eparm_1 - \tau w) \eparm_2^{-1}, \eparm_2^{-2})$
	with Cohen's $d$ denoted as $\tau$, the unadjusted standard error for 
	$\tau$ in a balanced trial with total sample size $N$ is 
	$
	\text{SE}(\tau) = \sqrt{\frac{2}{N}\left(\frac{\tau^2}{4} + 2\right)}.
	$
	\label{lem:semarginal}
\end{lem}

The standard error for Cohen's $d$ obtained from an analysis 
adjusting with respect to a single normally distributed covariate
$X_1$ using the multivariate transformation model is also available in
closed form.

\begin{lem}
	If $Y \mid W = w \sim \ND(-(\eparm_{11} - \tau w) \eparm_{12}^{-1}, \eparm_{12}^{-2})$ 
	and $X_1 \sim \ND(-\eparm_{21} \eparm_{22}^{-1}, \eparm_{22}^{-2})$
	whose joint distribution is given by a Gaussian copula with correlation
	$\rho = -\nicefrac{\lambda}{\sqrt{1 + \lambda^2}}$, where $\lambda := \lambda_{21}$ is the single unconstrained copula parameter in (\ref{eq:mLambda}), the adjusted standard error for 
	$\tau$ in a balanced trial with total sample size $N$ is
	$\text{SE}(\tau, \lambda) =$
	\added{
	$ \sqrt{\frac{2}{N}\frac{(1 + \lambda^2)\tau^2 + 8}{4 (1 + \lambda^2)}}
	$}.\\
    \label{lem:semmlt}
\end{lem}
Proofs for both Lemmata are provided in \App~\ref{app:seestimate}. The
ratio of the squared standard errors can be interpreted as the fraction of
the sample size of the unadjusted analysis that is required 
to achieve the same power in an adjusted analysis. Relevant reductions of
more than $25\%$ can be gained \added{for a true Cohen's $d \approx 1.0$} when a single prognostic covariate with
correlation of $0.5$ \deleted{with the outcome} can be incorporated in the \NAMI 
procedure proposed here (\AppFigure~\ref{fig:contse} in \App~\ref{app:seestimate}). The standard error
$\text{SE}(\widehat{\tau})$ is not always larger than the standard error
$\text{SE}(\widehat{\tau}, \widehat{\lambda})$ because the unadjusted 
marginal and the adjusted marginal parameter estimates typically differ
slightly. However, $\text{SE}(\tau, \lambda)$ is monotonically decreasing
with increasing values of $\mid \lambda \mid$ and thus prognostic strength of
$X_1$.

The normal assumption is rarely met in practice and can be relaxed by
allowing more flexible transformation functions $\h$ and $\h_j$.
\cite{Hothorn_Moest_Buehlmann_2017} identified polynomials in Bernstein form of order $M$ 
as a suitable choice since they approximate any function over a closed interval for a sufficiently large $M$ according to 
Weierstrass' approximation theorem \citep{Farouki_2012}.
Under $G = \Phi$ and a flexible, potentially nonlinear $h$, the interpretation of $\tau$ is not in terms of Cohen's $d$ but as the mean difference on the latent normal scale. 
\deleted{To obtain an effect interpretation that is more intuitive,}
Transforming $\tau$ to a probabilistic index 
$\Prob(Y_0 < Y_1) = \Phi(\nicefrac{\tau}{\sqrt{2}})$, where $Y_0$ is defined as the outcome under $W = 0$ and $Y_1$ as
an independent outcome under $W = 1$, \added{allows a more intuitive
interpretation.}

For the multivariate transformation model, no further transformation of $\rY$ in the standard normal world is required since we assume that this already happened in the marginal model using $h$. 
The conditional distribution of $\rY$ given $W$ and continuous $\rX$ is then given by 
\begin{equation}
	\Prob(\rY \le \ry \mid W = w, \rX = \rx) = \Phi\left[\sum_{j = 1}^{J-1} \omega_{Jj} \h_j(x_j) + \omega_{JJ}\Bigl\{\h(y) - \tau w\Bigr\}\right].
	\label{eq:condcont}
\end{equation}
By design, the treatment effect $\tau$ is collapsible in this conditional
model: Integrating over covariates $X_1, \dots, X_{J - 1}$ via the joint
Gaussian copula model produces the marginal model~(\ref{eq:marginal}) featuring $\tau$.
The conditional distribution function~(\ref{eq:condcont}) is identical to the one implemented by the linear
transformation model~(\ref{eq:ltm}) when all covariates are jointly
normal (with linear transformation functions $\h_j, j = 1, \dots, J - 1$);
the noncollapsible conditional treatment effect is given by $\tau_\rx = \omega_{JJ}
\times \tau$.  \deleted{In addition, the partially linear transformation
model~(\ref{eq:condcont}) suggests
an approach to model criticism. Setting $\omega_{Jj} = 1, j = 1, \dots, J$
and relaxing the monotonicity assumption on $\h_j, j = 1, \dots, J - 1$ (but not
on $\h$) makes the model estimable by additive transformation models
\mbox{\citep{tamasi_2025}}. Violations of monotonicity of the estimated smooth functions $\widehat{\h}_j, j = 1, \dots,
J - 1$ suggests lack of copula model fit \mbox{\citep{dette_2014}}.}

\paragraph{Binary and ordinal outcome}

For {discrete, ordered outcomes} $\samY = \Bigl\{y_1 < \dots < y_K \Bigr\}$, a natural choice of $G$ is the
inverse logit link function $\expit$, that is, the cumulative distribution
function of a standard logistic distribution.
The transformation function $\h(y_k) = \eparm_k$ is parameterized as a step function with steps at $y_k, k = 1, ..., K-1$ and 
$\eparm_K = \infty$. 
The parameterization of model~(\ref{eq:lincom}) is then $\parm = (\eparm_1, ..., \eparm_{K-1})^\top$ and 
$\basisy(y_k) = e_{K-1}(k)$, where $e_{K-1}(k)$ is a unit vector of length $K-1$, with its $k$th element being one.
For $K = 2$, this results in a binary logistic regression model
\begin{equation}
	\Prob(\rY \le \ry_1 \mid W = w) = \Prob(\rY = \ry_1 \mid W = w) = \expit\left(\eparm_1 - \tau w \right).
	\label{eq:logreg}
\end{equation}
For $K >2$, the proportional odds model
$
\Prob(\rY \le \ry_k \mid W = w) = \expit\left(\eparm_k - \tau w \right)
$
emerges.
In both models, $\tau$ is interpretable as a log-odds ratio.
It has long been known \citep{mckelvey_statistical_1975}, that the log-odds
ratio is noncollapsible: the effect estimate $\tau$ is influenced by the error variance such that conditioning on additional prognostic 
covariates (by adding a linear predictor $\widetilde{\rx}^\top \shiftparm$ as in
model~(\ref{eq:ltm})) results in biased estimates $\widehat{\tau}_\rx$ for $\tau$.

The conditional distribution for a binary $\rY$ given both $W$ and continuous $\rX$
is then 
\begin{eqnarray}
	\Prob(\rY \le \ry_k \mid W = w, \rX = \rx) = \Phi\left[\sum_{j = 1}^{J-1} \omega_{Jj} \h_j(x_j) + \omega_{JJ} \Phi^{-1}\Bigl\{\expit(\eparm_k - \tau w)\Bigr\}\right].
	\label{eq:condbin}
\end{eqnarray}
This model is mutually exclusive with binary logistic
regression (the linear transformation model~(\ref{eq:ltm})) in the sense that only one of them can be correct. The log-odds ratio $\tau$
in model~(\ref{eq:condbin}) is, by design, collapsible because its corresponding marginal model for $\rY$ given treatment is the simple binary logistic
regression in (\ref{eq:logreg}) featuring log-odds ratio $\tau$.

\paragraph{Survival outcome}

For a {time to event outcome} $Y \in \RR^{+}$ 
the choice $G = \cloglog^{-1}$ makes $\tau$ a log-hazard
ratio in the marginal proportional hazards model
\begin{equation}
	\Prob(Y \le \ry \mid W = w) = \cloglog^{-1}\bigl(\h(\ry) - \tau w\bigr)
	\label{eq:margsurv}
\end{equation} 
with corresponding survivor function $\exp(-\exp(\h(\ry) - \tau w))$ and
thus cumulative hazard function $\exp(\h(\ry)) / \exp(\tau w)$.
Different parameterizations of $\h(y) = \basisy(y)^\top \parm$ in \eqref{eq:margsurv} give rise to
different survival models.  The transformation function $\h(y) = \eparm_1 +
\eparm_2 \log(y)$ based on $\basisy(y) = \left(1, \log(y)\right)^\top$ and
$\parm = (\eparm_1, \eparm_2)^\top$ corresponds to a Weibull proportional
hazard model.  A nonlinear baseline log-cumulative hazard function $h$,
parameterized in terms of a polynomial in Bernstein form $\basisy(y)^\top
\parm$, gives rise to a fully parameterized proportional hazards model.  It
is known that the log-hazard ratio in the Cox model is noncollapsible
\citep[see for example,][]{martinussen_collapsibility_2013,aalen_does_2015,sjlander_etal_2016},
i.e., $\tau$ in a marginal Cox model differs in its
interpretation to $\tau_{\rx}$ in the conditional Cox model~(\ref{eq:ltm}) featuring $G = \cloglog^{-1}$.
The derived conditional model then reads
\begin{equation}
	\Prob(\rY \le \ry \mid W = w, \rX = \rx) = \Phi\left[\sum_{j = 1}^{J-1} \omega_{Jj} \h_j(x_j) + \omega_{JJ}
	\Phi^{-1}\Bigl\{\cloglog^{-1}(\h(\ry) - \tau w)\Bigr\}\right].
	\label{eq:condsurv}
\end{equation}
The parameter $\tau$ can be interpreted as a marginal log-hazard ratio comparing $W =
0$ to $W = 1$, that is, (\ref{eq:condsurv}) can be interpreted as a
collapsible version of the proportional hazards model where the
proportionality assumption only applies to the hazard ratio $\exp(\tau)$ for
the treatment but it does not apply to the effect of
covariates. Again, this model is mutually exclusive with a conditional proportional
hazards model (the linear transformation model~(\ref{eq:ltm})) in the sense that only one of them can be correct.



\subsection{Model diagnostics}
\label{sec:diagnostics}

\added{Model~(\ref{eq:joinedmultivar}) makes certain assumptions, all of which can be
criticized in light of data. The model does not make assumptions regarding the
marginal covariate distributions~(\ref{eq:marginalx}) because we can
write $h_j(x_j) = \Phi^{-1}(\Prob(X_j \le x_j))$ for any cumulative
distribution function $\Prob(X_j \le x_j)$. In contrast, the marginal
outcome model~(\ref{eq:marginal}) assumes a shift effect $\tau w$ on the scale of
$G$. This assumption can be checked by plotting treatment-specific
distribution functions transformed by $G^{-1}$. Checking proportional hazards by comparing 
Kaplan-Meier curves on the log-log scale is such a strategy.
The partially linear transformation
models~(\ref{eq:condcont}), (\ref{eq:condbin}), and (\ref{eq:condsurv}) suggest
an approach to criticize the copula structure. Setting $\omega_{Jj} = 1, j = 1, \dots, J$
and relaxing the monotonicity assumption on $\h_j, j = 1, \dots, J - 1$ (but not
on $\h$) makes these models estimable by additive transformation models
\mbox{\citep{tamasi_2025}}. Violations of monotonicity of the estimated smooth functions $\widehat{\h}_j, j = 1, \dots,
J - 1$ suggests lack of copula model fit \mbox{\citep{dette_2014}}. However,
one would assume to observe a monotone relationship between a
well-established prognostic score and outcome. In the lack thereof, combining covariates into
a score first, using independent data, and subsequent adjustment
to this novel score might be beneficial
\mbox{\citep{Schuler_Walsh_Hall_2022}}. Omitting informative covariates in
the model only affects the standard error of $\widehat{\tau}$ because the model
is closed under marginalization.}

\subsection{Parameterization and inference}
\label{sec:covariates}
We parameterize the monotone nondecreasing transformation function $\h_j, j = 1, \dots, J-1$ in the marginal model~(\ref{eq:marginalx})
for covariates $\rX$ from arbitrary sample spaces $\samX_j$ in the same way as explained for the marginal
transformation function $\h$.
The practical choices of $h$ for $\rY$ discussed in Section~\ref{subsec:outcometypes} 
also apply to $\rX$ reflected in the
transformation $\h_j(x_j) = \basisy_j(x_j)^\top \parm_j,$ allowing for
continuous, binary, categorical ordered and survival covariates.  \deleted{Unlike the
treatment effect $\tau$, whose interpretation is governed by the choice of
$G$, the probit link in the marginal cumulative distribution
functions~(\ref{eq:marginalx}) does not impose a restriction, because we can
write $h_j(x_j) = \Phi^{-1}(\Prob(X_j \le x_j))$ for any cumulative
distribution function $\Prob(X_j \le x_j)$.}




Model~(\ref{eq:joinedmultivar}) is fully specified by the parameter vector 
$
\Theta = \left( \parm_1, ..., \parm_{J-1}, \parm, \tau, \lambdavec \right)^\top.
$
Vectors $\parm_j, j = 1, ..., J-1$ reflect the parameters of the
transformation functions $h_j$ in the marginal model~(\ref{eq:marginalx}) 
for covariates $\rX$, $\parm$ and $\tau$ are the parameters of the
marginal model for $\rY$ where $\parm$ reflect the parameters for the
transformation function $h$ of $\rY$ and $\tau$ is the marginal effect
of interest from model~(\ref{eq:marginal}).  The vector $\lambdavec$
includes the copula parameters $\mLambda$ in (\ref{eq:mLambda}).

The log-likelihood and score functions for absolutely continuous variables
$\rX, \rY$, for discrete variables $\rX, \rY$ and for mixed discrete and
continuous observations are available from \cite{Hothorn_2024}. Except for
some special cases (Cohen's $d$ with marginally normal covariates), where the
negative log-likelihood function is convex in $\Theta$, the arising
optimization problems are either biconvex or nonconvex. The marginally
estimated parameters $\parm_1, \dots, \parm_{J-1}$ and $(\parm, \tau)$ provide
good starting values for simultaneous maximum likelihood estimation.
Exact and approximate algorithms for parameter estimation are presented in \cite{Hothorn_2024}.
Maximum likelihood standard errors for all parameters, especially for
$\tau$, are computed by inverting the observed Fisher information. Wald tests and confidence
intervals rely on the asymptotic normality of the maximum likelihood
estimators \citep{Klein_Hothorn_Barbanti_2020}.

\added{Outcome and covariates might be missing at random. In such a case, the
log-likelihood contribution is obtained by marginalizing out the missing
variables. For example, if all covariates are missing for one observation,
the log-likelihood contribution for this subject is simply the 
log-likelihood of the marginal model~(\ref{eq:marginal}).}
\added{Dependent censoring for survival outcomes can be incorporated by adding a
time-to-censoring covariate $X_1$ in addition to the time-to-event outcome
$\rY$. \mbox{\cite{Deresa_Keilegom_2023}} establish a copula model for time-to-event and
time-to-censoring, including a Gaussian copula under certain restrictions
for the time-to-censoring model (such as a log-linear transformation
function $\h_1$), which is a special case of
model~(\ref{eq:joinedmultivar}).} 

\section{Empirical evaluation}
\label{sec:simulation}

In in-silico experiments, we empirically investigated \NAMI (NAMI) with
respect to the following research questions: 
	(\textit{RQ~1}) ``Does NAMI produce unbiased estimates of the true marginal effect $\tau$?'';
	(\textit{RQ~2}) ``Does NAMI lead to reduced standard errors and increased power?'';
    (\textit{RQ~3}) ``How do prognostic strengths of covariates influence the performance of NAMI?'';
	(\textit{RQ~4}) ``How sensitive is NAMI to a larger number of noise covariables?'';
	\added{\textit{(RQ~5)} ``How is NAMI affected by misspecifications of the marginal model or copula structure?''}.
To answer these questions, we simulated data with known marginal effect $\tau$ for diverse outcome types
(Cohen's $d$ for normally distributed outcomes, log-odds ratios for binary
outcomes, and log-hazard ratios for survival outcomes) from a correctly
specified model~(\ref{eq:joinedmultivar}) and under model misspecification 
(\App~\ref{app:misspec}).

\paragraph{Experimental setup}


A treatment indicator $W \sim B(1, 0.5)$ reflected a balanced RCT and 
a potentially
prognostic covariate $X_1$ was $\chi^2_5$ distributed.
The outcome $\rY$ was generated according to conditional distribution
functions $F(\ry \mid W = w, \rX = \rx)$:
	\begin{eqnarray*}
        \begin{cases}
	    \text{normal:} &  \Phi\Bigl[\omega_{21} h_1(x_1) + \omega_{22} \left(\eparm_1 + \eparm_2 \ry - \tau w\right)\Bigr], \quad \ry \in \RR \\
            \text{binary:} &  \Phi\Bigl[\omega_{21} h_1(x_1) + \omega_{22} \Phi^{-1}\{\expit(\eparm_1 - \tau w)\}\Bigr], \quad \ry \in \{0, 1\} \\
	    \text{survival:} &  \Phi\Bigl[\omega_{21} h_1(x_1) + \omega_{22} \Phi^{-1}\{\cloglog^{-1}(\eparm_1 + \eparm_2 \log(\ry) - \tau w)\}\Bigr], \quad \ry \in \RR^+.
	\end{cases}
\end{eqnarray*}
We use $\eparm_1 = 0$ and $\eparm_2 = 1$ in all models, such that the first model defines a normally distributed $\rY \sim \ND(w \tau, 1)$, the second a binomial distributed 
$\rY$, and the third a Weibull distributed $\rY$.
For the Weibull distributed $\rY$, we additionally added \added{\textit{independent}} right-censoring of varying 
degrees (see in \App~\ref{app:simulation}).


The true marginal effect was $\tau = 0.5$ or $\tau = 0$. 
All simulations were performed under different values of
$\lambda$
reflecting scenarios with absent, weak, moderate and strong prognostic effects of
$X_1$ on $Y$. 
Specifically, $\lambda \in \{0, -0.314, -0.750, - 2.065\}$ represented correlations $\rho \in \{0, 0.3, 0.6, 0.9\}$ 
via the conversion formula $\lambda = \omega_{21} =
-\exp(\nicefrac{\logit(\rho^2)}{2})$ and $\omega_{22} = \sqrt{\lambda^2 + 1}$.
To study RQ 4, we additionally sampled correlated $t$ 
distributed covariates $X_2, ..., X_P$ \added{independently of $X_1$ and $\rY$}, such that the overall number of covariates was $P =
\{1, 5, 15\}$ (including $X_1$).
\added{For RQ 5, we focused on two scenarios: (M1) misspecification of the marginal model in case of a survival outcome and (M2) misspecification of the copula structure in case of a binary outcome.
For M1, we generate data based on the $\Gamma$-frailty model by \mbox{\cite{aalen_does_2015}} which 
features a time-dependent hazard if $\tau \neq 0$.
For M2, we let the prognostic effect of a normally distributed $X_1$ on a binary $\rY$ be quadratic.
Details are given in \App~\ref{app:misspec}.
}



The sample size was set to achieve 60\% power for testing $H_0: \tau = 0$ in an unadjusted marginal analysis under $\tau = 0.5$, resulting in $N = 82$ (continuous), $N = 322$ (binary), and $N = 262$ (survival) observations.
Details on the sample size calculations are given in \App~\ref{ap:samplesize}. 
For all experiments, we used \replaced{$10,000$}{$5000$} simulation replications.


Given the generated data, we estimated the parameter $\tau$ with NAMI and computed its standard
error as well as the $p$-value of a $5\%$ Wald test against $H_0: \tau = 0$. 
For all outcome types, we compared NAMI to an unadjusted marginal inference (MI) 
model ignoring all covariates and to a noncollapsible linear transformation model~(\ref{eq:ltm}) (LTM) estimating $\tau_\rx$ along
with regression coefficients in the linear predictor $\widetilde{\rx}^\top \shiftparm$.
\replaced{
For binary outcomes, we compared NAMI against the standardization approach of \cite{Zhang_Tsiatis_2008} (YSTD)
and targeted maximum likelihood estimation (TMLE) of \cite{vanderlaan_tmle_2006}  with known propensity scores and a logistic outcome model.
For M2 we also obtained results from a TMLE model where the logistic regression model is replaced by boosting (TMLEXGB).
For survival outcomes, we additionally obtained results from the standardization approach 
of \cite{Lu_Tsiatis_2008} (YSTD) and its recent extension \mbox{\citep[LRCL,][]{Ye_Shao_Yi_2024}}.
}{
We also obtained results from 
the standardisation approaches of \mbox{\cite{Zhang_Tsiatis_2008}} and \mbox{\cite{Lu_Tsiatis_2008}} 
for binary and survival outcomes (YSTD) and its recent extension for survival
\mbox{\citep[LRCL,][]{Ye_Shao_Yi_2024}}. 
} 
To the best of our knowledge, no method exists for direct estimation of marginal Cohen's $d$ with covariate adjustment.
%
%
%
%
Computational details are given in 
\if1\ARXIV
{Sections~\ref{sec:software}}
\else
{\App~\ref{sec:software}}
\fi
and~\ref{app:simulation}.


We compare the estimation procedures based on the distribution of the estimated
treatment effects of $W$, i.e., $\widehat{\tau}_\rx$
for linear transformation models and $\widehat{\tau}$ for all other methods, as well as the distribution of the corresponding standard
errors and $p$-values for a Wald test against $H_0: \tau = 0$.
We also estimated the power for $\tau = 0.5$ and the empirical size for $\tau = 0$.

\begin{figure}[t!]
\centering

\includegraphics[width=.9\textwidth]{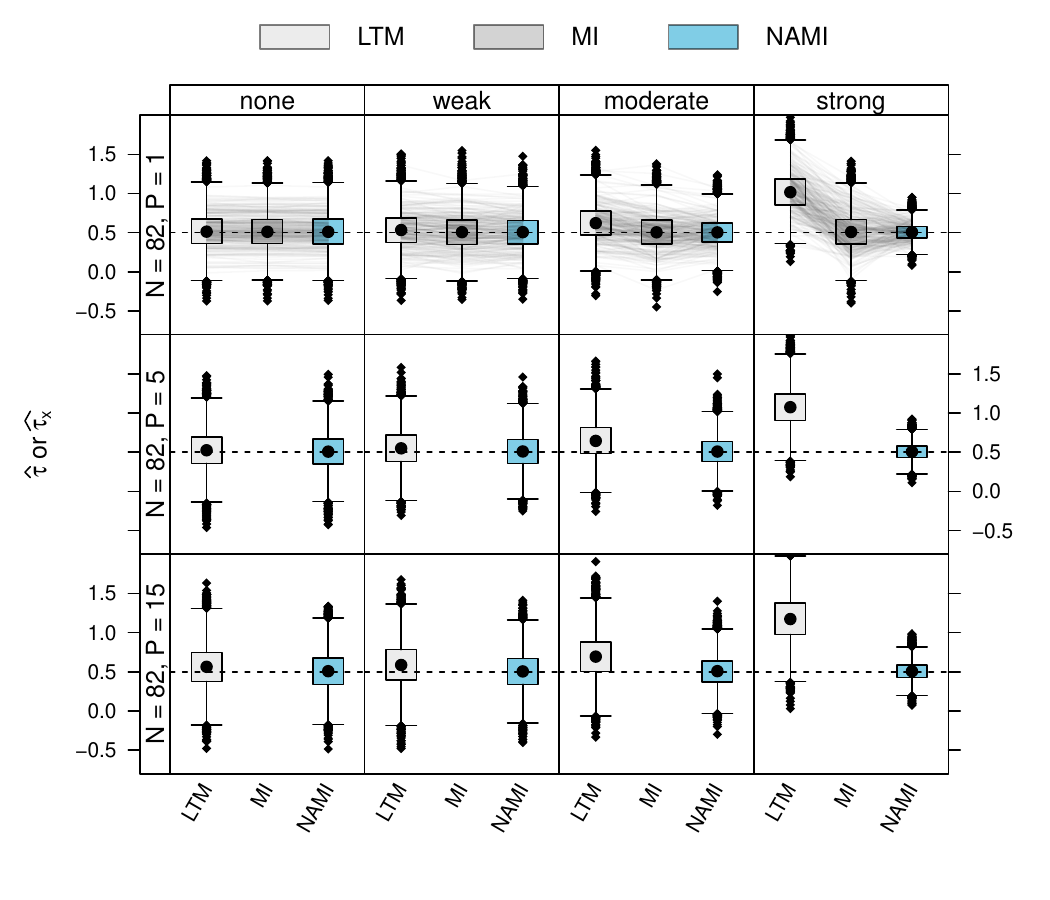} 

\caption{Empirical experiments for normally distributed outcome under $\tau = 0.5$ (dashed lines): Distribution of treatment effect estimates
$\widehat{\tau}$ of Cohen's $d$ obtained from unadjusted marginal inference (MI) and \NAMI (NAMI), and
effect estimates of $\widehat{\tau}_\rx$ by linear transformation models (LTM) 
under varying prognostic strengths of covariate $X_1$ (in columns) and increasing number of noise covariates
($P$, in rows). \label{fig:conttau05}}
\end{figure}

\paragraph{Results}
\added{Detailed simulation results are presented in
\App~\ref{app:simulation} and we illustrate general patterns based on the
distribution of treatment effect estimates for continuous outcomes presented 
in Figure~\ref{fig:conttau05}. Under non- or only weakly informative
prognostic covariates, marginal, conditional, and adjusted effect estimates
performed practically identical, regardless of the number of noise variables
in the model. In the presence of a moderate and strong prognostic covariate,
the bias expected due to noncollapsibility in the conditional estimate became
obvious. Compared to the marginal approach ignoring covariates, the
variability of NAMI estimates was smaller. This general pattern could also be
observed for binary and survival outcomes under difference censoring
schemes.}

\added{Regarding the empirical sizes in Table~\ref{tab:size}, only the marginal
approach and TMLE maintained the nominal level of $5\%$ in all scenarios
while all other forms of covariate adjustment exhibited size distortions of
different degrees. All conditional models were misspecified, and the
liberality of, for example, the proportional hazards model (with empirical size up
to $10\%$) can be explained by this fact. The liberality of YSTD, LRCL, and
NAMI were comparable and increased with the number of noise covariates. For
NAMI, this behavior can be explained by lack of quality of the normal
approximation to the distribution of the Wald statistic, because the
standard errors were in line with the standard deviation of the NAMI
estimates (\AppFigure~\ref{fig:contvcov00} and
\AppFigure~\ref{fig:contvcov05}), however, QQ-plots
(\AppFigure~\ref{fig:qq00}) indicated a lack of normality in the
tails. For larger sample sizes, the nominal level was maintained
(\AppTable~\ref{tab:sizeN400}).}
\added{The power gains obtained from leveraging prognostic information
(Table~\ref{tab:power}) were comparable between the different adjustment
procedures. All procedures roughly were in line with the planned $60\%$
power in absence of any covariate information. However, even in the presence of one
moderately strong prognostic factor, it was possible to increase the power
to $80\%$ for the continuous outcome, but not for binary or survival data.}

\afterpage{\clearpage}

\begin{table}[th!]
\caption{Empirical experiments: Empirical size for different outcome types obtained from linear transformation models (LTM), 
unadjusted marginal inference (MI), and \NAMI (NAMI)
under varying prognostic strength of covariate $X_1$ (in columns) and varying number of (noise) covariates ($P$, in rows).
For binary outcomes, analysis was also performed for the
standardization approach of \cite{Zhang_Tsiatis_2008} (YSTD) and for the targeted maximum likelihood estimator by \cite{vanderlaan_tmle_2006} (TMLE).
For survival outcomes, also results from \cite{Lu_Tsiatis_2008} (YSTD),
and \cite{Ye_Shao_Yi_2024} (LRCL) were obtained.
For survival outcomes, results are only shown for \textit{heavy} censoring. \added{All methods model the continuous outcome by a linear model with treatment effects as Cohen's $d$; the binary outcome by a logistic regression model with log-odds ratios as treatment effects and the survival outcome by a Cox proportional hazard model with log-hazard ratios as treatment effects.} 
\label{tab:size}}
\def\arraystretch{1}
\centering
\begin{tabular}{lllrrrr}
\hline
&&&\multicolumn{4}{c}{Size}\\
\cline{4-7}
DGP&Algorithm&P&\multicolumn{1}{c}{none}&\multicolumn{1}{c}{weak}&\multicolumn{1}{c}{moderate}&\multicolumn{1}{c}{strong}\\
\hline
continuous&MI&P = 1&0.053&     &     &     \\
&LTM&P = 1&0.058&0.055&0.054&0.051\\
&&P = 5&0.059&0.062&0.061&0.060\\
&&P = 15&0.083&0.083&0.079&0.084\\
&NAMI&P = 1&0.057&0.054&0.056&0.056\\
&&P = 5&0.060&0.061&0.065&0.061\\
&&P = 15&0.089&0.088&0.087&0.087\\[3pt]
binary&MI&P = 1&0.049&     &     &     \\
&LTM&P = 1&0.049&0.051&0.050&0.055\\
&&P = 5&0.051&0.052&0.051&0.052\\
&&P = 15&0.059&0.052&0.056&0.061\\
&NAMI&P = 1&0.049&0.053&0.050&0.058\\
&&P = 5&0.051&0.051&0.052&0.053\\
&&P = 15&0.061&0.053&0.058&0.066\\
&YSTD&P = 1&0.049&0.052&0.052&0.058\\
&&P = 5&0.052&0.054&0.053&0.054\\
&&P = 15&0.060&0.052&0.058&0.070\\
&TMLE&P = 1&0.047&0.050&0.050&0.055\\
&&P = 5&0.048&0.050&0.050&0.048\\
&&P = 15&0.053&0.045&0.050&0.049\\[3pt]
survival&MI&P = 1&0.052&     &     &     \\
&LTM&P = 1&0.052&0.056&0.068&0.112\\
&&P = 5&0.056&0.054&0.069&0.104\\
&&P = 15&0.067&0.069&0.090&0.113\\
&NAMI&P = 1&0.053&0.054&0.055&0.060\\
&&P = 5&0.056&0.051&0.056&0.059\\
&&P = 15&0.060&0.065&0.069&0.073\\
&LRCL&P = 1&0.049&0.050&0.051&0.053\\
&&P = 5&0.056&0.052&0.056&0.059\\
&&P = 15&0.064&0.068&0.069&0.076\\
&YSTD&P = 1&0.051&0.053&0.050&0.045\\
&&P = 5&0.062&0.060&0.061&0.072\\
&&P = 15&0.085&0.083&0.086&0.088\\
\hline
\end{tabular}
\end{table}

\begin{table}
\caption{Empirical experiments: Estimated power for different outcome types
for Wald tests obtained from linear transformation models (LTM), 
unadjusted marginal inference (MI), and \NAMI (NAMI) under varying prognostic strength of covariate $X_1$ (in columns) and varying number of (noise) covariates ($P$, in rows). 
For binary outcomes, analysis was also performed for the
standardization approach of \cite{Zhang_Tsiatis_2008} (YSTD) and for the targeted maximum likelihood estimator by \cite{vanderlaan_tmle_2006} (TMLE).
For survival outcomes, also results from \cite{Lu_Tsiatis_2008} (YSTD),
and \cite{Ye_Shao_Yi_2024} (LRCL) were obtained.
For survival outcomes, results are only shown for \textit{heavy} censoring. 
\added{All methods model the continuous outcome by a linear model with treatment effects as Cohen's $d$; the binary outcome by a logistic regression model with log-odds ratios as treatment effects and the survival outcome by a Cox proportional hazard model with log-hazard ratios as treatment effects.}
 \label{tab:power}}
\centering
\begin{tabular}{lllrrrr}
\hline
&&&\multicolumn{4}{c}{Power}\\
\cline{4-7}
DGP&Algorithm&P&\multicolumn{1}{c}{none}&\multicolumn{1}{c}{weak}&\multicolumn{1}{c}{moderate}&\multicolumn{1}{c}{strong}\\
\hline
continuous&MI&P = 1&0.624&     &     &     \\
&LTM&P = 1&0.625&0.659&0.787&0.994\\
&&P = 5&0.616&0.652&0.789&0.995\\
&&P = 15&0.610&0.637&0.770&0.991\\
&NAMI&P = 1&0.625&0.661&0.801&1.000\\
&&P = 5&0.613&0.657&0.804&0.999\\
&&P = 15&0.610&0.640&0.780&0.998\\[3pt]
binary&MI&P = 1&0.588&     &     &     \\
&LTM&P = 1&0.589&0.622&0.711&0.938\\
&&P = 5&0.594&0.606&0.706&0.931\\
&&P = 15&0.604&0.626&0.711&0.932\\
&NAMI&P = 1&0.589&0.625&0.724&0.944\\
&&P = 5&0.594&0.610&0.718&0.939\\
&&P = 15&0.603&0.630&0.718&0.940\\
&YSTD&P = 1&0.591&0.625&0.717&0.941\\
&&P = 5&0.599&0.614&0.715&0.938\\
&&P = 15&0.616&0.637&0.726&0.942\\
&TMLE&P = 1&0.585&0.619&0.712&0.937\\
&&P = 5&0.583&0.598&0.701&0.927\\
&&P = 15&0.579&0.603&0.691&0.916\\[3pt]
survival&MI&P = 1&0.595&     &     &     \\
&LTM&P = 1&0.596&0.622&0.722&0.960\\
&&P = 5&0.595&0.609&0.705&0.956\\
&&P = 15&0.583&0.612&0.701&0.948\\
&NAMI&P = 1&0.600&0.627&0.738&0.980\\
&&P = 5&0.600&0.620&0.718&0.977\\
&&P = 15&0.587&0.620&0.726&0.972\\
&LRCL&P = 1&0.576&0.592&0.643&0.732\\
&&P = 5&0.583&0.592&0.645&0.759\\
&&P = 15&0.585&0.617&0.664&0.769\\
&YSTD&P = 1&0.599&0.603&0.612&0.629\\
&&P = 5&0.627&0.633&0.648&0.721\\
&&P = 15&0.655&0.669&0.670&0.734\\
\hline
\end{tabular}

\end{table}
 
\added{Model misspecification in form of data sampled from a correctly specified conditional model (M1) led to conservatism for NAMI while MI and LRCL, although marginally misspecified, maintained the
level correctly. However, NAMI showed higher power than LRCL (\AppFigure~\ref{fig:misspec-pval}). An incorrectly specified copula (M2),
due to a quadratic impact of $X_1$ on $\rY$, could not be handled
correctly by any of the procedures under test. All parameter estimates were
identical to the marginal one, including TMLE with boosting
(\AppFigure~\ref{fig:misspecbin-pval}).}

\section{Applications}
\label{sec:application}

We present three applications \added{in detail} \deleted{of \NAMI for Cohen's $d$ in an equivalence trial, 
for odds ratios in $2 \times 2$ tables, and for hazard ratios in survival
analysis} in the \texttt{NAMI} vignette of the \pkg{tram} package and only report
main findings here.


\paragraph{Continuous outcome: Immunotoxicity study on Chloramine}
\label{subsec:appl-cont}

\deleted{Data on the effect of Chlor\-amine-dosed water on the weight of female mice
were reported as part of the 
\href{https://tools.niehs.nih.gov/cebs3/views/index.cfm?action=main.dataReview&bin_id=1179}{National
Toxicology Program}.
Repeated measurements were conducted on
days $1$, $8$, $15$, $22$, and $29$ in five dose groups ($0$, $2$, $10$,
$20$, $100$ mg/kg). We focus on the comparison of the highest dose ($W = 1, N_1 =
40$) with the control group ($W = 0, N_0 = 40$) with respect to outcome $Y$, the weight
at day $29$. The weight on day $1$ is used as the only covariate $X_1$. 
We test the equivalence hypothesis ``\emph{no} effect of Chloramine on
weight'' formulated in terms of Cohen's $d$ as the marginal treatment effect
$\tau$. The two-sided alternative states that $\tau$ is in
the equivalence interval $(-\delta, \delta)$ with corresponding 
$H_0: |\tau| \ge \delta$. We
reject $H_0$ at $5\%$ when the $95\%$ confidence interval for Cohen's
$d$ is completely contained in the equivalence interval. Because Cohen's $d$
does not depend on the measurement scale of the outcome, recommendations for
$\delta$ exist \mbox{\citep[with $\delta = 0.36$ we follow Table 1.1 given
by][acknowledging this oversimplification in our choice of $\delta$]{wellek_equiv_2010}}.
The unadjusted estimate of Cohen's $d$ is 
$\widehat{\tau} = 0.048$. The standard errors computed from the
observed Fisher information and the expected Fisher information
(Lemma~\ref{lem:semarginal}) evaluated
at $\widehat{\tau}$ are identical ($\text{SE}(\widehat{\tau}) = 0.224$)
and result in a $95\%$ Wald interval $( -0.390 , 0.486 )$. 
The unadjusted analysis therefore does not lead to a rejection of $H_0$.}

\added{Repeated measurements on the effect of Chlor\-amine-dosed water on the weight of female mice
were conducted on days $1$, $8$, $15$, $22$, and $29$ in five dose groups ($0$, $2$, $10$,
$20$, $100$ mg/kg). We compared the highest dose ($W = 1, N_1 =
40$) with the control group ($W = 0, N_0 = 40$) with respect to outcome $Y$, the weight
at day $29$, using weight on day $1$ as the only covariate $X_1$. 
We tested the equivalence hypothesis ``\emph{no} effect of Chloramine on
weight'' formulated in terms of Cohen's $d$ being in
the equivalence interval $(-\delta, \delta)$ with corresponding 
$H_0: |\tau| \ge \delta$. With $\delta = 0.36$
\mbox{\citep[Table 1.1,][]{wellek_equiv_2010}} one can 
reject $H_0$ at $5\%$ when the $95\%$ confidence interval for Cohen's
$d$ is completely contained in the equivalence interval. 
The unadjusted estimate of Cohen's $d$ was 
$\widehat{\tau} = 0.048$. The standard errors computed from the
observed and expected Fisher information 
(Lemma~\ref{lem:semarginal}) evaluated
at $\widehat{\tau}$ were identical ($\text{SE}(\widehat{\tau}) = 0.224$)
and resulted in a $95\%$ Wald interval $( -0.390 , 0.486 )$ and thus lack
of evidence against $H_0$.
Adjusting for weight at baseline in \NAMI (with $\h_1$ in Bernstein form), we
obtained
$\widehat{\tau} = -0.002$. The standard error (observed and
expected evaluated at $\widehat{\tau}$ and $\widehat{\lambda} = -0.800$)
of $0.175$ led to the Wald interval
$( -0.344 , 0.341 )$. This interval was completely contained in the
equivalence interval and thus the absence of an effect of Chloramine on
weight could be inferred. 
The reduction in standard error came from the high association between
the outcome $\rY$ and the covariate $X_1$ ($\rho(\widehat{\lambda}) = 0.625$).
The coefficient of determination \replaced{$R^2 = 0.390$}{ $R^2 = 0.512$} 
suggested an improvement of the conditional over the marginal model.}


\paragraph{Binary outcome: Efficacy study on new combined chemotherapy}

\deleted{\mbox{\cite{roedel_oxaliplatin_2012}} analysed the pathological complete response
in rectal cancer patients as an early
endpoint comparing Fluorouracil-based standard of care ($W = 0$, $N_0 =
623$) with a combination therapy adding
Oxaliplatin ($W = 1$, $N_1 = 613$). The binary outcome $Y$ was 
defined by the absence of viable tumour cells in the primary tumour
and lymph nodes after surgery. \mbox{\cite{roedel_oxaliplatin_2012}} reported an odds
ratio of $1.40$ with $95\%$ confidence interval $(1.02, 1.92)$ based on a
Cochran-Mantel-Haenszel $\chi^2$ test 
stratified for lymph node involvement 
(positive vs.~negative) and clinical T category (1–3 vs.~4). 
For data from the completed trial, 
the unadjusted marginal log-odds ratio estimated by a binary logistic regression is
$0.352$ with corresponding $95\%$ Wald interval
$( 0.036 , 0.668 )$.}

\added{Analysing the pathological complete response in rectal cancer patients as an early
endpoint comparing Fluorouracil-based standard of care ($W = 0$, $N_0 =
623$) with a combination therapy adding
Oxaliplatin ($W = 1$, $N_1 = 613$), with a binary outcome 
defined by the absence of viable tumor cells in the primary tumor
and lymph nodes after surgery, resulted in an 
unadjusted marginal log-odds ratio of
$0.352$ with corresponding $95\%$ Wald interval $( 0.036 , 0.668 )$.}

\added{When adjusting for six potentially prognostic covariates (age, sex, ECOG
performance status, distance to the anal verge of the tumor and the two
stratum variables lymph node involvement and clinical T category, with in
total $62$ missing values not requiring exclusion from the estimation
procedure, see Section~\ref{sec:covariates}), the
marginal log-odds ratio $0.352$ with $95\%$
Wald interval $( 0.036 , 0.667 )$ was almost identical to the unadjusted
result. Adjusting for covariates did neither improve fit ($R^2 =
0.030$) nor precision, however,
adding six variables carrying little information also did not increase the
standard error.}

\paragraph{Survival outcome: Longevity study of male fruit flies}

\deleted{\mbox{\cite{partridge_flies_1981}} assessed whether sexual
activity affects the lifespan of male fruit flies.
A total of $N = 125$ flies were randomly
divided into five groups of $25$: males forced to live alone, males assigned
to live with one or eight receptive females, and males assigned to live with
one or eight nonreceptive females.  For the sake of simplicity, we 
focus on the analysis of the two groups with
eight female flies added that were either all nonreceptive ($W = 0$) or
receptive ($W = 1$).  The primary outcome $Y$ was the survival time of 
male flies in days.  The thorax length $X_1$ of the male flies was also
measured -- a covariate that is strongly associated with longevity.}

\deleted{The marginal Cox proportional hazards model for time to death, with baseline log-cumulative
hazard function $\h$ in Bernstein form of order six, defines the 
marginal treatment effect as log-hazard ratio $\tau$. For thorax length, $\h_1$
was also parameterised in Bernstein form with order six.
The joint distribution of both variables was expressed by a Gaussian
copula.}

\added{We compared time to death between 
two groups of male flies living in company of 
eight female flies that were either all nonreceptive ($W = 0, N_0 = 25$) or
receptive ($W = 1, N_1 = 25$) based on data from an experiment 
assessing whether sexual
activity affects the lifespan of male fruit flies.}

The unadjusted marginal log-hazard ratio was $\widehat{\tau} =
2.157$ with $95\%$ Wald interval $( 1.343 , 2.970 )$.  
Adjusting for thorax length, a covariate that is strongly associated with longevity,
the log-hazard ratio was $2.048$, with shorter
$95\%$ Wald interval $( 1.431 , 2.666 )$.  The coefficient of
determination $R^2 = 0.671$ 
indicated that thorax length is highly prognostic. 

\begin{figure}[ht]
\begin{knitrout}
\definecolor{shadecolor}{rgb}{0.969, 0.969, 0.969}\color{fgcolor}
\includegraphics[width=1\linewidth]{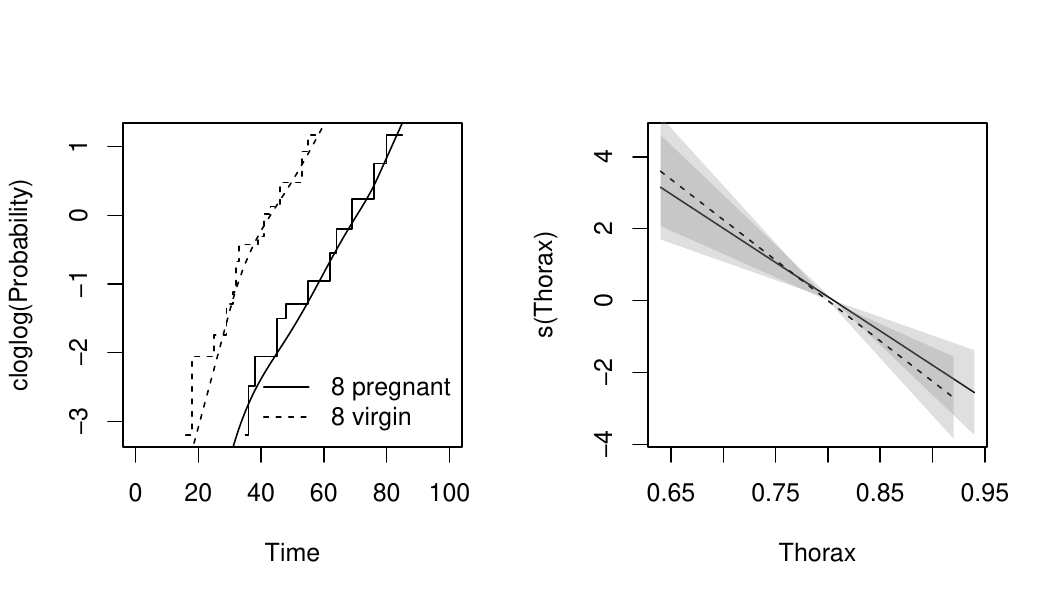} 
\end{knitrout}
\caption{Model diagnosis plots for fruit flies data. Left: Empirical cumulative distribution
functions for time-to-death, separately estimated for the two groups, on the
complementary log-log scale, overlayed with $\widehat{h}(y) - \widehat{\tau} w$
corresponding to the marginal Cox model.
Right: Smooth spline functions $s$ of thorax length for both groups estimated using a
conditional additive transformation model.
Monotonicity does not indicate lack of fit of the copula model.
\label{fig:surv_diag}}
\end{figure}

\added{Figure~\ref{fig:surv_diag} presents a diagnostic assessment as
suggested in Section~\ref{sec:diagnostics}. Assuming proportional hazards
seemed justified because the empirical cumulative distribution functions were
parallel on the complementary log-log scale. Regressing time-to-death on
a smooth unconstrained function of thorax length in model~(\ref{eq:condsurv})
did not suggest violations of monotonicity and thus the assumed copula
structure.}



\section{Discussion}
\label{sec:discussion} 

Nonparanormal adjusted marginal inference introduces covariate adjustment for the
estimation of noncollap\-sible marginal treatment effect parameters \added{for diverse outcome types}, most
importantly of standardized differences in means,
odds ratios and hazard ratios, under censoring and missingness. 
The resulting marginal treatment effect estimates are less
variable than their unadjusted counterparts when relevant prognostic
information is available in baseline covariates. 
The effects can be compared between different studies, for example for meta-analyses, in research syntheses, or in replication studies.
\added{Due to its parametric nature, \NAMI offers a ranking of covariates according to their prognostic strengths.}
The combination of marginal and copula models can be criticized in light of
data. Section~\ref{sec:diagnostics} suggests strategies for model criticism
and diagnosis which we applied for the analysis of the fruit flies data.

\deleted{The reduced standard errors might help to design smaller trials -- 
the sample size reduction factors presented in
\AppFigure~\ref{fig:contse}, for example, demonstrate the potential of \NAMI
to reduce the necessary sample size in equivalence studies.
However, a high level of evidence regarding the strength of prognostic variables is
necessary a priori.}
\added{One referee raised the question whether one would really be willing to gamble on possible power gains during sample size planning. The modest gains in
theoretical precision (\AppFigure~\ref{fig:contse}) and empirical power
(Table~\ref{tab:power}) combined with liberality (Table~\ref{tab:size}, all
procedures except TMLE)
and theoretical findings in idealized settings \mbox{\citep{siegfried_relevance_2023}} indeed suggest that a high level of evidence regarding the strength of,
preferably one or a few, prognostic variables would be necessary a priori.
In a trial powered for an unadjusted analysis, NAMI might be a helpful
parametric approach to compensate for power losses, due to noncompliance or other
issues. The empirical experiments suggest that
\NAMI performed either on par (in the binary setting) or outperformed
established semiparametric adjusted marginal inference procedures. Although
the latter procedures have been derived in theoretically general terms
\mbox{\citep{Zhang_Tsiatis_2008}}, implementations are currently tailored to
specific cases (log-odds and log-hazard ratios). 
The reference
implementation of \NAMI allows the estimation of marginal treatment effects for
general transformation models for continuous and discrete outcomes under
several forms of censoring and missingness. Thus, \NAMI provides an interesting alternative to the already established methods discussed in 
\mbox{\cite{van_lancker_covariate_2024}}.}

\deleted{Classical model fit criteria can be applied to the marginal models,
for example a plot of $\cloglog$-transformed Kaplan-Meier curves for
assessing parallelism when the treatment effect was defined as a log-hazard
ratio.  Conceptually, one could refrain from defining a scalar treatment
effect in the marginal model by fitting two separate transformations, and thus distribution functions $F_w(\ry) = \Phi(w \h(\ry \mid 1) + (1 - w)
\h(\ry \mid 0))$, to the outcomes in both treatment groups \mbox{\citep[similar in spirit
to][]{kennedy_2023}}, maybe followed by an omnibus test based on a Smirnov
statistic $\sup_{\ry \in \samY} \mid F_1(\ry) - F_0(\ry) \mid$. 
Additive transformation models provide an alternative way of
estimating the novel partially linear conditional transformation
models~(\ref{eq:condcont}), (\ref{eq:condbin}), or (\ref{eq:condsurv}) where lack
of monotonicity of the covariate effects indicates a lack of fit in the
Gaussian copula structure. 
Parameter estimation in \NAMI can be sensitive to model misspecification. 
\App~\ref{app:misspec} demonstrates this for a survival outcome where \NAMI
depends on a misspecified marginal Cox proportional hazards model.  Because
\NAMI ensures that the treatment effect is collapsible, omitted variables do
not affect the marginal effect.  Also goodness of fit remains unaffected: If
the Gaussian copula structure is appropriate given all covariates, it can be
assumed to fit well for any subset of covariates.}

\deleted{We evaluated our method based on its parameter estimates and standard
errors, and we assessed power and size. These experiments suggest that
\NAMI performed either on par (in the binary setting) or outperformed
established semiparametric adjusted marginal inference procedures. Although
the latter procedures have been derived in theoretically general terms
\mbox{\citep{Zhang_Tsiatis_2008}}, currently only implementations tailored to
specific cases (log-odds and log-hazard ratios) are available. The reference
implementation of \NAMI allows estimation of marginal treatment effects for
general transformation models for continuous and discrete outcomes under
several forms of censoring. Thus, \NAMI provides an interesting alternative
to these standard methods discussed in 
\mbox{\cite{van_lancker_covariate_2024}}.}

\deleted{With a reference software implementation
being available in the \href{https://doi.org/10.32614/CRAN.package.tram}{\pkg{tram}}
add-on package to the \proglang{R} system
for statistical computing, \NAMI awaits further scrutiny by practitioners.}

\section{Computational details}
\label{sec:software}

All computations were performed using
\proglang{R}~version 4.5.0  \citep{R}.
A reference implementation of marginal and multivariate transformation models is available
in the \proglang{R}~add-on package \pkg{tram} \citep{pkg:tram}.
The semiparametric standardisation approach of \cite{Zhang_Tsiatis_2008} and \cite{Lu_Tsiatis_2008} for binary and survival outcomes (referred to as YSTD in Section~\ref{sec:simulation}) 
is implemented in the \pkg{speff2trial} package \citep{pkg:speff2trial}. 
The approach of \cite{Ye_Shao_Yi_2024} for survival outcomes (referred to as LRCL in Section~\ref{sec:simulation}) is available in \pkg{RobinCar} \citep{pkg:RobinCar}.
The targeted maximum likelihood approach for binary outcomes in available in the \pkg{tmle} package \citep{pkg:tmle}.
The simulation study was run in parallel with the \pkg{batchtools} package \citep{pkg:batchtools}.
The code to reproduce the results discussed in Section~\ref{sec:simulation} can be found at
\url{https://gitlab.uzh.ch/susanne.dandl1/marginal_noncollapsibility}.
Performing \NAMI in \textsf{R} is relatively straightforward. The core of the Chloramine analysis
in Section~\ref{subsec:appl-cont} is
\begin{knitrout}
\definecolor{shadecolor}{rgb}{0.969, 0.969, 0.969}\color{fgcolor}\begin{kframe}
\begin{alltt}
\hlkwd{library}\hldef{(}\hlsng{"tram"}\hldef{)}
\hlcom{## marginal normal model feat. Cohen's d; unadjusted marginal inference}
\hlkwd{confint}\hldef{(m0} \hlkwb{<-} \hlkwd{Lm}\hldef{(y} \hlopt{~} \hldef{w,} \hlkwc{data} \hldef{= d))}
\hlcom{#      2.5 %   97.5 %}
\hlcom{# -0.3901557 0.486494}
\hldef{m1} \hlkwb{<-} \hlkwd{BoxCox}\hldef{(x} \hlopt{~} \hlnum{1}\hldef{,} \hlkwc{data} \hldef{= d)} \hlcom{## marginal model for baseline weight}
\hlkwd{confint}\hldef{(}\hlkwd{mmlt}\hldef{(m0, m1,} \hlkwc{formula} \hldef{=} \hlopt{~} \hlnum{1}\hldef{,} \hlkwc{data} \hldef{= d))} \hlcom{## adjusted marginal inference}
\hlcom{#      2.5 %     97.5 % }
\hlcom{# -0.3442240  0.3407635 }
\end{alltt}
\end{kframe}
\end{knitrout}
The complete analyses presented in Section~\ref{sec:application} are
reproducible from within \textsf{R} via:
\begin{knitrout}
\definecolor{shadecolor}{rgb}{0.969, 0.969, 0.969}\color{fgcolor}\begin{kframe}
\begin{alltt}
\hlkwd{library}\hldef{(}\hlsng{"tram"}\hldef{)}
\hlkwd{vignette}\hldef{(}\hlsng{"NAMI"}\hldef{,} \hlkwc{package} \hldef{=} \hlsng{"tram"}\hldef{)}
\end{alltt}
\end{kframe}
\end{knitrout}
This code also demonstrates appropriateness of marginal and copula model assumptions for the
datasets via additive transformation models \citep{tamasi_2025}.
 
\paragraph{Acknowledgements}

Financial support by Swiss National Science Foundation, grant number
200021\_219384, is acknowledged.

\bibliography{NAMI,packages}

\newpage

\renewcommand\thefigure{S.~\arabic{figure}}    
\renewcommand\thetable{S.~\arabic{table}}
\renewcommand\thesection{A.\arabic{section}}
\setcounter{figure}{0}
\setcounter{table}{0}
\setcounter{section}{0}

\appendix

\section{Derivation of standard errors}
\label{app:seestimate}

The following provides the proofs for Lemma~\ref{lem:semarginal} and Lemma~\ref{lem:semmlt} of Section~\ref{subsec:outcometypes}.

\subsection{Proof of Lemma 1}
\label{subsec:semarginal}

Consider a normally distributed outcome $\rY \mid W = w \sim \ND(-(\eparm_1 + \tau (w - 0.5)) \eparm_2^{-1}, 
\eparm_2^{-2})$ with binary treatment indicator $W \in \{0, 1\}$, 
$W \sim B(1, 0.5)$.
Compared to the linear marginal model
$
	F_w(\ry) = \Phi\left(\eparm_1 + \eparm_2 \ry - \tau w \right),
$
we model $\rY$ with a positive instead of a negative shift term
and a centered treatment indicator without loss of generality. The
conditional distribution function reads
\begin{equation*}
F(\ry \mid W = w) = \Phi\left(\eparm_1 + \eparm_2 \ry + \tau (w - 0.5) \right).
\end{equation*}
We have $\Ex(\rY \mid W = 0) = \frac{0.5\,\tau-\eparm_{1}}{\eparm_{2}}$,
$\Ex(\rY \mid W = 1) = \frac{-\eparm_{1}-0.5\,\tau}{\eparm_{2}}$, $\Ex(\rY^2
\mid W = 0) = \eparm_2^{-2} + \Ex(\rY \mid W = 0)^2$ and $\Ex(\rY^2
\mid W = 1) = \eparm_2^{-2} + \Ex(\rY \mid W = 1)^2$.
In the following, we define 
$\basisy(\rY) = (1, \rY, w - 0.5)^\top$, $\parm = (\eparm_{1}, \eparm_{2}, \tau)^\top$ such that 
$
\basisy(\rY)^\top \parm = \eparm_1 + \eparm_2 \rY + \tau (w - 0.5).
$
The derivative of the basis function is $\basisy^\prime(\rY) = (0, 1,
0)^\top$.
Consequently, the log-likelihood contribution of a single observation $(\rY, w)$ is 
\begin{eqnarray*}
	\ell(\parm; \rY, w) & = & \log\left[\phi\{\basisy(\rY)^\top\parm\} \cdot \basisy^\prime(\rY)^{\top}\parm\right] \\
	& \propto & -\frac{1}{2}\Bigl(\eparm_1 + \eparm_2 \rY + \tau (w - 0.5)\Bigr)^2 + \log(\eparm_2).
\end{eqnarray*}

The negative Hessian for a control subject is 
\begin{eqnarray*}
\calH_0 = 	-\dfrac{\partial^2 \ell(\parm; \rY, 0)}{\partial^2 \parm} = 
\begin{pmatrix}
 1 & Y & -\frac{1}{2} \\ 
 Y & \frac{\rY^2\, \eparm_{2}^2+1}{\eparm_{2}^2} & -\frac{Y}{2} \\
 -\frac{1}{2} & -\frac{\rY}{2} & \frac{1}{4} \\ 
\end{pmatrix}
\end{eqnarray*}
with corresponding expected Fisher information 
\begin{eqnarray*}
\mH_0 := \Ex(\calH_0) = \begin{pmatrix}
1 & -\frac{2\,\eparm_{1}-\tau}{2\,\eparm_{2}} & -\frac{1}{2} \\
-\frac{2\,\eparm_{1}-\tau}{2\,\eparm_{2}} & \frac{4\,\eparm_{1}^2-4\,\tau\,\eparm_{1}+\tau^2+8}{4\,\eparm_{2}^2} & \frac{2\,\eparm_{1}-\tau}{4\,\eparm_{2}} \\
-\frac{1}{2} & \frac{2\,\eparm_{1}-\tau}{4\,\eparm_{2}} & \frac{1}{4} \\
\end{pmatrix}.
\end{eqnarray*}

For a subject in the treated group, we have
\begin{eqnarray*}
\calH_1 =      -\dfrac{\partial^2 \ell(\parm; \rY, 1)}{\partial^2 \parm} =
\begin{pmatrix}
1 & \rY & \frac{1}{2} \\ 
\rY & \frac{\rY^2\,\eparm_{2}^2+1}{\eparm_{2}^2} & \frac{\rY}{2} \\ 
\frac{1}{2} & \frac{\rY}{2} & \frac{1}{4} \\ 
\end{pmatrix}
\end{eqnarray*}
with expected Fisher information
\begin{eqnarray*}
\mH_1 := \Ex(\calH_1) = \begin{pmatrix}
1 & -\frac{2\,\eparm_{1}+\tau}{2\,\eparm_{2}} & \frac{1}{2} \\
-\frac{2\,\eparm_{1}+\tau}{2\,\eparm_{2}} & \frac{4\,\eparm_{1}^2+4\,\tau\,\eparm_{1}+\tau^2+8}{4\,\eparm_{2}^2} & -\frac{2\,\eparm_{1}+\tau}{4\,\eparm_{2}} \\
\frac{1}{2} & -\frac{2\,\eparm_{1}+\tau}{4\,\eparm_{2}} & \frac{1}{4} \\
\end{pmatrix}.
\end{eqnarray*}
The expected Fisher information $\mH = \mH_0 + \mH_1$ is then
\begin{eqnarray*}
\begin{pmatrix}
2 & -\frac{2\,\eparm_{1}}{\eparm_{2}} & 0 \\
-\frac{2\,\eparm_{1}}{\eparm_{2}} & \frac{4\,\eparm_{1}^2+\tau^2+8}{2\,\eparm_{2}^2} & -\frac{\tau}{2\,\eparm_{2}} \\
0 & -\frac{\tau}{2\,\eparm_{2}} & \frac{1}{2} \\
\end{pmatrix}
\end{eqnarray*}
with inverse
\begin{eqnarray*}
\mH^{-1} = \begin{pmatrix}
\frac{\eparm_{1}^2+2}{4} & \frac{\eparm_{1}\,\eparm_{2}}{4} & \frac{\tau\,\eparm_{1}}{4} \\ 
\frac{\eparm_{1}\,\eparm_{2}}{4} & \frac{\eparm_{2}^2}{4} & \frac{\tau\,\eparm_{2}}{4} \\ 
\frac{\tau\,\eparm_{1}}{4} & \frac{\tau\,\eparm_{2}}{4} & \frac{\tau^2+8}{4} \\
\end{pmatrix}
\end{eqnarray*}
leading to a standard error of $\text{SE}(\tau) =
\sqrt{\frac{2}{N}\left(\frac{\tau^2}{4} + 2\right)}$ for a balanced trial with total sample
size $N$.

\qed

\subsection{Proof of Lemma 2}

Consider a normally distributed outcome $\rY \mid W = w \sim \ND(-(\eparm_1 - \tau (w - 0.5))/\eparm_2, 1/\eparm_2^2)$ with binary treatment indicator $W \in \{0, 1\}$, 
$W \sim B(1, 0.5)$, and a normally distributed covariate $X \sim \ND(-\eparm_{21}/\eparm_{22}, (1 + \lambda^2) / \eparm_{22}^2)$
with covariance $-\lambda / (\eparm_{12} \eparm_{22})$, in other words, the model
\begin{eqnarray*}
\mLambda(\lambda)^{-1} \begin{pmatrix}
\eparm_{11} + \eparm_{12} \rY + \tau (w - 0.5) \\
\eparm_{21} + \eparm_{22} X
\end{pmatrix} \sim \ND_2(\nullvec_2, \mI_2).
\end{eqnarray*}
Because we are interested in $\tau$ only, it is sufficient to work 
with $\mLambda$ rather than $\mOmega$ because only
the interpretation of the parameters $\eparm_{21}$ and $\eparm_{22}$ would
be affected \citep{Hothorn_2024}.

With 
$$\mSigma = \mLambda^{-1} \mLambda^{-\top} = 
\begin{pmatrix}[1]
1 & -\lambda \\
-\lambda & 1 + \lambda^2 
\end{pmatrix}$$
we obtain $\Ex(\rY \mid W = 0) =
\frac{0.5\,\tau-\eparm_{11}}{\eparm_{12}}$, $\Ex(\rY \mid W =
1) = \frac{-\eparm_{11}-0.5\,\tau}{\eparm_{12}}$, $\Ex(\rY^2
\mid W = 0) =
\frac{\left(0.5\,\tau-\eparm_{11}\right)^2}{\eparm_{12}^2}+\frac{1}{\eparm_{12}^2}$,
$\Ex(\rY^2 \mid W = 1) =
\frac{\left(-\eparm_{11}-0.5\,\tau\right)^2}{\eparm_{12}^2}+\frac{1}{\eparm_{12}^2}$.
Furthermore, we have $\Ex(X) = -\frac{\eparm_{21}}{\eparm_{22}}$,
$\Ex(X^2) =
\frac{\eparm_{21}^2}{\eparm_{22}^2}+\frac{\lambda^2+1}{\eparm_{22}^2}$, 
$\Cov(\rY, X) = -\frac{\lambda}{\eparm_{12}\,\eparm_{22}}$
and thus $\Ex(\rY X \mid W = 0) =
-\frac{\left(0.5\,\tau-\eparm_{11}\right)\,\eparm_{21}}{\eparm_{12}\,\eparm_{22}}-\frac{\lambda}{\eparm_{12}\,\eparm_{22}}$
and 
$\Ex(\rY X \mid W = 1) =
-\frac{\left(-\eparm_{11}-0.5\,\tau\right)\,\eparm_{21}}{\eparm_{12}\,\eparm_{22}}-\frac{\lambda}{\eparm_{12}\,\eparm_{22}}$.

The log-likelihood for a parameter vector $\Theta = (\eparm_{11},
\eparm_{12}, \tau, \eparm_{21}, \eparm_{22}, \lambda)^\top$ is
\begin{eqnarray*}
\ell(\Theta; \rY, X, w) & \propto & -\frac{1}{2}\Bigl(\eparm_{11} + \eparm_{12} \rY + \tau (w - 0.5)\Bigr)^2 + \log(\eparm_{12}) \\
& & -\frac{1}{2}\left[\lambda \Bigl\{\eparm_{11} + \eparm_{12} \rY + \tau (w - 0.5)\Bigr\} + \eparm_{21} + \eparm_{22} X\right]^2 +
\log(\eparm_{22}).
\end{eqnarray*}

For a single subject in the control or treated group, we obtain the observed and expected
Fisher information matrices, $\calH^{(0)}$, $\calH^{(1)}$ and $\mH_0$,
$\mH_1$ as
\begin{landscape}
\begin{eqnarray*}
& & \calH^{(0)} =  -\dfrac{\partial^2 \ell(\Theta; \rY, X, 0)}{\partial^2 \Theta} = \\
& & \begin{pmatrix}\lambda^2+1 & \rY\,\lambda^2+\rY & -\frac{\lambda^2+1}{2} & \lambda & X\,\lambda & X\,\eparm_{22}+\eparm_{21}+2\,\rY\,\lambda\,\eparm_{12}+2\,\lambda\,\eparm_{11}-\lambda\,\tau \\ 
\rY\,\lambda^2+\rY & \frac{\left(\rY^2\,\lambda^2+\rY^2\right)\,\eparm_{12}^2+1}{\eparm_{12}^2} & -\frac{\rY\,\lambda^2+\rY}{2} & \rY\,\lambda & X\,\rY\,\lambda & X\,\rY\,\eparm_{22}+\rY\,\eparm_{21}+2\,\rY^2\,\lambda\,\eparm_{12}+2\,\rY\,\lambda\,\eparm_{11}-\rY\,\lambda\,\tau \\
 -\frac{\lambda^2+1}{2} & -\frac{\rY\,\lambda^2+\rY}{2} & \frac{\lambda^2+1}{4} & -\frac{\lambda}{2} & -\frac{X\,\lambda}{2} & -\frac{X\,\eparm_{22}+\eparm_{21}+2\,\rY\,\lambda\,\eparm_{12}+2\,\lambda\,\eparm_{11}-\lambda\,\tau}{2} \\
 \lambda & \rY\,\lambda & -\frac{\lambda}{2} & 1 & X & \frac{2\,\rY\,\eparm_{12}+2\,\eparm_{11}-\tau}{2} \\
 X\,\lambda & X\,\rY\,\lambda & -\frac{X\,\lambda}{2} & X & \frac{X^2\,\eparm_{22}^2+1}{\eparm_{22}^2} & \frac{2\,X\,\rY\,\eparm_{12}+2\,X\,\eparm_{11}-X\,\tau}{2} \\
\calH^{(0)}_{61} & \calH^{(0)}_{62} & \calH^{(0)}_{63} & \calH^{(0)}_{64}  & \calH^{(0)}_{65} & \calH^{(0)}_{66} \\
\end{pmatrix} \\
& & \calH^{(0)}_{61}  =   X\,\eparm_{22}+\eparm_{21}+2\,\rY\,\lambda\,\eparm_{12}+2\,\lambda\,\eparm_{11}-\lambda\,\tau \\
& & \calH^{(0)}_{62}  =  X\,\rY\,\eparm_{22}+\rY\,\eparm_{21}+2\,\rY^2\,\lambda\,\eparm_{12}+2\,\rY\,\lambda\,\eparm_{11}-\rY\,\lambda\,\tau \\
& & \calH^{(0)}_{63}  =   -\frac{X\,\eparm_{22}+\eparm_{21}+2\,\rY\,\lambda\,\eparm_{12}+2\,\lambda\,\eparm_{11}-\lambda\,\tau}{2} \\
& & \calH^{(0)}_{64}  =   \frac{2\,\rY\,\eparm_{12}+2\,\eparm_{11}-\tau}{2} \\
& & \calH^{(0)}_{65}  =   \frac{2\,X\,\rY\,\eparm_{12}+2\,X\,\eparm_{11}-X\,\tau}{2} \\
& & \calH^{(0)}_{66}  =    \frac{4\,\rY^2\,\eparm_{12}^2+\left(8\,\rY\,\eparm_{11}-4\,\rY\,\tau\right)\,\eparm_{12}+4\,\eparm_{11}^2-4\,\tau\,\eparm_{11}+\tau^2}{4} \\
& & \mH_0 := \Ex(\calH^{(0)}) = \\
& & \begin{pmatrix}\lambda^2+1 & -\frac{\left(2\,\lambda^2+2\right)\,\eparm_{11}+\left(-\lambda^2-1\right)\,\tau}{2\,\eparm_{12}} & -\frac{\lambda^2+1}{2} & \lambda & -\frac{\lambda\,\eparm_{21}}{\eparm_{22}} & 0 \\ 
-\frac{\left(2\,\lambda^2+2\right)\,\eparm_{11}+\left(-\lambda^2-1\right)\,\tau}{2\,\eparm_{12}} & \frac{\left(4\,\lambda^2+4\right)\,\eparm_{11}^2+\left(-4\,\lambda^2-4\right)\,\tau\,\eparm_{11}+\left(\lambda^2+1\right)\,\tau^2+4\,\lambda^2+8}{4\,\eparm_{12}^2} & \frac{\left(2\,\lambda^2+2\right)\,\eparm_{11}+\left(-\lambda^2-1\right)\,\tau}{4\,\eparm_{12}} & -\frac{2\,\lambda\,\eparm_{11}-\lambda\,\tau}{2\,\eparm_{12}} & \frac{\left(2\,\lambda\,\eparm_{11}-\lambda\,\tau\right)\,\eparm_{21}-2\,\lambda^2}{2\,\eparm_{12}\,\eparm_{22}} & \frac{\lambda}{\eparm_{12}} \\
 -\frac{\lambda^2+1}{2} & \frac{\left(2\,\lambda^2+2\right)\,\eparm_{11}+\left(-\lambda^2-1\right)\,\tau}{4\,\eparm_{12}} & \frac{\lambda^2+1}{4} & -\frac{\lambda}{2} & \frac{\lambda\,\eparm_{21}}{2\,\eparm_{22}} & 0 \\
 \lambda & -\frac{2\,\lambda\,\eparm_{11}-\lambda\,\tau}{2\,\eparm_{12}} & -\frac{\lambda}{2} & 1 & -\frac{\eparm_{21}}{\eparm_{22}} & 0 \\
 -\frac{\lambda\,\eparm_{21}}{\eparm_{22}} & \frac{\left(2\,\lambda\,\eparm_{11}-\lambda\,\tau\right)\,\eparm_{21}-2\,\lambda^2}{2\,\eparm_{12}\,\eparm_{22}} & \frac{\lambda\,\eparm_{21}}{2\,\eparm_{22}} & -\frac{\eparm_{21}}{\eparm_{22}} & \frac{\eparm_{21}^2+\lambda^2+2}{\eparm_{22}^2} & -\frac{\lambda}{\eparm_{22}} \\
 0 & \frac{\lambda}{\eparm_{12}} & 0 & 0 & -\frac{\lambda}{\eparm_{22}} & 1 \\ 
\end{pmatrix}
\end{eqnarray*}
\end{landscape}

\begin{landscape}
\begin{eqnarray*}
& & \calH^{(1)} =  -\dfrac{\partial^2 \ell(\Theta; \rY, X, 1)}{\partial^2 \Theta} = \\
& & \begin{pmatrix}\lambda^2+1 & Y\,\lambda^2+Y & \frac{\lambda^2+1}{2} & \lambda & X\,\lambda & X\,\eparm_{22}+\eparm_{21}+2\,Y\,\lambda\,\eparm_{12}+2\,\lambda\,\eparm_{11}+\lambda\,\tau \\ 
Y\,\lambda^2+Y & \frac{\left(Y^2\,\lambda^2+Y^2\right)\,\eparm_{12}^2+1}{\eparm_{12}^2} & \frac{Y\,\lambda^2+Y}{2} & Y\,\lambda & X\,Y\,\lambda & X\,Y\,\eparm_{22}+Y\,\eparm_{21}+2\,Y^2\,\lambda\,\eparm_{12}+2\,Y\,\lambda\,\eparm_{11}+Y\,\lambda\,\tau \\
 \frac{\lambda^2+1}{2} & \frac{Y\,\lambda^2+Y}{2} & \frac{\lambda^2+1}{4} & \frac{\lambda}{2} & \frac{X\,\lambda}{2} & \frac{X\,\eparm_{22}+\eparm_{21}+2\,Y\,\lambda\,\eparm_{12}+2\,\lambda\,\eparm_{11}+\lambda\,\tau}{2} \\
 \lambda & Y\,\lambda & \frac{\lambda}{2} & 1 & X & \frac{2\,Y\,\eparm_{12}+2\,\eparm_{11}+\tau}{2} \\
 X\,\lambda & X\,Y\,\lambda & \frac{X\,\lambda}{2} & X & \frac{X^2\,\eparm_{22}^2+1}{\eparm_{22}^2} & \frac{2\,X\,Y\,\eparm_{12}+2\,X\,\eparm_{11}+X\,\tau}{2} \\
\calH^{(1)}_{61} & \calH^{(1)}_{62} & \calH^{(1)}_{63} & \calH^{(1)}_{64}  & \calH^{(1)}_{65} & \calH^{(1)}_{66} \\
\end{pmatrix} \\
& & \calH^{(1)}_{61}  =   X\,\eparm_{22}+\eparm_{21}+2\,Y\,\lambda\,\eparm_{12}+2\,\lambda\,\eparm_{11}+\lambda\,\tau \\
& & \calH^{(1)}_{62}  =  X\,Y\,\eparm_{22}+Y\,\eparm_{21}+2\,Y^2\,\lambda\,\eparm_{12}+2\,Y\,\lambda\,\eparm_{11}+Y\,\lambda\,\tau \\
& & \calH^{(1)}_{63}  =  \frac{X\,\eparm_{22}+\eparm_{21}+2\,Y\,\lambda\,\eparm_{12}+2\,\lambda\,\eparm_{11}+\lambda\,\tau}{2} \\
& & \calH^{(1)}_{64}  =  \frac{2\,Y\,\eparm_{12}+2\,\eparm_{11}+\tau}{2} \\
& & \calH^{(1)}_{65}  =  \frac{2\,X\,Y\,\eparm_{12}+2\,X\,\eparm_{11}+X\,\tau}{2} \\
& & \calH^{(1)}_{66}  =  \frac{4\,Y^2\,\eparm_{12}^2+\left(8\,Y\,\eparm_{11}+4\,Y\,\tau\right)\,\eparm_{12}+4\,\eparm_{11}^2+4\,\tau\,\eparm_{11}+\tau^2}{4} \\ 
& & \mH_1 := \Ex(\calH_1) = \\
& & \begin{pmatrix}\lambda^2+1 & -\frac{\left(2\,\lambda^2+2\right)\,\eparm_{11}+\left(\lambda^2+1\right)\,\tau}{2\,\eparm_{12}} & \frac{\lambda^2+1}{2} & \lambda & -\frac{\lambda\,\eparm_{21}}{\eparm_{22}} & 0 \\
 -\frac{\left(2\,\lambda^2+2\right)\,\eparm_{11}+\left(\lambda^2+1\right)\,\tau}{2\,\eparm_{12}} & \frac{\left(4\,\lambda^2+4\right)\,\eparm_{11}^2+\left(4\,\lambda^2+4\right)\,\tau\,\eparm_{11}+\left(\lambda^2+1\right)\,\tau^2+4\,\lambda^2+8}{4\,\eparm_{12}^2} & -\frac{\left(2\,\lambda^2+2\right)\,\eparm_{11}+\left(\lambda^2+1\right)\,\tau}{4\,\eparm_{12}} & -\frac{2\,\lambda\,\eparm_{11}+\lambda\,\tau}{2\,\eparm_{12}} & \frac{\left(2\,\lambda\,\eparm_{11}+\lambda\,\tau\right)\,\eparm_{21}-2\,\lambda^2}{2\,\eparm_{12}\,\eparm_{22}} & \frac{\lambda}{\eparm_{12}} \\
 \frac{\lambda^2+1}{2} & -\frac{\left(2\,\lambda^2+2\right)\,\eparm_{11}+\left(\lambda^2+1\right)\,\tau}{4\,\eparm_{12}} & \frac{\lambda^2+1}{4} & \frac{\lambda}{2} & -\frac{\lambda\,\eparm_{21}}{2\,\eparm_{22}} & 0 \\
 \lambda & -\frac{2\,\lambda\,\eparm_{11}+\lambda\,\tau}{2\,\eparm_{12}} & \frac{\lambda}{2} & 1 & -\frac{\eparm_{21}}{\eparm_{22}} & 0 \\
 -\frac{\lambda\,\eparm_{21}}{\eparm_{22}} & \frac{\left(2\,\lambda\,\eparm_{11}+\lambda\,\tau\right)\,\eparm_{21}-2\,\lambda^2}{2\,\eparm_{12}\,\eparm_{22}} & -\frac{\lambda\,\eparm_{21}}{2\,\eparm_{22}} & -\frac{\eparm_{21}}{\eparm_{22}} & \frac{\eparm_{21}^2+\lambda^2+2}{\eparm_{22}^2} & -\frac{\lambda}{\eparm_{22}} \\
 0 & \frac{\lambda}{\eparm_{12}} & 0 & 0 & -\frac{\lambda}{\eparm_{22}} & 1 \\ 
\end{pmatrix}
\end{eqnarray*}
\end{landscape}

The expected Fisher information from a pair of one control and one treated
subject is then
\begin{eqnarray*}
& & \mH  =  \mH_0 + \mH_1 = \\
& & \begin{pmatrix}
2\,\lambda^2+2 & -\frac{\left(2\,\lambda^2+2\right)\,\eparm_{11}}{\eparm_{12}} & 0 & 2\,\lambda & -\frac{2\,\lambda\,\eparm_{21}}{\eparm_{22}} & 0 \\ 
-\frac{\left(2\,\lambda^2+2\right)\,\eparm_{11}}{\eparm_{12}} & \frac{\left(4\,\lambda^2+4\right)\,\eparm_{11}^2+\left(\lambda^2+1\right)\,\tau^2+4\,\lambda^2+8}{2\,\eparm_{12}^2} & -\frac{\left(\lambda^2+1\right)\,\tau}{2\,\eparm_{12}} & -\frac{2\,\lambda\,\eparm_{11}}{\eparm_{12}} & \frac{2\,\lambda\,\eparm_{11}\,\eparm_{21}-2\,\lambda^2}{\eparm_{12}\,\eparm_{22}} & \frac{2\,\lambda}{\eparm_{12}} \\
 0 & -\frac{\left(\lambda^2+1\right)\,\tau}{2\,\eparm_{12}} & \frac{\lambda^2+1}{2} & 0 & 0 & 0 \\
 2\,\lambda & -\frac{2\,\lambda\,\eparm_{11}}{\eparm_{12}} & 0 & 2 & -\frac{2\,\eparm_{21}}{\eparm_{22}} & 0 \\
 -\frac{2\,\lambda\,\eparm_{21}}{\eparm_{22}} & \frac{2\,\lambda\,\eparm_{11}\,\eparm_{21}-2\,\lambda^2}{\eparm_{12}\,\eparm_{22}} & 0 & -\frac{2\,\eparm_{21}}{\eparm_{22}} & \frac{2\,\eparm_{21}^2+2\,\lambda^2+4}{\eparm_{22}^2} & -\frac{2\,\lambda}{\eparm_{22}} \\
 0 & \frac{2\,\lambda}{\eparm_{12}} & 0 & 0 & -\frac{2\,\lambda}{\eparm_{22}} & 2 \\ 
\end{pmatrix}.
\end{eqnarray*}

\begin{figure}[t]
\begin{center}

\includegraphics[width=.8\textwidth]{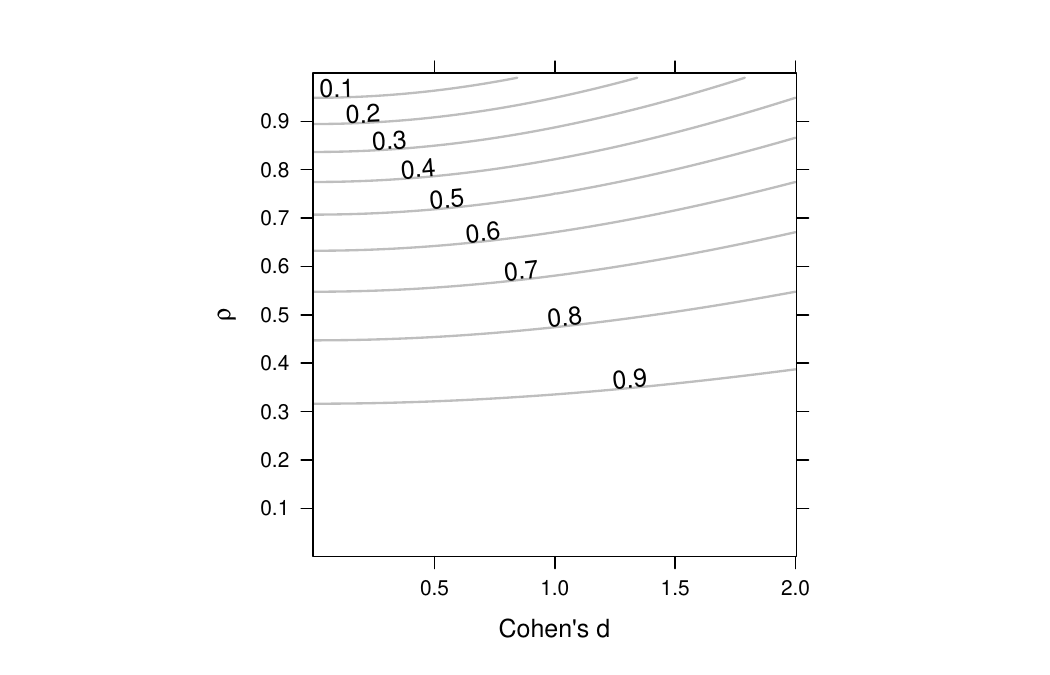} 

\caption{Theoretical fraction $\nicefrac{\text{SE}(\tau, \lambda)^2}{\text{SE}(\tau)^2}$
of the adjusted and unadjusted squared standard errors for
Cohen's $d$ (the treatment effect $\tau$) in a normal model with one prognostic
variable whose correlation to the outcome is given by $\rho =
-\nicefrac{\lambda}{\sqrt{1 + \lambda^2}}$.  The numbers
can be interpreted as the fraction of the sample size required in a trial
adjusting for prognostic information relative to the sample size required
for a trial without such an adjustment.  \label{fig:contse}}
\end{center}
\end{figure}

Partitioning this matrix in four $3 \times 3$ matrices ($\mA, \mB, \mB^\top, \mD$) and application of
the Schur complement, we obtain the first three rows and columns $\mA_1$ of the
inverse expected Fisher information as follows:
\begin{eqnarray*}
\mH  & = & \begin{pmatrix}
\mA & \mB \\
\mB^\top & \mD \\
\end{pmatrix} \\
\mD^{-1} & = & \begin{pmatrix}\frac{\eparm{21}^2+2}{4} & \frac{\eparm{21}\,\eparm{22}}{4} & \frac{\lambda\,\eparm{21}}{4} \\
 \frac{\eparm{21}\,\eparm{22}}{4} & \frac{\eparm{22}^2}{4} & \frac{\lambda\,\eparm{22}}{4} \\
 \frac{\lambda\,\eparm{21}}{4} & \frac{\lambda\,\eparm{22}}{4} & \frac{\lambda^2+2}{4} \\ 
\end{pmatrix} \\
\mH^{-1} & = & \begin{pmatrix}
\mA_1 & \mB_1 \\
\mB_1^\top & \mD_1 \\
\end{pmatrix} \\
\mA - \mB \mD^{-1} \mB^\top & = & 
\begin{pmatrix}2 & -\frac{2\,\eparm_{11}}{\eparm_{12}} & 0 \\ 
-\frac{2\,\eparm_{11}}{\eparm_{12}} & \frac{4\,\eparm_{11}^2+\left(\lambda^2+1\right)\,\tau^2+8}{2\,\eparm_{12}^2} & -\frac{\left(\lambda^2+1\right)\,\tau}{2\,\eparm_{12}} \\
 0 & -\frac{\left(\lambda^2+1\right)\,\tau}{2\,\eparm_{12}} & \frac{\lambda^2+1}{2} \\ 
\end{pmatrix} \\
\mA_1 & = & (\mA - \mB \mD^{-1} \mB^\top)^{-1} = 
\begin{pmatrix}\frac{\eparm_{11}^2+2}{4} & \frac{\eparm_{11}\,\eparm_{12}}{4} & \frac{\tau\,\eparm_{11}}{4} \\
 \frac{\eparm_{11}\,\eparm_{12}}{4} & \frac{\eparm_{12}^2}{4} & \frac{\tau\,\eparm_{12}}{4} \\
 \frac{\tau\,\eparm_{11}}{4} & \frac{\tau\,\eparm_{12}}{4} &
\frac{\left(\lambda^2+1\right)\,\tau^2+8}{4\,\lambda^2+4} \\ \end{pmatrix}. 
\end{eqnarray*} Thus, for a balanced trial with in total $N$ observations,
the standard error for $\tau$ as a function of $\tau$ and $\lambda$ is
$\text{SE}(\tau, \lambda) = \sqrt{\frac{2}{N}\frac{(1 + \lambda^2) \tau^2 +
8}{4 (1 + \lambda^2)}}$.  The squared fraction of the adjusted and
unadjusted standard errors is presented in \Figure~\ref{fig:contse}.

\qed

\newpage

\section{Simulation study}
\label{app:simulation}

\subsection{Sample size calculation}
\label{ap:samplesize}

We derived the sample size for unadjusted marginal tests against 
null hypothesis $H_0: \tau = 0$ vs. $H_1: \tau \neq 0$ for continuous (normally distributed), 
binary (binomially distributed) and survival (Weibull distributed) outcomes 
with true treatment effect $\tau = 0.5$, size $\alpha = 0.05$ and power $1 - \beta =
0.6$.

\subsubsection{Continuous outcome}
For continuous outcomes, the sample size calculation for each treatment group was based on 
$$
N_{1, \text{cont}} = N_{0, \text{cont}} = \left(\frac{\text{SE}(\tau)\left(z_{\beta} - z_{1-\tfrac{\alpha}{2}}\right)}{\tau}\right)^2 = 
\left(\frac{\text{SE}(0.5)\left(z_{0.4} - z_{0.975}\right)}{0.5}\right)^2 = 41
$$
for each control/treatment group.
The derivation is based on maximum likelihood theory, according to which the distribution of the maximum likelihood estimator $\widehat{\tau}$ is normally distributed with  
$$
\sqrt{N} \frac{\widehat{\tau}}{\text{SE}(\tau)} \sim \ND\left(\frac{\tau}{\text{SE}(\tau)}, \frac{1}{N}\right).
$$
Since we aim for equally sized treatment/control groups, it is suitable to derive $N_1 = N_0$ for a single group by obtaining the value via the one-sided null hypothesis $H_0: \tau \le 0$ vs. $H_1: \tau > 0$:
\begin{eqnarray*}
	 1 - \beta &=& 1 - \Prob\left(\frac{\widehat{\tau}}{\text{SE}(\tau)} \le \frac{z_{1-\frac{\alpha}{2}}}{\sqrt{N_1}}\right) \\
	 \beta &=& \Prob\left(\frac{\widehat{\tau}}{\text{SE}(\tau)} \le \frac{z_{1-\frac{\alpha}{2}}}{\sqrt{N_1}}\right) \\
	\beta &=& \Phi\left(\sqrt{N_1}\left(\frac{z_{1-\frac{\alpha}{2}}}{\sqrt{N_1}}- \frac{\tau}{\text{SE}(\tau)}\right)\right) \\
	\Phi^{-1}(\beta) - z_{1-\frac{\alpha}{2}} &=& -\sqrt{N_1}
	\frac{\tau}{\text{SE}(\tau)} \\
	N_1 &=& \left(\frac{\text{SE}(\tau) \left(\Phi^{-1} \left(\beta\right) - z_{1-\frac{\alpha}{2}}\right)}{\tau}\right)^2.
\end{eqnarray*}
The formula to compute the standard error for the setting without covariate adjustment $\text{SE}(\tau)$ is given in Lemma~\ref{lem:semarginal}.


\subsubsection{Binary outcome}
For binary outcomes, the sample calculation was based on Formula (4.14) in \cite{Fleiss_2003}
$$
N_{1, \text{binary}} = \frac{\left(z_{\frac{\alpha}{2}} \sqrt{(p_1 + p_2)\left(1-\frac{p_1 + p_2}{r+1}\right)} + z_{\beta}\sqrt{r p_1 (1-p_1) + p_2 (1-p_2)}\right)^2}{r(p_1 - p_2)^2},
$$
where $r$  is the proportion of treated to control patients (here, $1$), 
$p_1 = \tfrac{p_2 \exp(\tau)}{(1 + p_2 (\exp(\tau) - 1))}$ with $p_2$ as the proportion of controls with outcome $Y = 1$. A value of $p_2 = 0.5$ was obtained based on a small experiment with 100 observations, sampled from the above-described data-generating process for binary data.  
With $\tau = 0.5$, $1 - \beta = 0.6$ and $\alpha = 0.05$, we obtained a sample size of 
$
N_{1, \text{binary}} = 161
$
for each control/treatment group.

\subsubsection{Survival outcome}
The sample size calculation for survival outcomes was based on Formula (3) in \cite{Wu_2015}
$$
N_{1, \text{surv}} = \frac{(r + 1)^2}{r} \frac{(z_{1-\frac{\alpha}{2}} + z_{1 - \beta})^2}{\tau^2 (p_0 + r p_1)},
$$
where $r$ is the proportion of treated to control patient (here, $1$), $p_0$ is the noncensoring probability for a control observation and $p_1$ is the noncensoring probability for a treated observation. 
For a noncensoring probability of 0.3 -- a worst case scenario we indeed considered for the survival outcome --
we obtained a sample size of $N_{1, \text{surv}} = 131$ for each group (given $\tau = 0.5$, $1 - \beta = 0.6$ and $\alpha = 0.05$).


\subsection{Details on the study setup}
For the Weibull distributed $\rY$, we additionally added right-censoring by sampling censoring times $C$ from the conditional distribution function
$
\Prob(C \le c \mid W = w, \rX = \rx) =	\Phi\left[\omega_{21} h_1(x_1) + 
\omega_{22} \Phi^{-1}\bigl\{\cloglog^{-1}(\eparm_1 + \eparm_2 \log(c) - \gamma - \tau w)\bigr\}\right]
$
with $\eparm_1 = 0$ and $\eparm_2 = 1$.
An observation was not censored if $\rY < C$. 
The parameter $\gamma$ defines the probabilistic index, that is, the noncensoring
probability $P(T < C \mid W = W, \rX = \rx) =
\expit(\gamma)$ \citep[see~Table~1 in][]{Sewak_Hothorn_2023}.
In our experiments, we employed values of $\gamma$ corresponding to 
noncensoring probabilities of $0.3$ (heavy censoring) and $0.7$ (mild censoring).

NAMI and MI relied on the following, correctly
specified, marginal models for $\rY$: The linear
model $
F_w(\ry) = \Phi\left(\eparm_1 + \eparm_2 \ry - \tau w \right)
$
was used for continuous (normally distributed) outcomes; 
a logistic regression model~(\ref{eq:logreg}) for binary outcomes; 
and a Cox proportional hazards model for survival outcomes (i.e.,
model~(\ref{eq:margsurv}) with $h(y)$
parameterized by a polynomial in Bernstein form of order six).
For NAMI, the marginal distribution functions
$\Phi\bigl(\h_j(x_j)\bigr)$ of the $P = J - 1$ covariates were parametrized by a polynomial in Bernstein form of order six.


For NAMI, the marginal distribution functions
$\Phi(\h_j(x_j))$ of the $P = J - 1$ covariates were parametrized by a polynomial in Bernstein form of order six.
Thus, NAMI always relied on the misspecified marginal
model for $X_1$ and was overparameterized in the presence of noise
variables. 
Because only one out of $P = J - 1$ covariates was potentially
prognostic, only one out of $\nicefrac{J (J - 1)}{2}$ parameters in $\lambdavec$
was potentially nonzero. The LTM was
overparameterized in the setup with nonprognostic covariate $X_1$ (that is, with
$\shiftparm = \nullvec$) and misspecified in all other setups.
%

\subsection{Details on model misspecification}
\label{app:misspec}

We illustrate the effect of model misspecification in two scenarios:
when the marginal model is misspecified and when the copula structure is misspecified.

\subsubsection{Misspecification of the marginal model}

Model misspecification is demonstrated using the $\Gamma$-frailty model
by \cite{aalen_does_2015}. For one $\Gamma$-distributed covariate $X_1$, the
survival time $\rY$ is given by a frailty model being identical to a
conditional Weibull model:
\begin{eqnarray*}
	X_1 & \sim & \Gamma(\eta, \eta), \quad \Ex(X_1) = 1, \Var(X_1) = \eta^{-1} \\
	\Prob(\rY \le \ry \mid W = w, X_1 = x_1) & = & 
	1 - \exp\Bigl(-x_1 \exp(\eparm_1 + \eparm_2 \log(\ry) + \tau_\rx w)\Bigr) \\
	& = & \cloglog^{-1}\Bigl(\eparm_1 + \eparm_2 \log(\ry) + \tau_\rx w + \log(x_1)\Bigr).
\end{eqnarray*}

For conditional log-hazard ratios $\tau_\rx \neq 0$, 
the marginal distribution $\rY \mid W = w$ features a time-dependent hazard
ratio function which, as a function of the variance $\Var(X_1) = \eta^{-1}
\rightarrow \infty$ tends to one \citep{aalen_does_2015}. Thus, a marginal Cox proportional hazards
model $\cloglog^{-1}(\h(\ry) + \tau w)$ is misspecified.

For $\tau_\rx = 0$, the marginal Cox model $\cloglog^{-1}(\h(\ry) + \tau w)$ is
overparameterized, as $\tau = 0$. The joint distribution of $\h_1(X_1)$ and
$\h(\rY)$ can, however, not be expressed by a Gaussian copula. Therefore,
\NAMI is misspecified under both situations.

\subsubsection{Lack of monotonicity of the prognostic effect}

\added{
To simulate a setting where the prognostic effect is not monotone, 
we generate a standard normally distributed $X_1 \sim \ND(0, 1)$ that has a
quadratic effect on a binary $\rY$, i.e.
$$
	\Prob(\rY = 1 \mid W = w, X_1 = x_1) = \expit\left(0 - \tau_{\rx} w - x_1^2 \right).
$$
}

\added{
The Gaussian structure can only capture monotonicity of the prognostic effect,
thus, NAMI is misspecified in this setting.
Also the LTM and TMLE used to answer RQ 1 to 4 in Section~\ref{sec:simulation}
are misspecified because 
they assume linearity on the scale of the linear predictor. 
TMLE with a gradient-boosting-based outcome model (TMLEXGB) should operate under weaker assumptions.
}

\subsection{Results}

\paragraph{RQ 1: Unbiased estimation?}

\begin{figure}[th!]
\centering

\includegraphics[width=.9\textwidth]{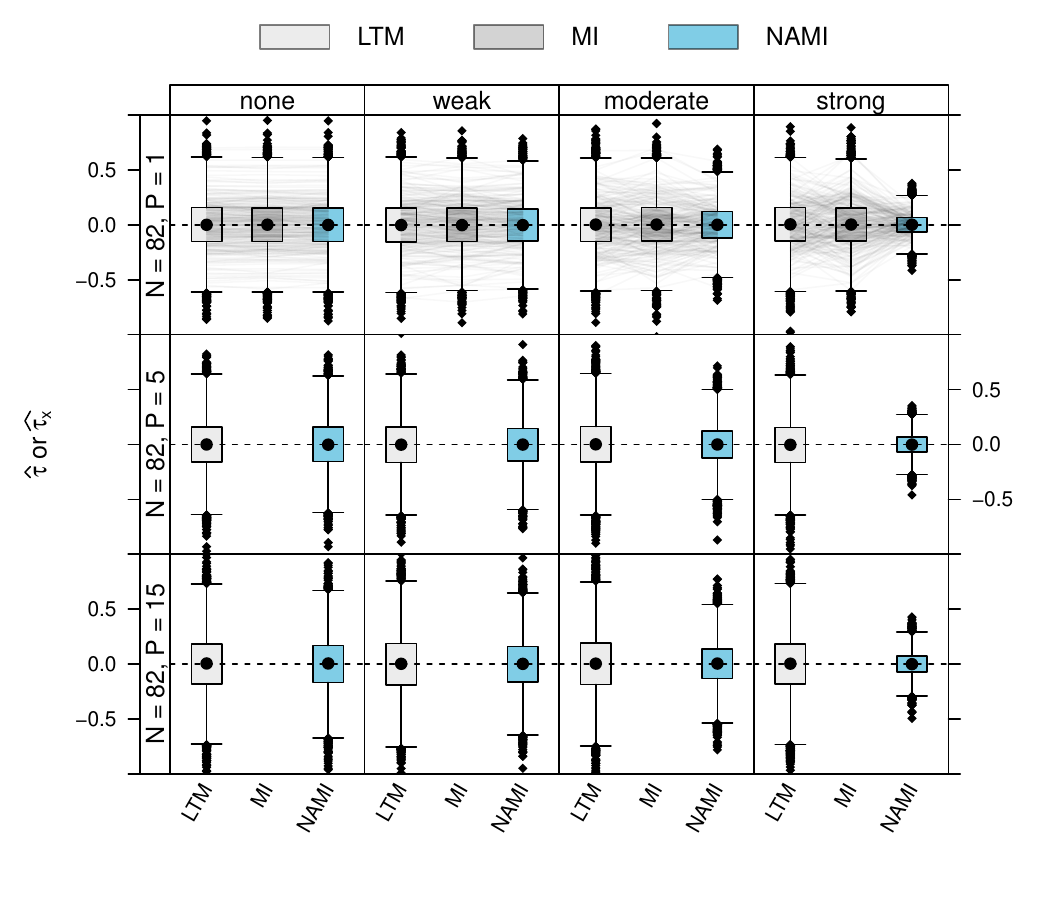} 

\caption{Empirical experiments for normally distributed outcome under $\tau = 0$ (dashed lines): Distribution of treatment effect estimates
$\widehat{\tau}$ of Cohen's $d$ obtained from unadjusted marginal inference (MI) and \NAMI (NAMI), and
effect estimates of $\widehat{\tau}_\rx$ by linear transformation models (LTM)
under varying prognostic strengths of covariate $X_1$ (in columns) and increasing number of noise covariates
($P$, in rows).
\label{fig:conttau00}}
\end{figure}

Overall, MI and NAMI produced nearly unbiased parameter estimates of the true marginal
effect under all conditions (\Figure~\ref{fig:conttau05}~and~\AppFigure~\ref{fig:conttau00}, \App~\ref{app:simulation} for binary and survival outcomes). The corresponding boxplots
of $\widehat{\tau}$ were symmetrically distributed around $\tau$. In contrast,
adjusting for covariates led to much
larger values $\widehat{\tau}_\rx$ for LTM whenever the number of covariates was large or
at least one of them weakly informative, illustrating the effect of
noncollapsibility. 
The estimates were highly correlated in all scenarios.
When adjusting for a single noninformative covariate $X_1$
(\Figure~\ref{fig:conttau05}, top-left
panel), NAMI, MI and LTM  performed on par. 
For binary and survival outcomes, competing methods \added{TMLE,} YSTD and LRCL also obtained nearly unbiased
marginal estimates. 

\paragraph{RQ 2: Reduced standard errors?}

\begin{figure}[th!]
\centering

\includegraphics[width=.9\textwidth]{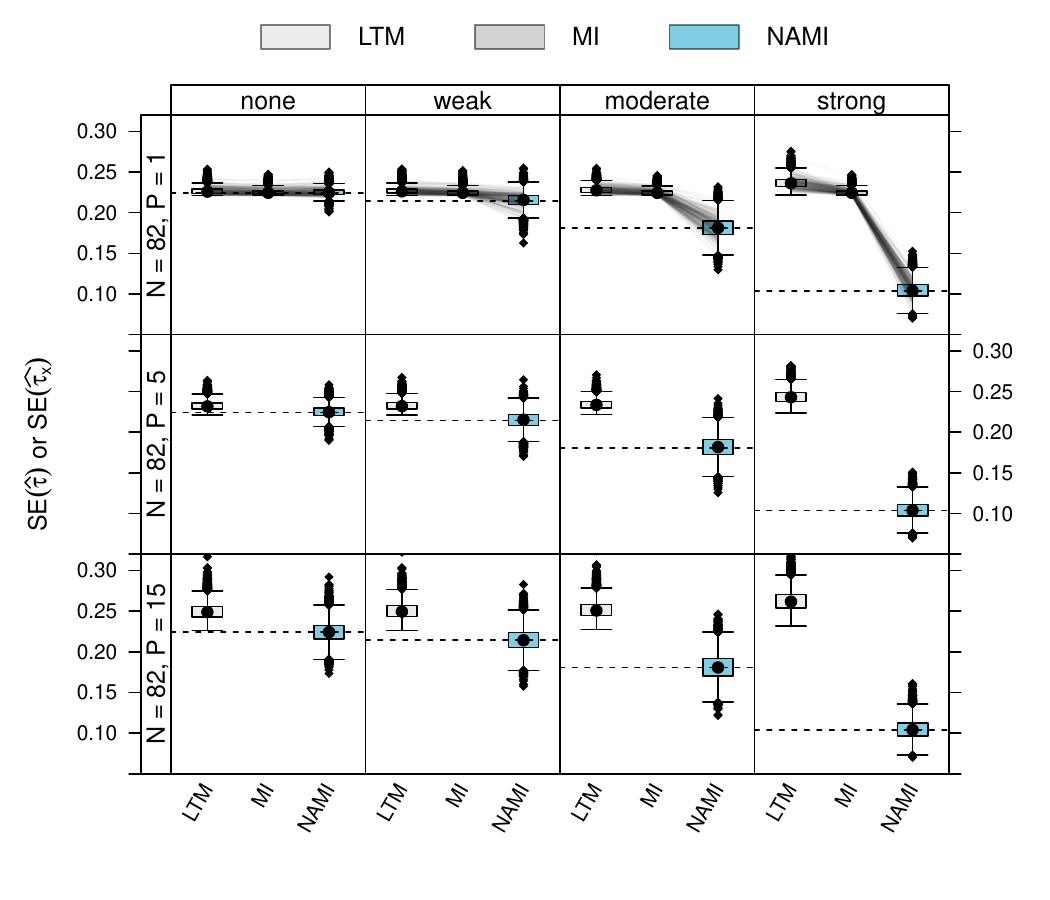} 

\caption{Empirical experiments for normally distributed outcome under $\tau = 0.5$: 
Distribution of standard errors of treatment effect estimates
$\widehat{\tau}$ of Cohen's $d$ obtained from unadjusted marginal inference (MI) and \NAMI (NAMI), and
effect estimates of $\widehat{\tau}_\rx$ by linear transformation models (LTM) 
under varying prognostic strengths of covariate $X_1$ (in columns) and increasing number of noise covariates
($P$, in rows). Standard
errors were computed by inverting the numerically determined negative
Hessian.
The dashed lines correspond to the theoretical standard errors $\text{SE}(\tau, \lambda)$ given true $\tau = 0$ and differing true $\lambda$ (Lemma~\ref{lem:semmlt}).
\label{fig:contvcov05}}
\end{figure}

\begin{figure}[th!]
\centering

\includegraphics[width=.9\textwidth]{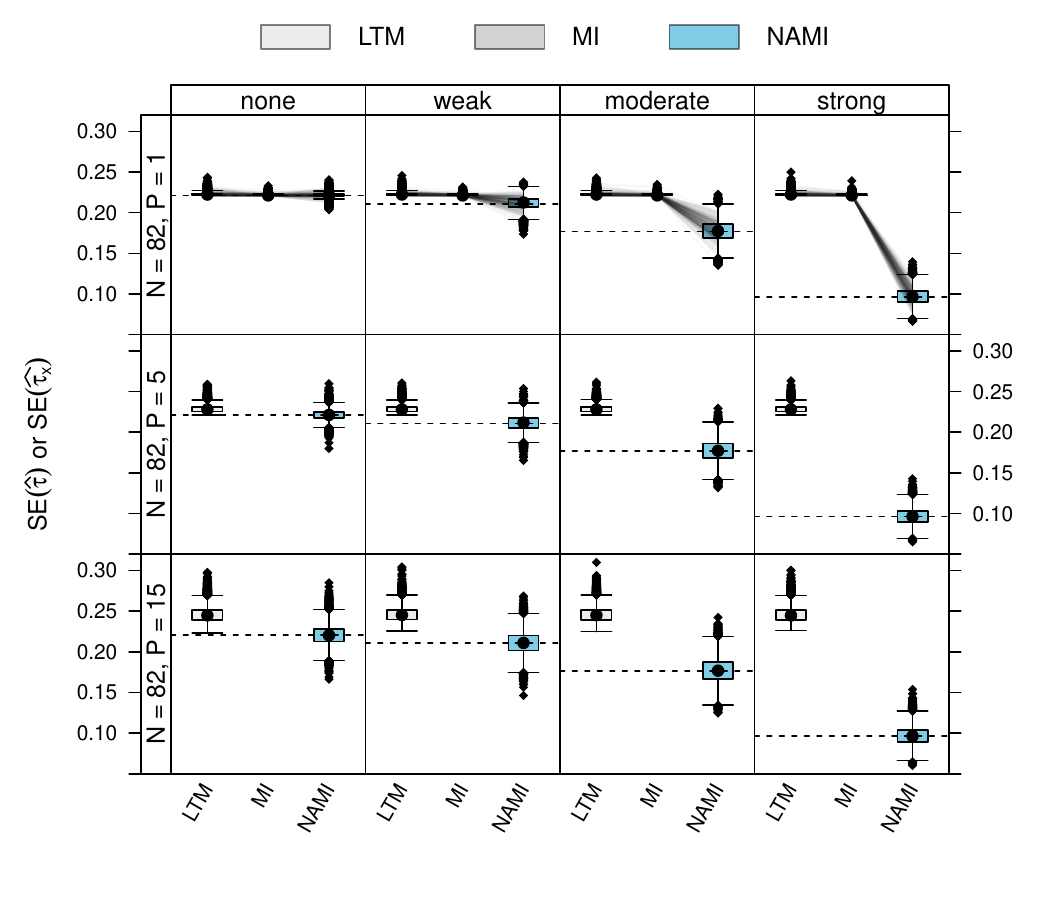} 

\caption{Empirical experiments for normally distributed outcome under $\tau = 0$: 
Distribution of standard errors of treatment effect estimates
$\widehat{\tau}$ of Cohen's $d$ obtained from unadjusted marginal inference (MI) and \NAMI (NAMI), and
effect estimates of $\widehat{\tau}_\rx$ by linear transformation models (LTM)
under varying prognostic strengths of covariate $X_1$ (in columns) and increasing number of noise covariates
($P$, in rows). Standard
errors were computed by inverting the numerically determined negative
Hessian.
The dashed lines correspond to the theoretical standard errors $\text{SE}(\tau, \lambda)$ given true $\tau = 0$ and differing true $\lambda$ (Lemma~\ref{lem:semmlt}).
\label{fig:contvcov00}}
\end{figure}

The first columns in \AppFigures~\ref{fig:contvcov05}~and~\ref{fig:contvcov00} show that the standard errors of NAMI were slightly larger
compared to the standard errors obtained from MI when
adjusting for one or more \textit{noninformative} covariates. This reflected the
increased variability of the NAMI parameter estimates visible in the first
column of \Figure~\ref{fig:conttau05}~and~\AppFigure~\ref{fig:conttau00}. 
When adjusting for one moderate or strong prognostic covariate, 
smaller standard errors for
NAMI corresponded to the decreased variability of the corresponding
parameter estimates.
Similar results were obtained for binary and survival outcomes.
Distributions of standard errors were similar for \added{TMLE,} YSTD and NAMI for binary outcomes; for survival outcomes, standard errors of NAMI were lower than for YSTD when $X_1$ was at least moderate prognostic.
\Table~\ref{tab:power} shows that higher prognostic strength leads to higher power. 
YSTD and NAMI performed similar in this regard for binary outcomes \added{closely followed by TMLE.}
\deleted{, while p} Power was higher for NAMI compared to YSTD and LRCL for survival outcomes. 

\paragraph{RQ 3: Influence of prognostic strength?}

Increasing the prognostic strength of $X_1$ led to less variable
estimates $\widehat{\tau}$ (first row in \Figure~\ref{fig:conttau05}~and~\AppFigure~\ref{fig:conttau00}), smaller standard
errors (first row in \AppFigures~\ref{fig:contvcov05}~and~\ref{fig:contvcov00}), and higher power
(\Table~\ref{tab:power}) for NAMI in comparison to MI.
%
For binary outcomes, YSTD performed similarly to NAMI\added{, closely followed by TMLE.} 
\deleted{, while f} For survival outcomes, NAMI had less variable estimates at higher prognostic levels, leading to smaller standard errors and larger power compared to YSTD and LRCL.
\added{The LTM is misspecified for the survival outcome, and this misspecification 
	becomes more severe the stronger the prognostic strength of the covariate, which is reflected in increased empirical sizes.}

\paragraph{RQ 4: Sensitivity to noise variables?}

\begin{figure}

\includegraphics[width=.9\textwidth]{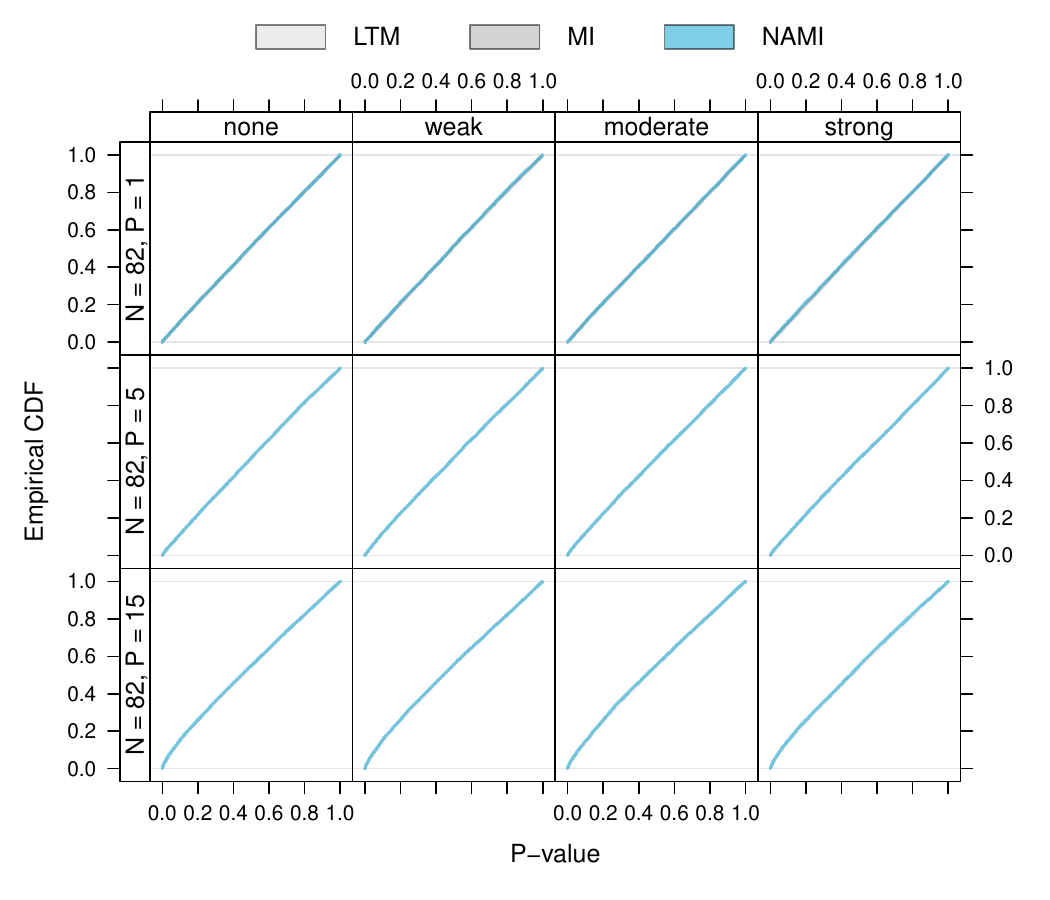} 

\caption{Empirical experiments for normally distributed outcome under $\tau = 0$: 
P-value distribution for test against null hypothesis $H_0: \tau = \tau_\rx = 0$ obtained from unadjusted marginal inference (MI), \NAMI (NAMI), 
or from linear transformation models (LTM)
under varying prognostic strengths of covariate $X_1$ (in columns) and increasing number of noise covariates
($P$, in rows). \label{fig:contpval00}}
\end{figure}

Adding noise variables had surprisingly little influence on the performance of
NAMI (first column of \AppFigures~\ref{fig:contvcov05}~and~\ref{fig:contvcov00}) and only induced bias for the LTM. 
However, there was an increase in the variability of the $\tau$ estimates obtained by NAMI, 
especially for $P = 15$ covariates.
These patterns were also visible for binary and survival outcomes.
YSTD and LRCL behaved similarly to NAMI in this respect (little influence, slightly increased variance).
Under $P = 15$, no reliable inference by LTM and NAMI can be guaranteed, as the
empirical sizes in \Table~\ref{tab:size} exceeded the nominal size 
and the empirical distributions of the $p$-values in \AppFigures~\ref{fig:contpval00}~and~\ref{fig:contpval05} deviated from the uniform distribution.
\added{\AppTable~\ref{tab:sizeN400} in the Appendix shows that the size distortions are diminished with an increased sample size of $N = 800$.}
\added{Compared to NAMI, TMLE was more robust, with empirical sizes remaining stable as the number of noise variables increased.}
Liberality of YSTD and LRCL under heavy censoring is shown in
\Table~\ref{tab:size}.

\begin{figure}

\includegraphics[width=\maxwidth]{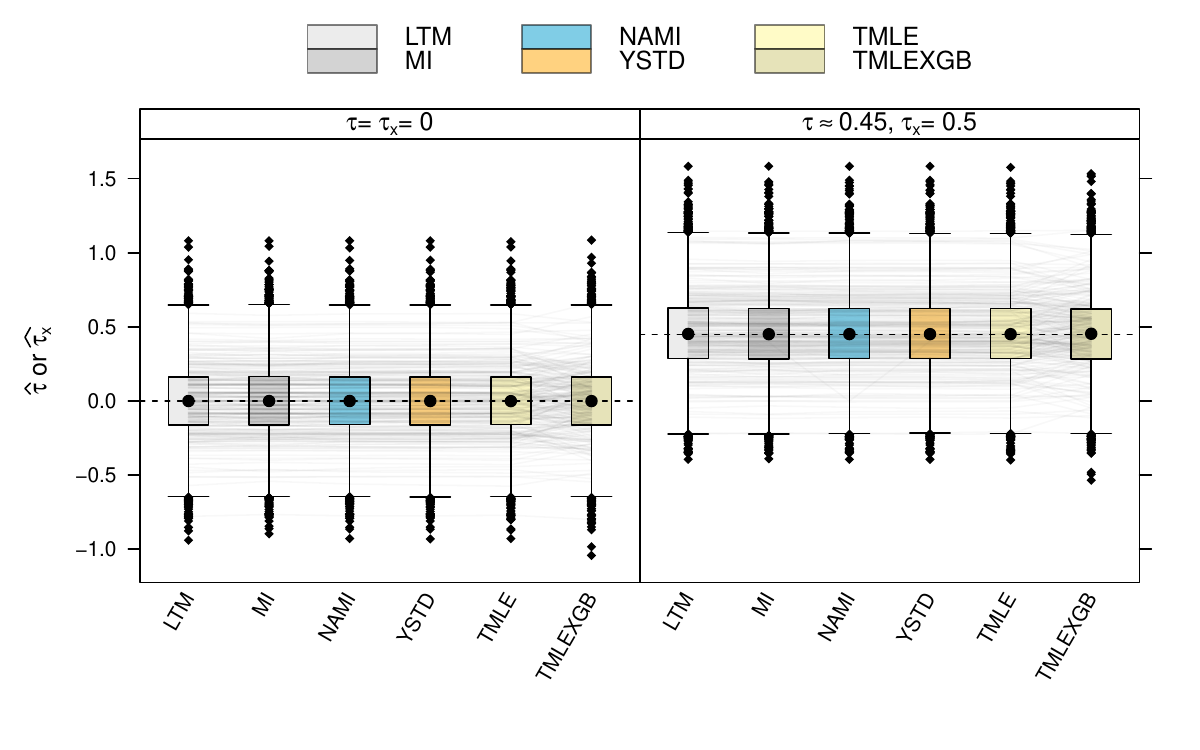} 

\caption{
Empirical experiments for M2 (misspecified Copula structure):
Distribution of log-hazard ratio treatment effect estimates $\widehat{\tau}$ 
from unadjusted marginal inference (MI), \NAMI (NAMI),
the standardization approach of \cite{Ye_Shao_Yi_2024} (YSTD), a
linear transformation model (LTM) and targetted maximum likelihood  
estimation \citep{vanderlaan_tmle_2006} with a (misspecified) logistic regression model (TMLE)
and a more flexible gradient-boosting-based model (TMLEXGB) to model the outcome.
Additional, conditional effect estimates $\widehat{\tau}_\rx$ estimated by linear transformation models (LTM)
are shown.
\label{fig:misspecbin-cf}}
\end{figure}

\paragraph{RQ 5: Impact of misspecifications?}
\added{
For M1 (misspecified marginal model) and $\tau_\rx = 0$, all procedures are nearly unbiased, and the parameter variances of NAMI and LRCL are reduced.
The left panel of \Figure~\ref{fig:misspec-pval} reveals that compared to the other approaches, 
the $p$-value distributions for testing $H_0 : \tau = \tau_{\rx} = 0$ of the misspecified NAMI is stochastically too large, leading to a conservative test.
For $\tau_\rx = 0.5$, MI, NAMI, and LRCL are biased towards zero. In contrast, the conditional estimate by the LTM is right on target.
However, the right panel in \Figure~\ref{fig:misspec-pval} shows that NAMI is more powerful than MI and LRCL. 
For M2 (misspecified Copula structure), all marginal methods are unbiased. 
Because the prognostic effect is underestimated, NAMI, the parametric TMLE, and LTM do not have any benefits compared to MI
(\Figure~\ref{fig:misspecbin-cf}) reflected in similar effects estimates, variances, and $p$-value distributions. 
No variance reductions are observed for TMLEXGB, despite its reliance on fewer assumptions regarding the outcome model and, therefore, the prognostic effect. This may be attributable to the small sample size and slow convergence rates.
}

\clearpage
\newpage

\subsection{Continuous outcome}

\begin{figure}

\includegraphics[width=.9\textwidth]{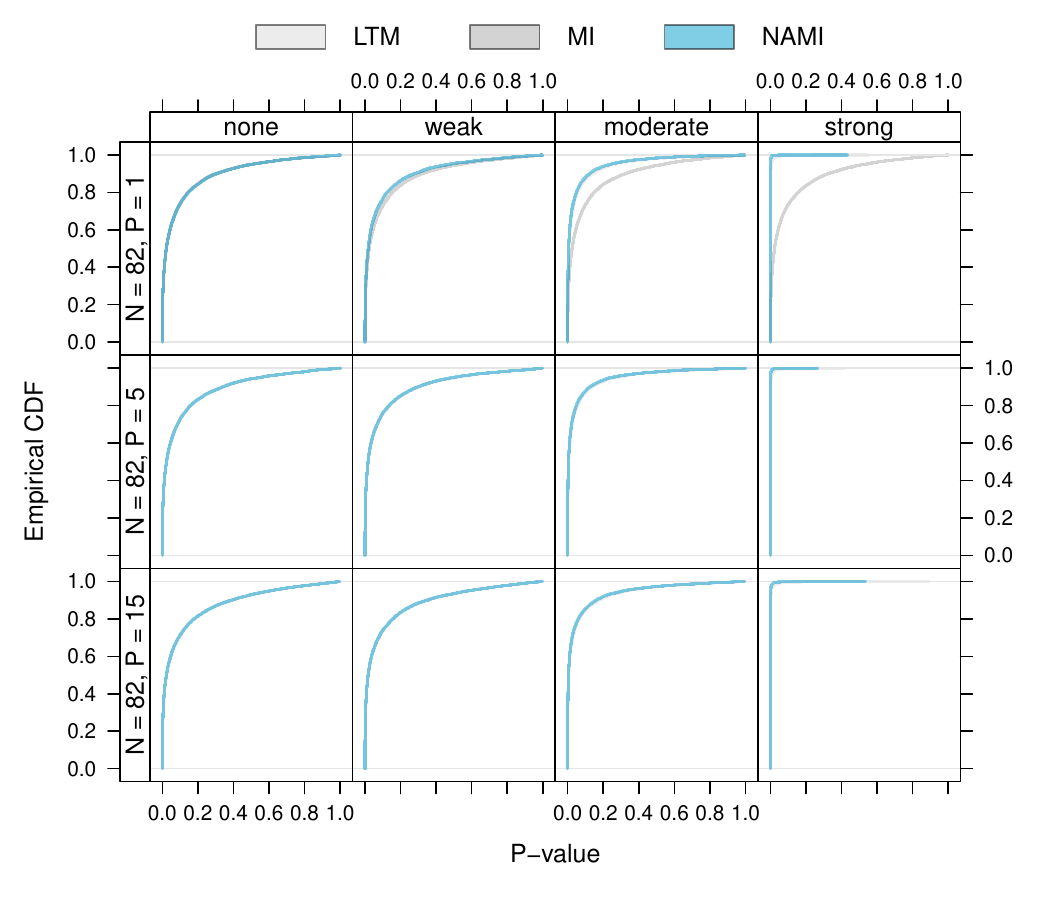} 

\caption{Empirical experiments for normally distributed outcome under $\tau = 0.5$: 
P-value distribution for test against null hypothesis $H_0: \tau = \tau_\rx = 0$ obtained from unadjusted marginal inference (MI), \NAMI (NAMI), 
or from linear transformation models (LTM)
under varying prognostic strengths of covariate $X_1$ (in columns) and increasing number of noise covariates
($P$, in rows). \label{fig:contpval05}}
\end{figure}

\clearpage

\begin{table}[th!]
\caption{Empirical size for normally distributed outcome 
obtained from \NAMI (NAMI), for $N = 800$, $P = 5$ covariates, 
under varying prognostic strength of covariate $X_1$ (in columns).
\label{tab:sizeN400}}
\def\arraystretch{1}
\centering
\begin{tabular}{lllrrrr}
\hline
&&&\multicolumn{4}{c}{Size}\\
\cline{4-7}
DGP&Algorithm&P&\multicolumn{1}{c}{none}&\multicolumn{1}{c}{weak}&\multicolumn{1}{c}{moderate}&\multicolumn{1}{c}{strong}\\
\hline
continuous&NAMI&P = 5&0.049&0.054&0.056&0.052\\
\hline
\end{tabular}

\end{table}

\begin{figure}[th!]

\includegraphics[width=.9\textwidth]{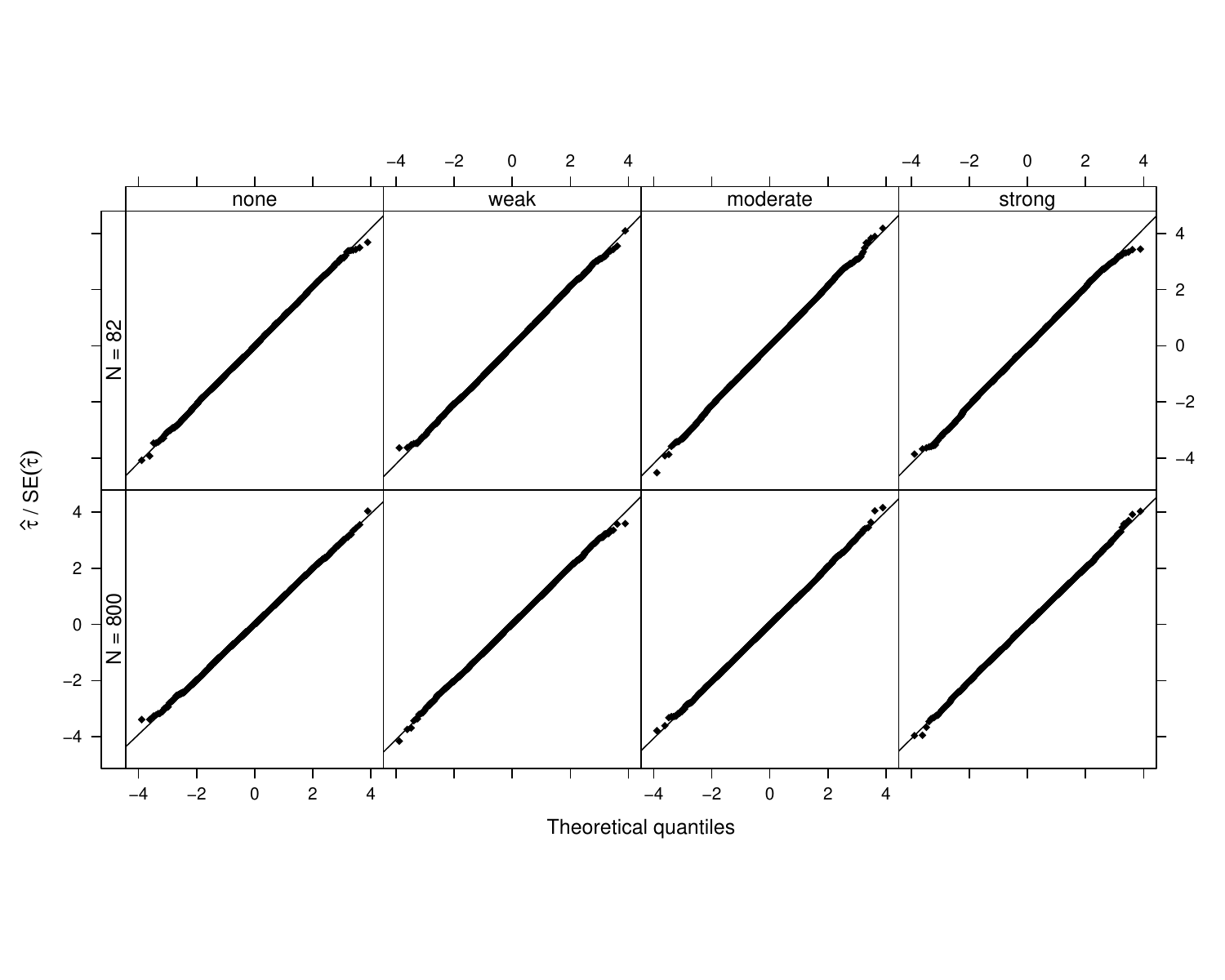} 

\caption{Empirical experiments for normally distributed outcome under $\tau = 0$: 
Q-Q-plot that compares quantiles of the unstandardized effect estimate $\widehat{\tau}$ 
of \NAMI (NAMI) against the theoretical quantiles of a normal distribution
for $P = 5$, varying sample sizes ($N$, in rows) and varying prognostic 
strengths of covariate $X_1$ (in columns). \label{fig:qq00}}
\end{figure}

\clearpage
\newpage
 
\clearpage
\newpage

\subsection{Binary outcome}

\begin{figure}[th!]
\centering

\includegraphics[width=.9\textwidth]{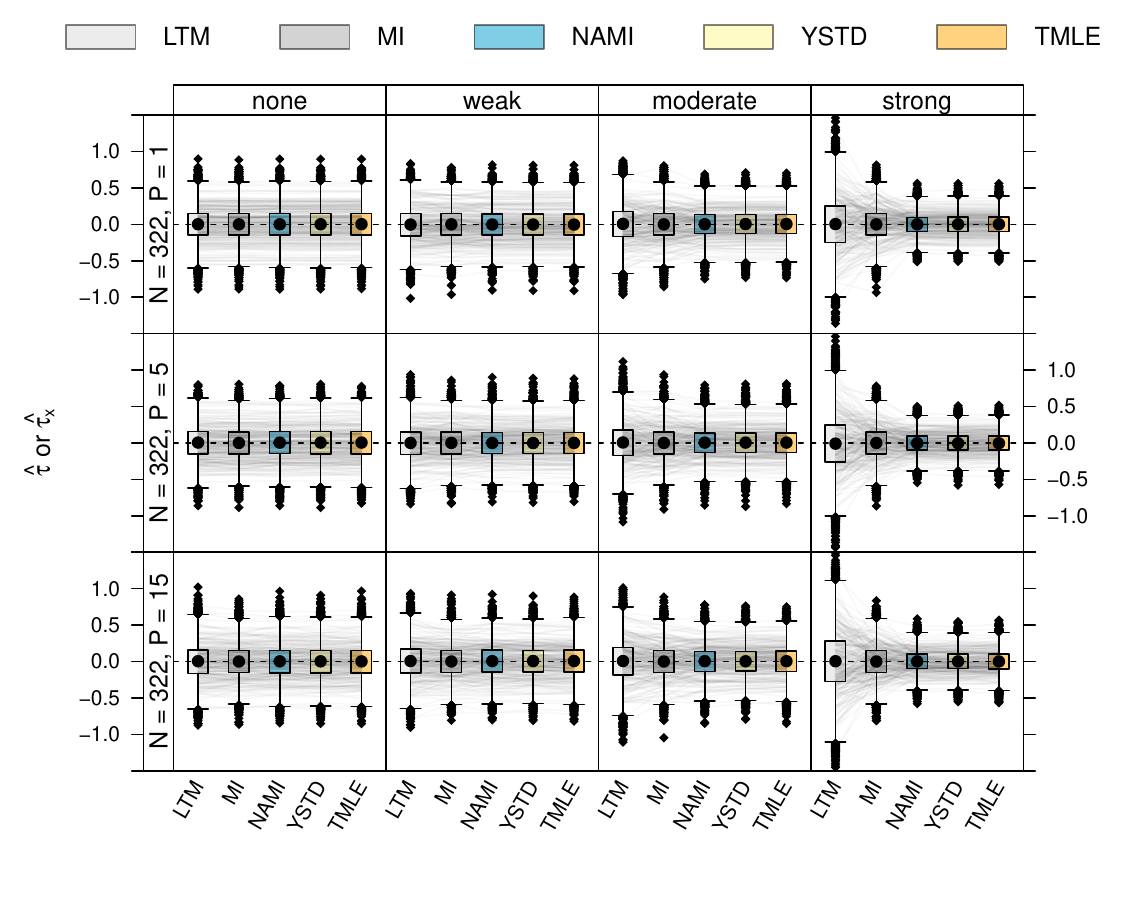} 

\caption{
Empirical experiments for binary outcome and $\tau = 0$ (dashed lines): Distribution of log-odds ratio
treatment effect estimates $\widehat{\tau}$ obtained from unadjusted marginal inference (MI), \NAMI (NAMI), the standardization approach of \cite{Zhang_Tsiatis_2008} (YSTD), or 
the targeted maximum likelihood estimator of \cite{vanderlaan_tmle_2006} (TMLE) and
effect estimates of $\widehat{\tau}_\rx$ by linear transformation models (LTM) 
under varying prognostic strengths of covariate $X_1$ (in columns) and increasing number of (noise) covariates
($P$, in rows).
\label{fig:taubinary0}}
\end{figure}

\begin{figure}
\centering

\includegraphics[width=.9\textwidth]{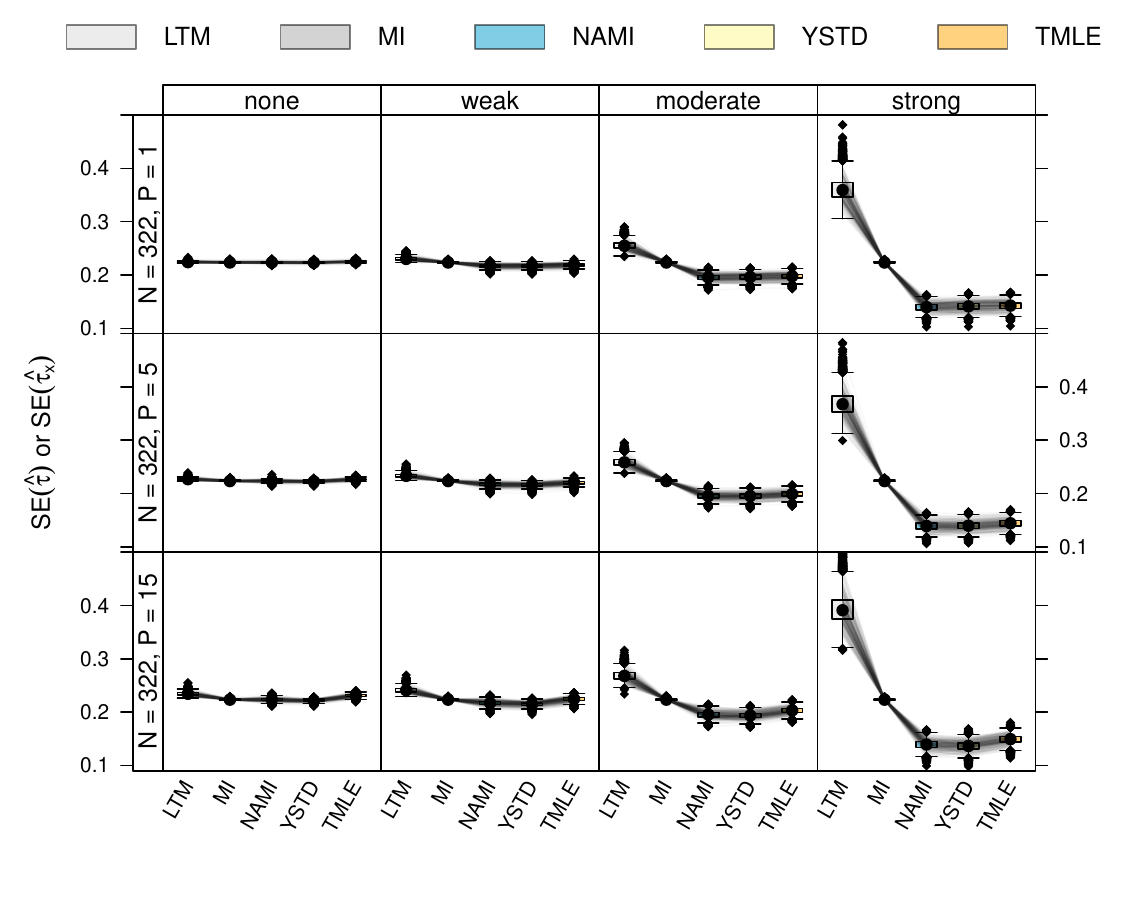} 

\caption{Empirical experiments for binary outcome and $\tau = 0$: Distribution of standard errors of
log-odds ratio treatment effect estimates $\widehat{\tau}$ obtained from unadjusted marginal inference (MI), \NAMI (NAMI), 
the standardization approach of \cite{Zhang_Tsiatis_2008} (YSTD), or 
the targeted maximum likelihood estimator of \cite{vanderlaan_tmle_2006} (TMLE) and
effect estimates of $\widehat{\tau}_\rx$ by linear transformation models (LTM) 
under varying prognostic strengths of covariate $X_1$ (in columns) and increasing number of (noise) covariates
($P$, in rows). Standard
errors were computed by inverting the numerically determined negative
Hessian.
\label{fig:sebinary0}}
\end{figure}

\begin{figure}
\centering

\includegraphics[width=.9\textwidth]{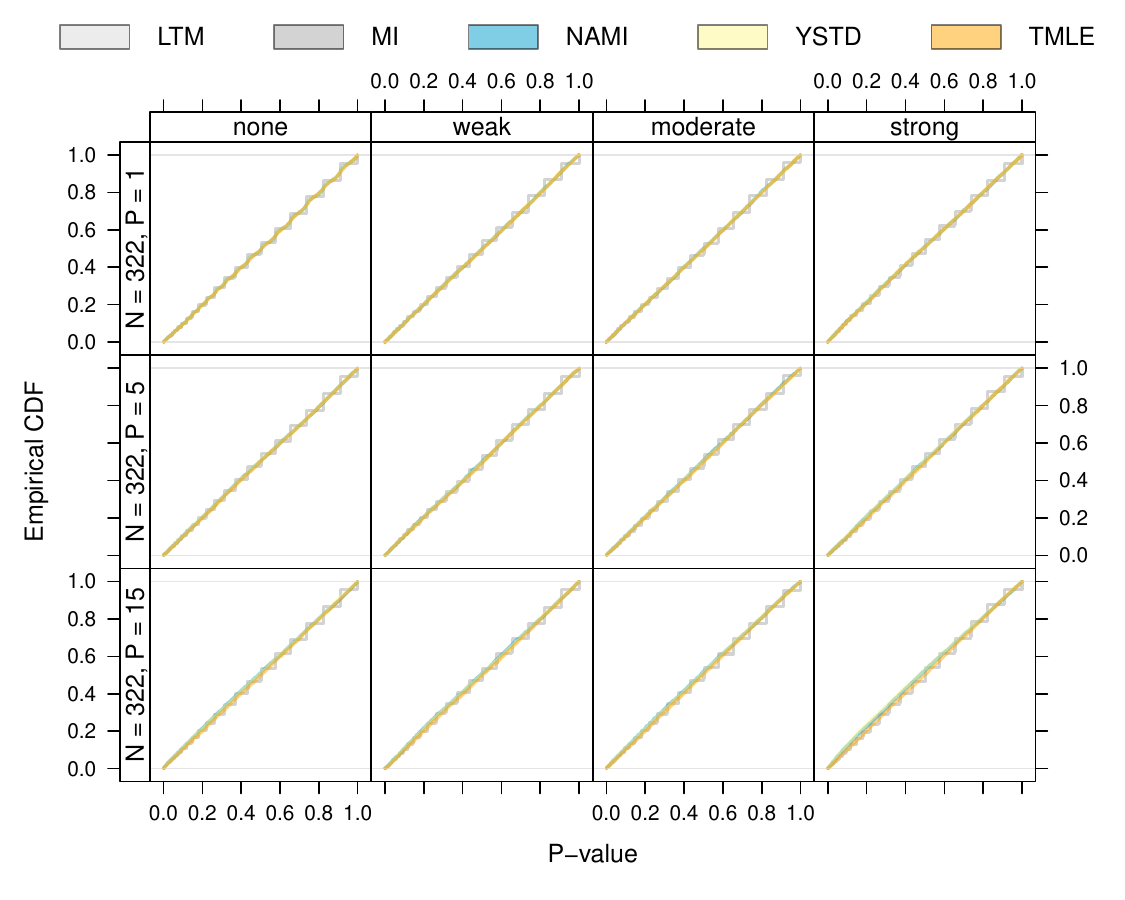} 

\caption{Empirical experiments for binary outcome and $\tau = 0$: 
P-value distribution for test of null hypothesis $H_0: \tau = \tau_\rx = 0$ obtained from unadjusted marginal inference (MI), \NAMI (NAMI),
the standardization approach of \cite{Zhang_Tsiatis_2008} (YSTD), the targeted maximum likelihood estimator of \cite{vanderlaan_tmle_2006} (TMLE), 
or from linear transformation models (LTM)
under varying prognostic strengths of covariate $X_1$ (in columns) and increasing number of noise covariates
($P$, in rows). \label{fig:binpval00}}
\end{figure}

\begin{figure}[ht!]
\centering

\includegraphics[width=.9\textwidth]{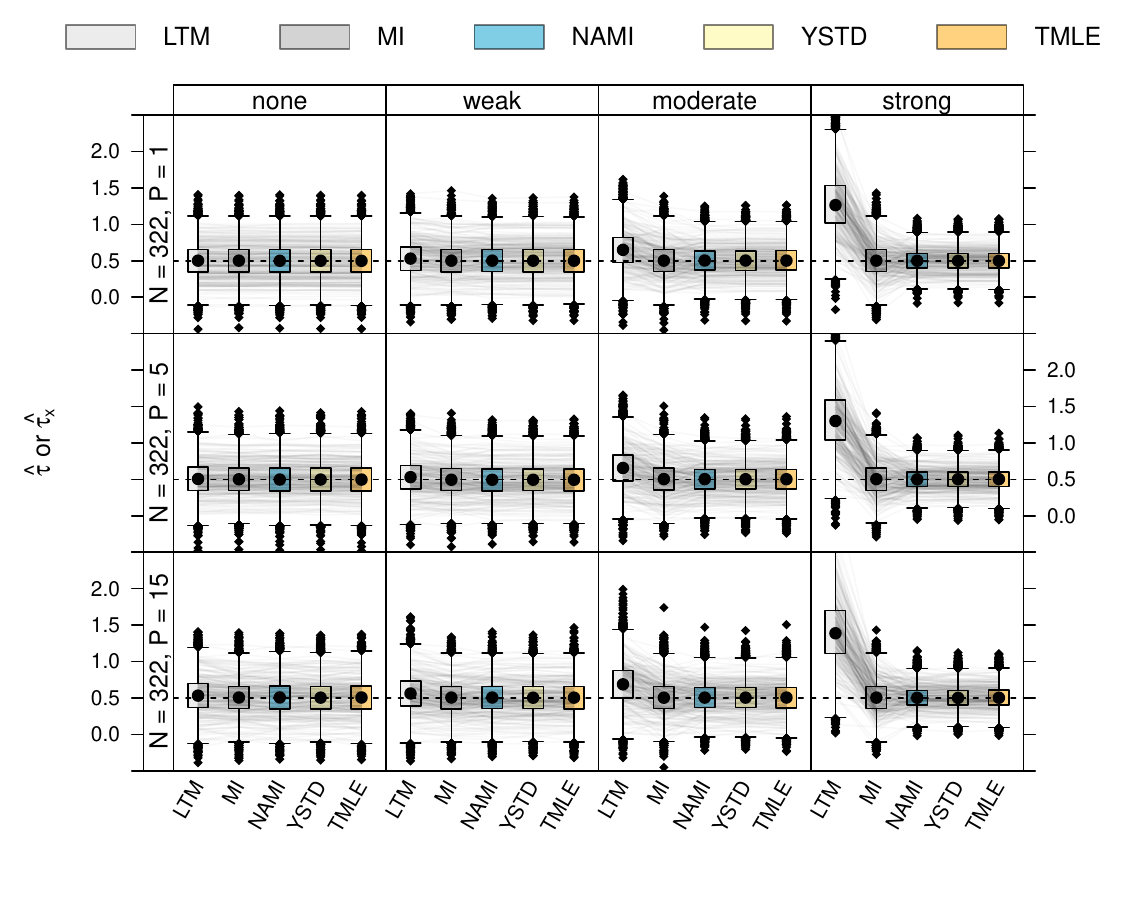} 

\caption{
Empirical experiments for binary outcome and $\tau = 0.5$ (dashed lines): Distribution of log-odds ratio
treatment effect estimates $\widehat{\tau}$ obtained from unadjusted marginal inference (MI), \NAMI (NAMI), 
the standardization approach of \cite{Zhang_Tsiatis_2008} (YSTD), or 
the targeted maximum likelihood estimator of \cite{vanderlaan_tmle_2006} (TMLE), and
effect estimates of $\widehat{\tau}_\rx$ by linear transformation models (LTM) 
under varying prognostic strengths of covariate $X_1$ (in columns) and increasing number of (noise) covariates
($P$, in rows).
\label{fig:taubinary5}}
\end{figure}

\begin{figure}
\centering

\includegraphics[width=.9\textwidth]{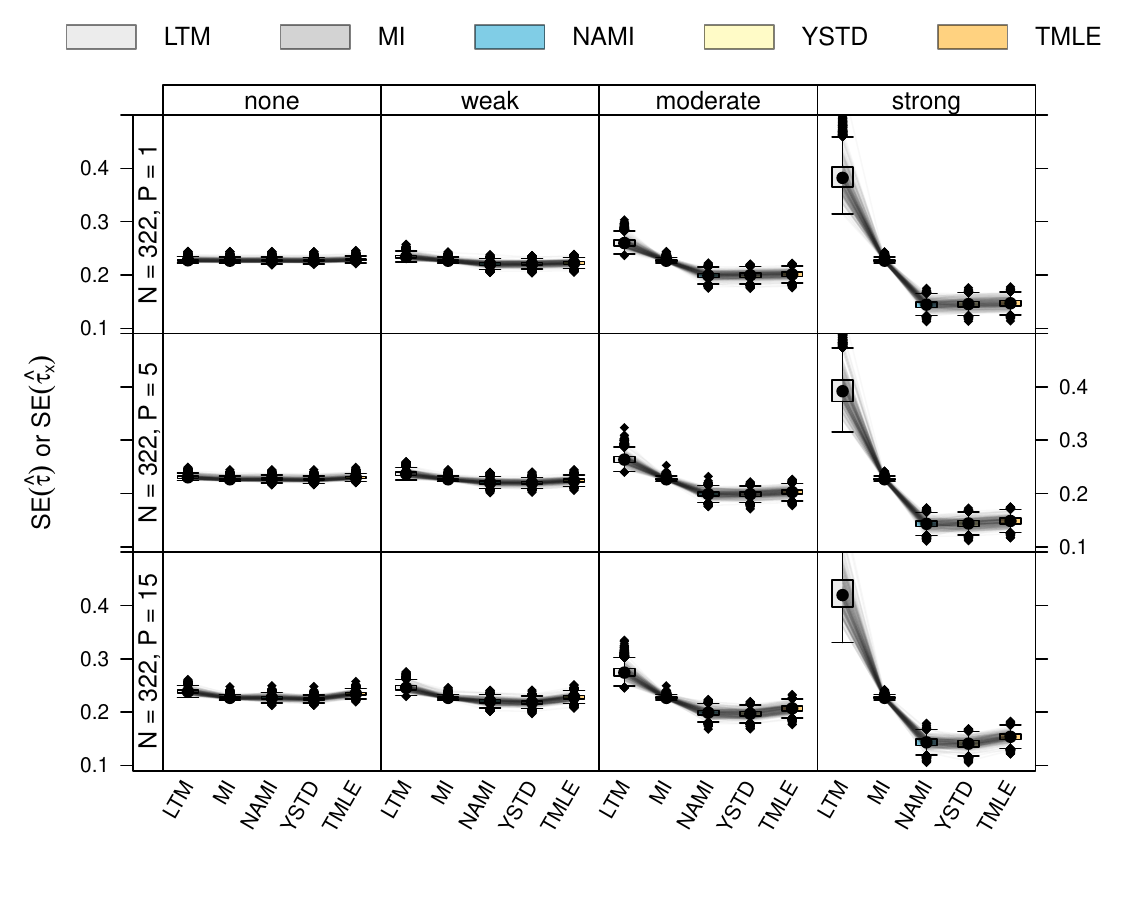} 

\caption{Empirical experiments for binary outcome and $\tau = 0.5$: Distribution of standard errors of
log-odds ratio treatment effect estimates $\widehat{\tau}$ obtained from unadjusted marginal inference (MI), \NAMI (NAMI), 
the standardization approach of \cite{Zhang_Tsiatis_2008} (YSTD), or 
the targeted maximum likelihood estimator of \cite{vanderlaan_tmle_2006} (TMLE), and
effect estimates of $\widehat{\tau}_\rx$ by linear transformation models (LTM)  
under varying prognostic strengths of covariate $X_1$ (in columns) and increasing number of (noise) covariates
($P$, in rows). Standard
errors were computed by inverting the numerically determined negative
Hessian.
\label{fig:sebinary5}}
\end{figure}

\begin{figure}
\centering

\includegraphics[width=.9\textwidth]{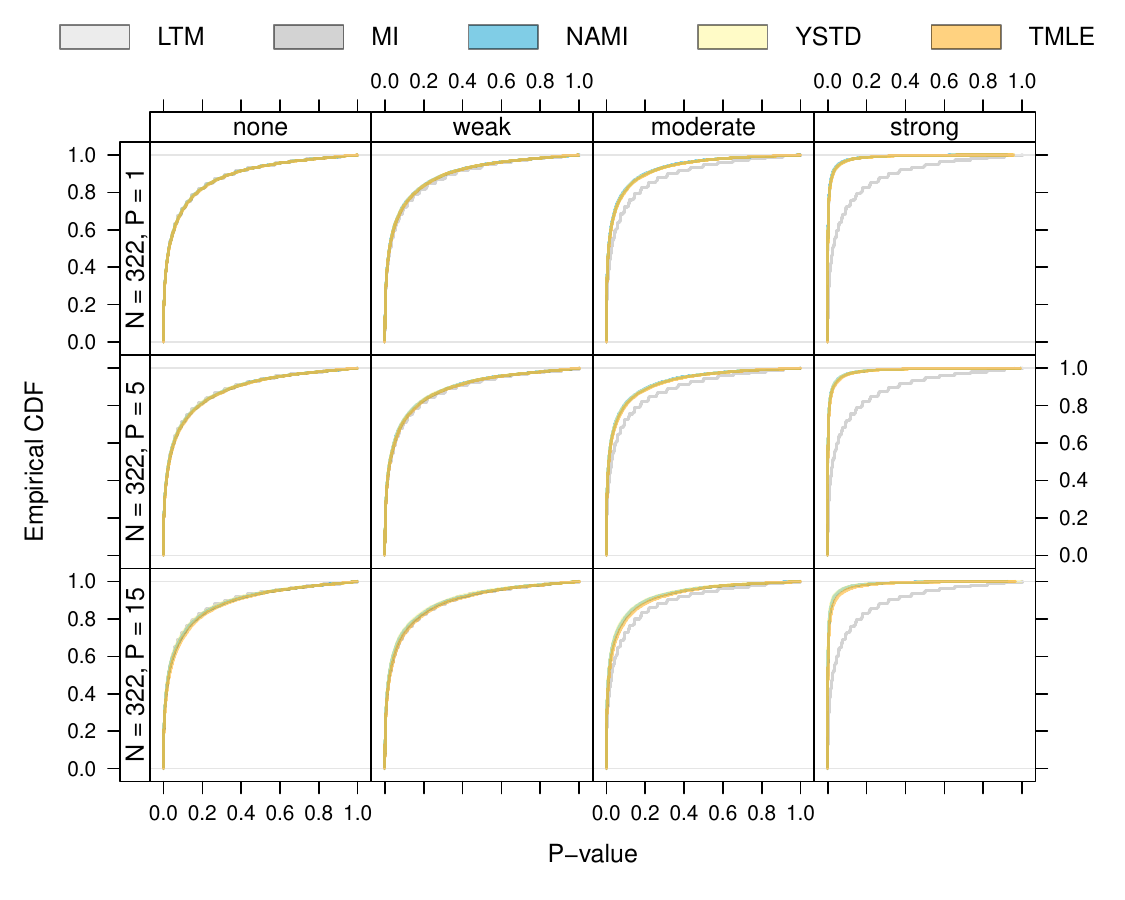} 

\caption{Empirical experiments for binary outcome and $\tau = 0.5$: 
P-value distribution for test of null hypothesis $H_0: \tau = \tau_\rx = 0$ obtained from unadjusted marginal inference (MI), \NAMI (NAMI),
the standardization approach of \cite{Zhang_Tsiatis_2008} (YSTD), 
the targeted maximum likelihood estimator of \cite{vanderlaan_tmle_2006} (TMLE), or from linear transformation models (LTM)
under varying prognostic strengths of covariate $X_1$ (in columns) and increasing number of noise covariates
($P$, in rows). \label{fig:binpval05}}
\end{figure}

\clearpage
\newpage
 
\clearpage
\newpage

\subsection{Survival outcome}

\begin{figure}[th!]
\centering

\includegraphics[width=.9\textwidth]{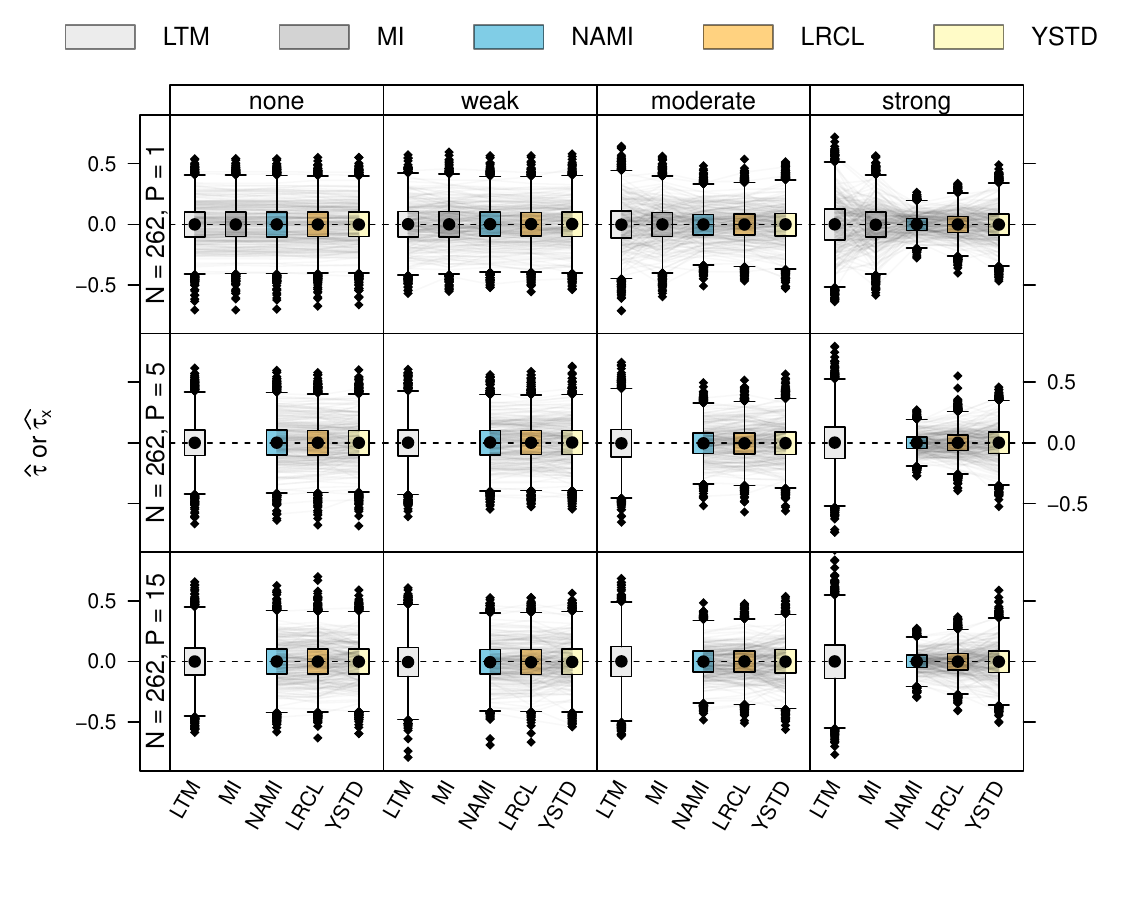} 

  \caption{Empirical experiments for survival outcome under \textit{mild} censoring and $\tau = 0$ (dashed lines): Distribution of standard errors of
log-hazard ratio treatment effect estimates $\widehat{\tau}$ obtained from unadjusted marginal inference (MI), \NAMI (NAMI), 
or the standardization approaches of \cite{Ye_Shao_Yi_2024} (LRCL) and \cite{Lu_Tsiatis_2008} (YSTD), and
effect estimates of $\widehat{\tau}_\rx$ by linear transformation models (LTM)  
under varying prognostic strengths of covariate $X_1$ (in columns) and increasing number of noise covariates
($P$, in rows). Standard
errors were computed by inverting the numerically determined negative
Hessian.
    \label{fig:tausurv07_0}}
\end{figure}

\begin{figure}
\centering

\includegraphics[width=.9\textwidth]{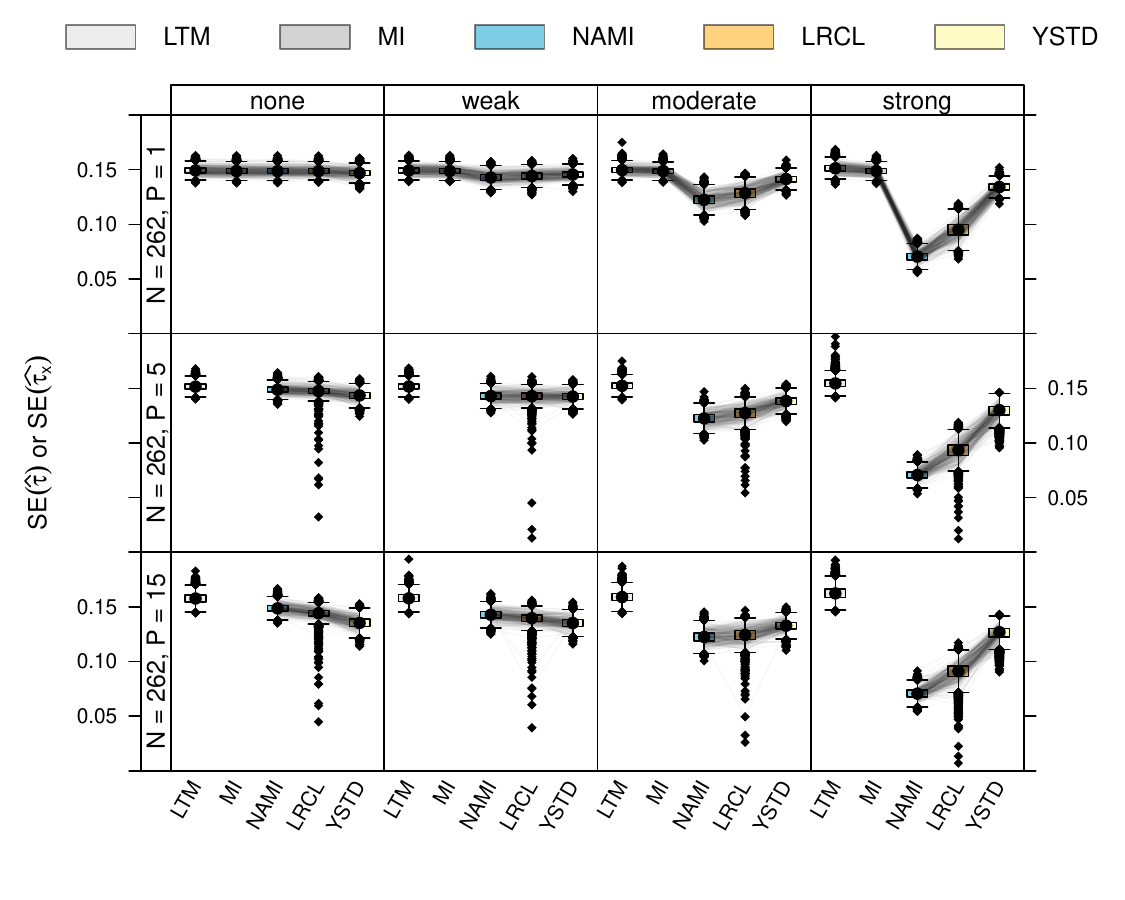} 

  \caption{Empirical experiments for survival outcome under \textit{mild} censoring and $\tau = 0$: Distribution of standard errors of
log-hazard ratio treatment effect estimates $\widehat{\tau}$ obtained from unadjusted marginal inference (MI), \NAMI (NAMI), 
or the standardization approaches of \cite{Ye_Shao_Yi_2024} (LRCL) and \cite{Lu_Tsiatis_2008} (YSTD), and
effect estimates of $\widehat{\tau}_\rx$ by linear transformation models (LTM)  
under varying prognostic strengths of covariate $X_1$ (in columns) and increasing number of noise covariates
($P$, in rows). Standard
errors were computed by inverting the numerically determined negative
Hessian.
    \label{fig:sesurv07_0}}
\end{figure}

\begin{figure}
\centering

\includegraphics[width=.9\textwidth]{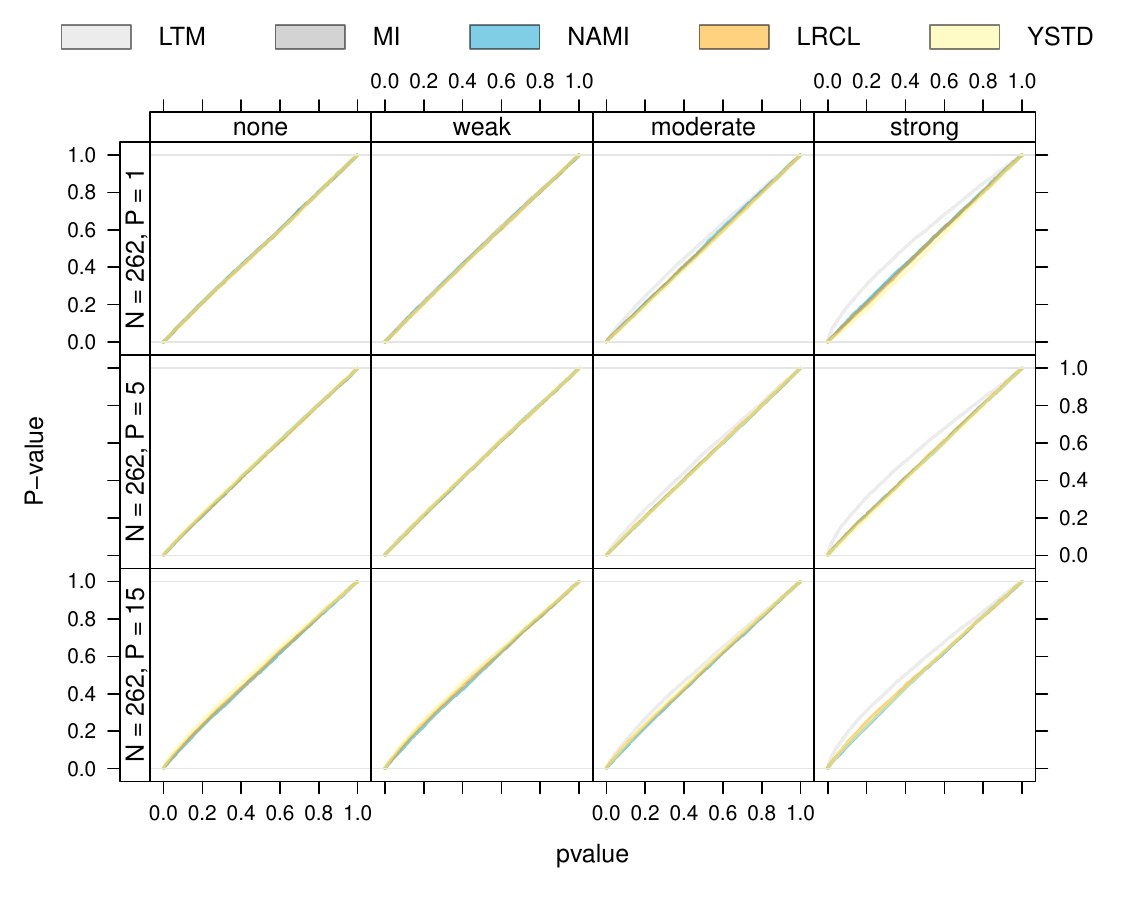} 

\caption{Empirical experiments for survival outcome under \textit{mild} censoring and $\tau = 0$: 
P-value distribution for test against null hypothesis $H_0: \tau = \tau_\rx = 0$ obtained from 
unadjusted marginal inference (MI), \NAMI (NAMI), 
or the standardization approaches of \cite{Ye_Shao_Yi_2024} (LRCL) and \cite{Lu_Tsiatis_2008}
(YSTD), and effect estimates of $\widehat{\tau}_\rx$ by linear transformation models (LTM)
under varying prognostic strengths of covariate $X_1$ (in columns) and increasing number of noise covariates
($P$, in rows). \label{fig:surv07pval0}}
\end{figure}

\begin{figure}
\centering

\includegraphics[width=.9\textwidth]{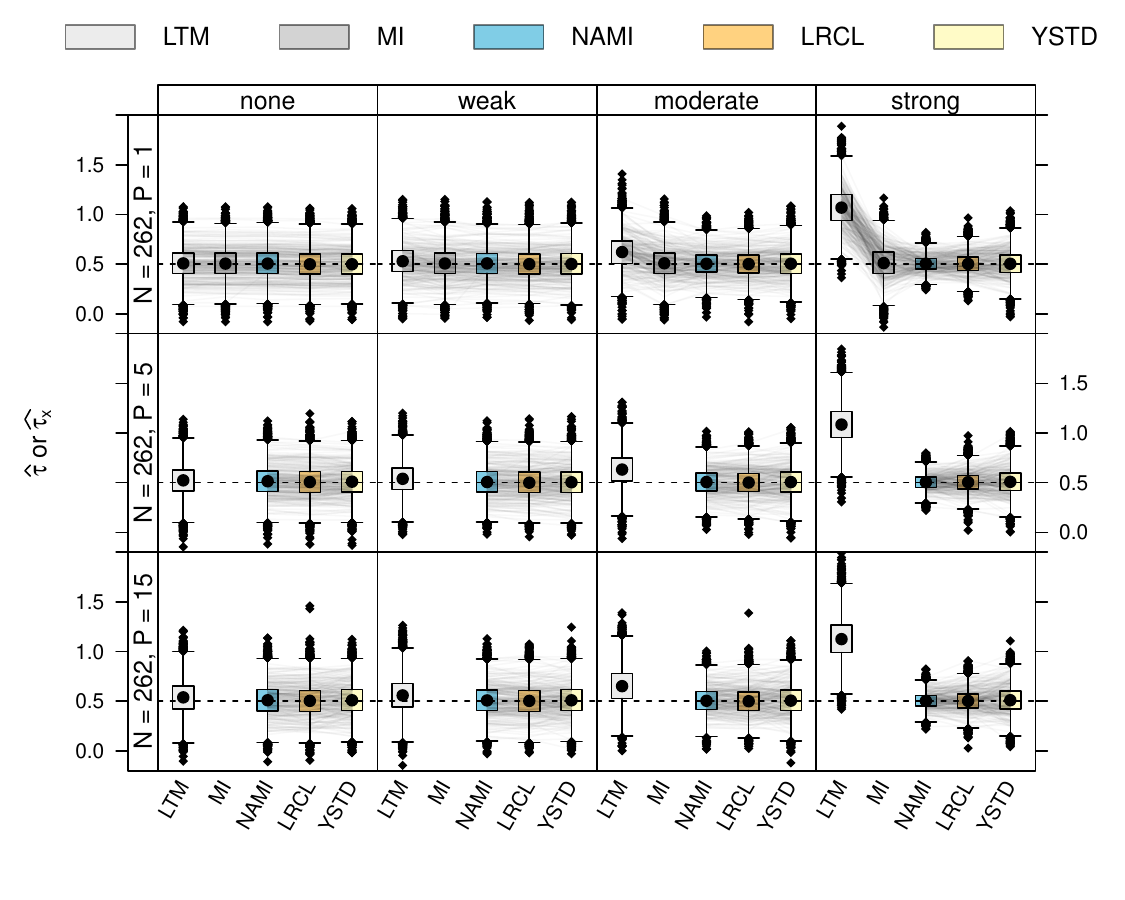} 

  \caption{
Empirical experiments for survival outcome under \textit{mild} censoring and  $\tau = 0.5$ (dashed lines): Distribution of log-hazard ratio
treatment effect estimates $\widehat{\tau}$ obtained from unadjusted marginal inference (MI), \NAMI (NAMI), 
or the standardization approaches of \cite{Ye_Shao_Yi_2024} (LRCL) and \cite{Lu_Tsiatis_2008} (YSTD), and
effect estimates of $\widehat{\tau}_\rx$ by linear transformation models (LTM)  
under varying prognostic strengths of covariate $X_1$ (in columns) and increasing number of noise covariates
($P$, in rows).
    \label{fig:tausurv07_05}}
\end{figure}

\begin{figure}
\centering

\includegraphics[width=.9\textwidth]{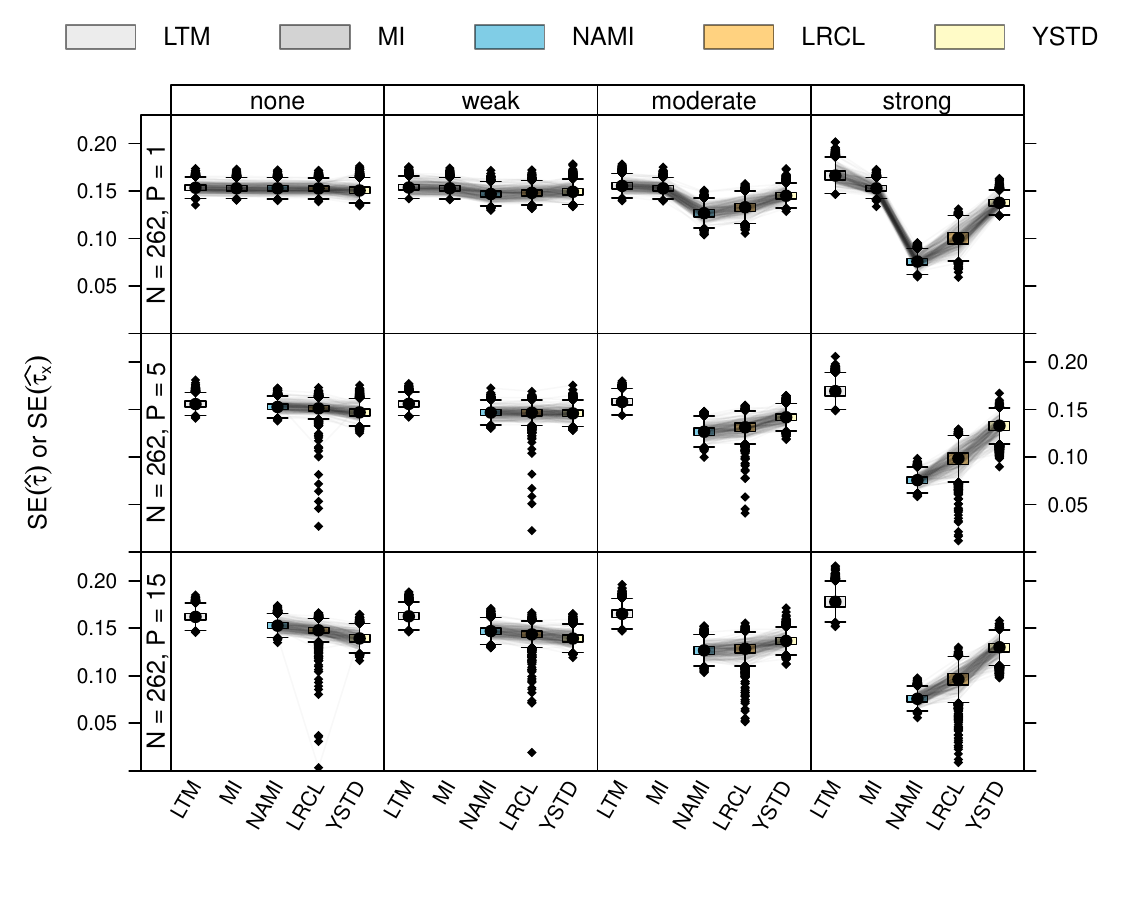} 

  \caption{Empirical experiments for survival outcome under \textit{mild} censoring and $\tau = 0.5$: Distribution of standard errors of
log-hazard ratio treatment effect estimates $\widehat{\tau}$ obtained from unadjusted marginal inference (MI), \NAMI (NAMI), 
or the standardization approaches of \cite{Ye_Shao_Yi_2024} (LRCL) and \cite{Lu_Tsiatis_2008} (YSTD), and
effect estimates of $\widehat{\tau}_\rx$ by linear transformation models (LTM)  
under varying prognostic strengths of covariate $X_1$ (in columns) and increasing number of noise covariates
($P$, in rows). Standard
errors were computed by inverting the numerically determined negative
Hessian.
    \label{fig:sesurv07_05}}
\end{figure}

\begin{figure}

\includegraphics[width=.9\textwidth]{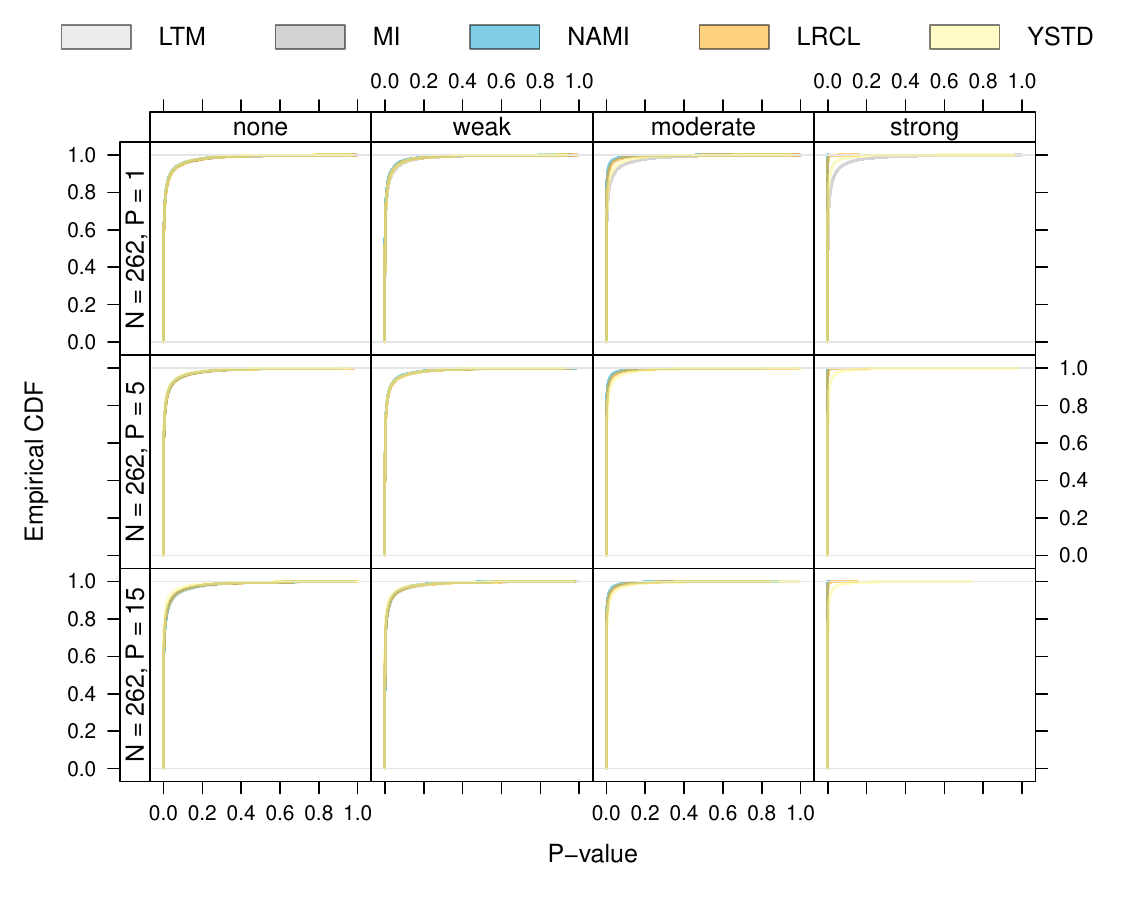} 

\caption{Empirical experiments for survival outcome under \textit{mild} censoring and $\tau = 0.5$: 
P-value distribution for test against null hypothesis $H_0: \tau = \tau_\rx = 0.5$ obtained from 
unadjusted marginal inference (MI), \NAMI (NAMI), 
or the standardization approaches of \cite{Ye_Shao_Yi_2024} (LRCL) and \cite{Lu_Tsiatis_2008}
(YSTD), and effect estimates of $\widehat{\tau}_\rx$ by linear transformation models (LTM)
under varying prognostic strengths of covariate $X_1$ (in columns) and increasing number of noise covariates
($P$, in rows). \label{fig:surv07pval05}}
\end{figure}

\begin{figure}
\centering

\includegraphics[width=.9\textwidth]{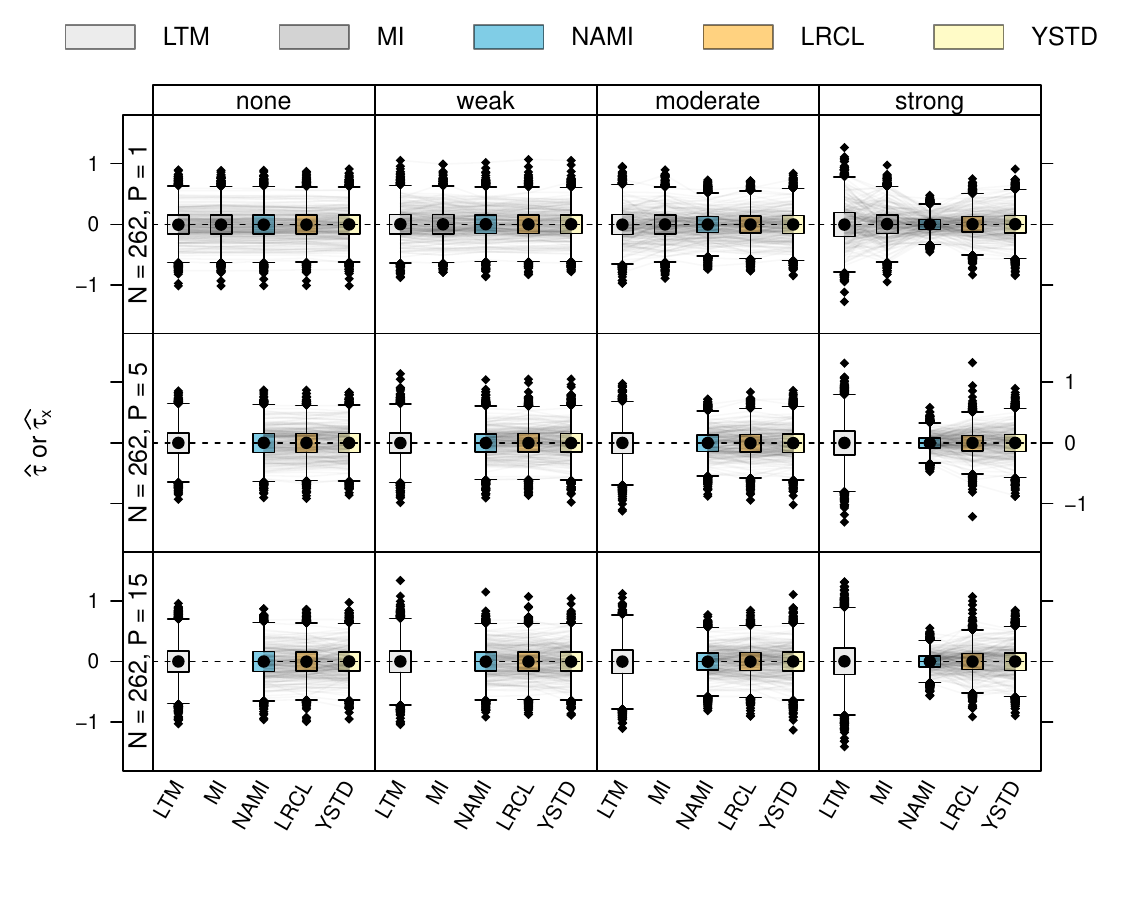} 

  \caption{Empirical experiments for survival outcome under \textit{heavy} censoring and $\tau = 0$ (dashed lines): Distribution of
log-hazard ratio treatment effect estimates $\widehat{\tau}$ obtained from unadjusted marginal inference (MI), \NAMI (NAMI), 
or the standardization approaches of \cite{Ye_Shao_Yi_2024} (LRCL) and \cite{Lu_Tsiatis_2008} (YSTD), and
effect estimates of $\widehat{\tau}_\rx$ by linear transformation models (LTM)  
under varying prognostic strengths of covariate $X_1$ (in columns) and increasing number of noise covariates
($P$, in rows).
    \label{fig:tausurv_05_0}}
\end{figure}

\begin{figure}
\centering

\includegraphics[width=.9\textwidth]{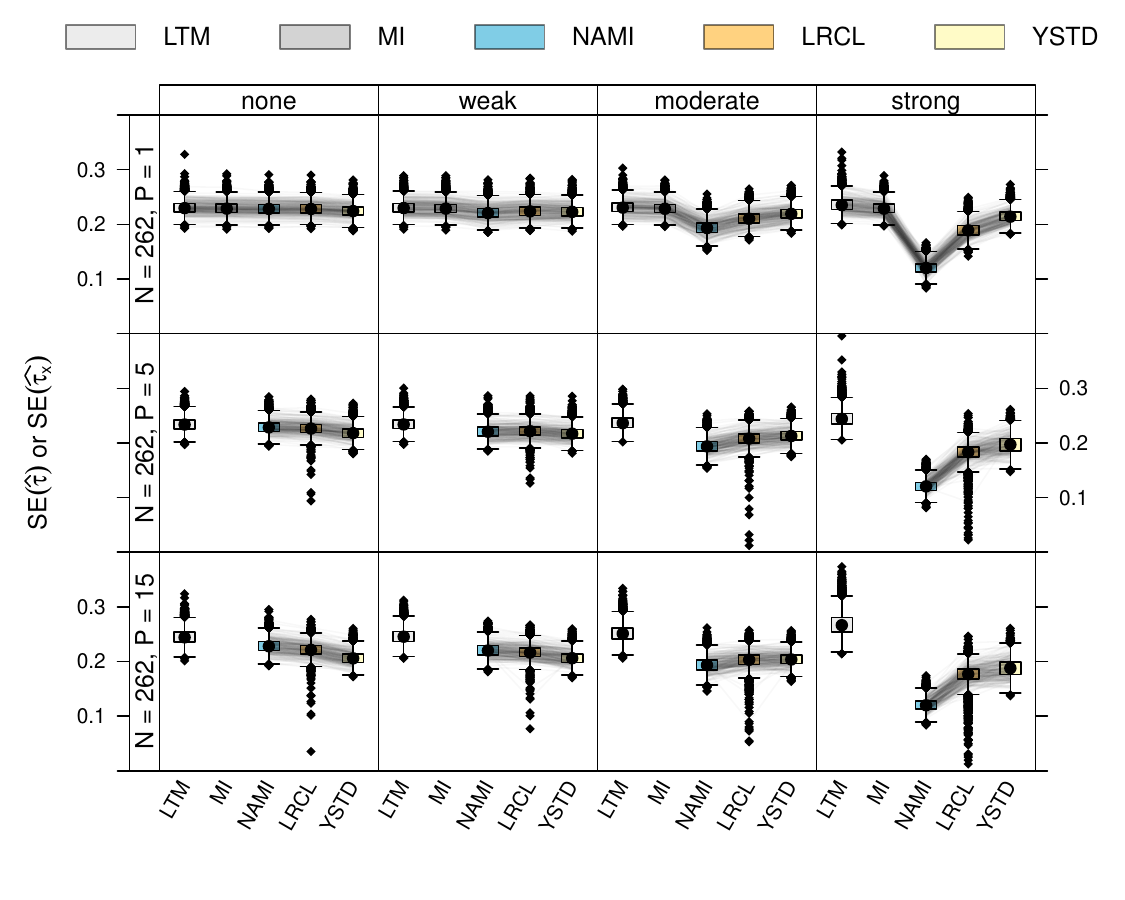} 

  \caption{Empirical experiments for survival outcome under \textit{heavy} censoring and $\tau = 0$: Distribution of standard errors of
log-hazard ratio treatment effect estimates $\widehat{\tau}$ obtained from unadjusted marginal inference (MI), \NAMI (NAMI), 
or the standardization approaches of \cite{Ye_Shao_Yi_2024} (LRCL) and \cite{Lu_Tsiatis_2008} (YSTD), and
effect estimates of $\widehat{\tau}_\rx$ by linear transformation models (LTM)  
under varying prognostic strengths of covariate $X_1$ (in columns) and increasing number of noise covariates
($P$, in rows). Standard
errors were computed by inverting the numerically determined negative
Hessian.
    \label{fig:sesurv03_0}}
\end{figure}

\begin{figure}
\centering

\includegraphics[width=.9\textwidth]{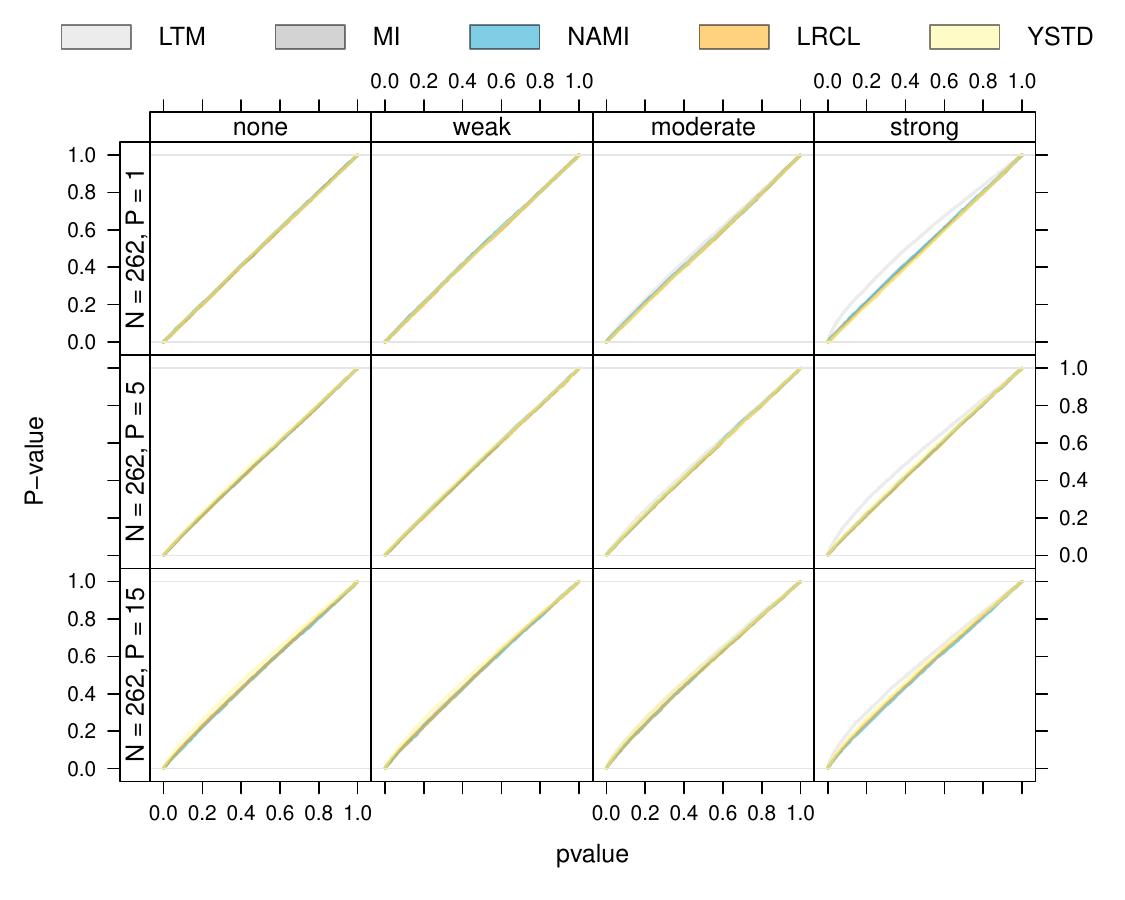} 

\caption{Empirical experiments for survival outcome under \textit{heavy} censoring and $\tau = 0$: 
P-value distribution for test against null hypothesis $H_0: \tau = \tau_\rx = 0$ obtained from 
unadjusted marginal inference (MI), \NAMI (NAMI), 
or the standardization approaches of \cite{Ye_Shao_Yi_2024} (LRCL) and \cite{Lu_Tsiatis_2008}
(YSTD), and effect estimates of $\widehat{\tau}_\rx$ by linear transformation models (LTM)
under varying prognostic strengths of covariate $X_1$ (in columns) and increasing number of noise covariates
($P$, in rows). \label{fig:surv03pval0}}
\end{figure}

\begin{figure}
\centering

\includegraphics[width=.9\textwidth]{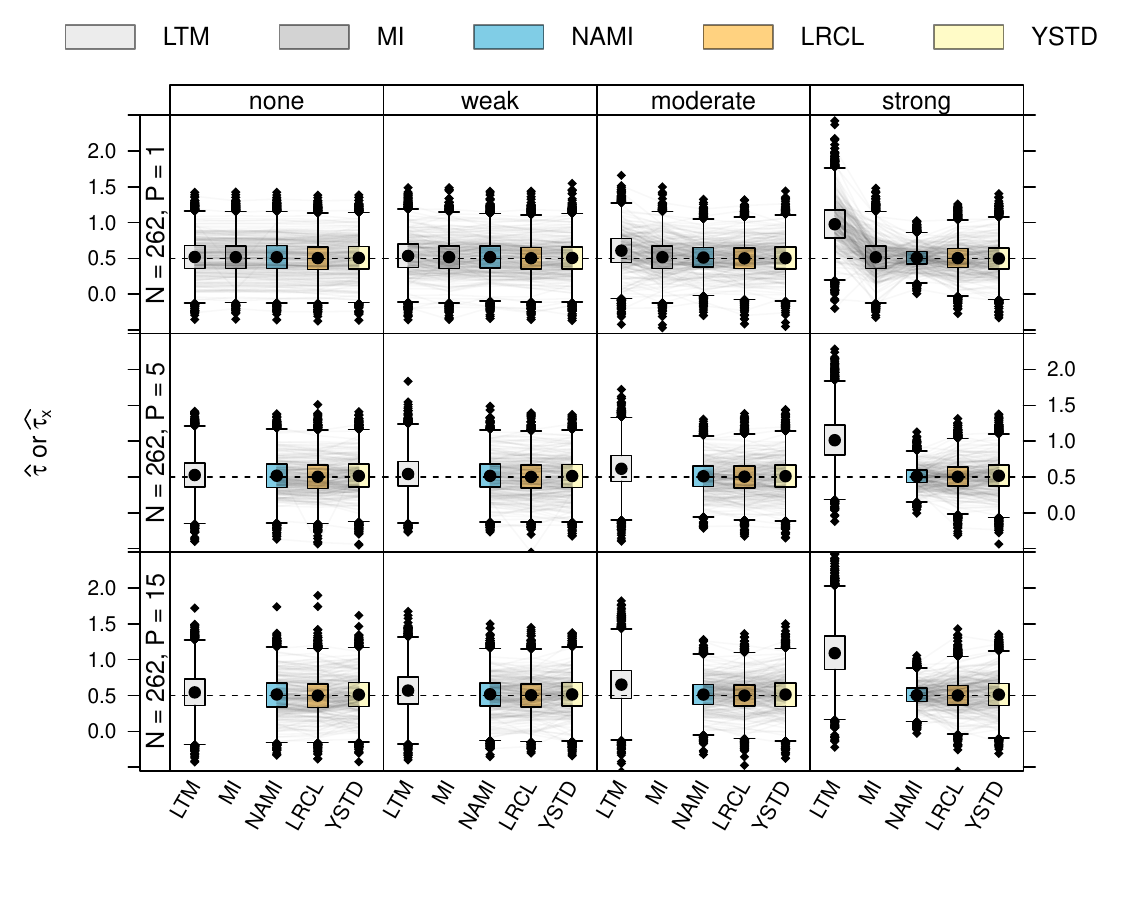} 

  \caption{
Empirical experiments for survival outcome under \textit{heavy} censoring and  $\tau = 0.5$ (dashed lines): Distribution of log-hazard ratio
treatment effect estimates $\widehat{\tau}$ obtained from unadjusted marginal inference (MI), \NAMI (NAMI), 
or the standardization approaches of \cite{Ye_Shao_Yi_2024} (LRCL) and \cite{Lu_Tsiatis_2008} (YSTD), and
effect estimates of $\widehat{\tau}_\rx$ by linear transformation models (LTM)  
under varying prognostic strengths of covariate $X_1$ (in columns) and increasing number of noise covariates
($P$, in rows).
    \label{fig:tausurv03_05}}
\end{figure}

\begin{figure}
\centering

\includegraphics[width=.9\textwidth]{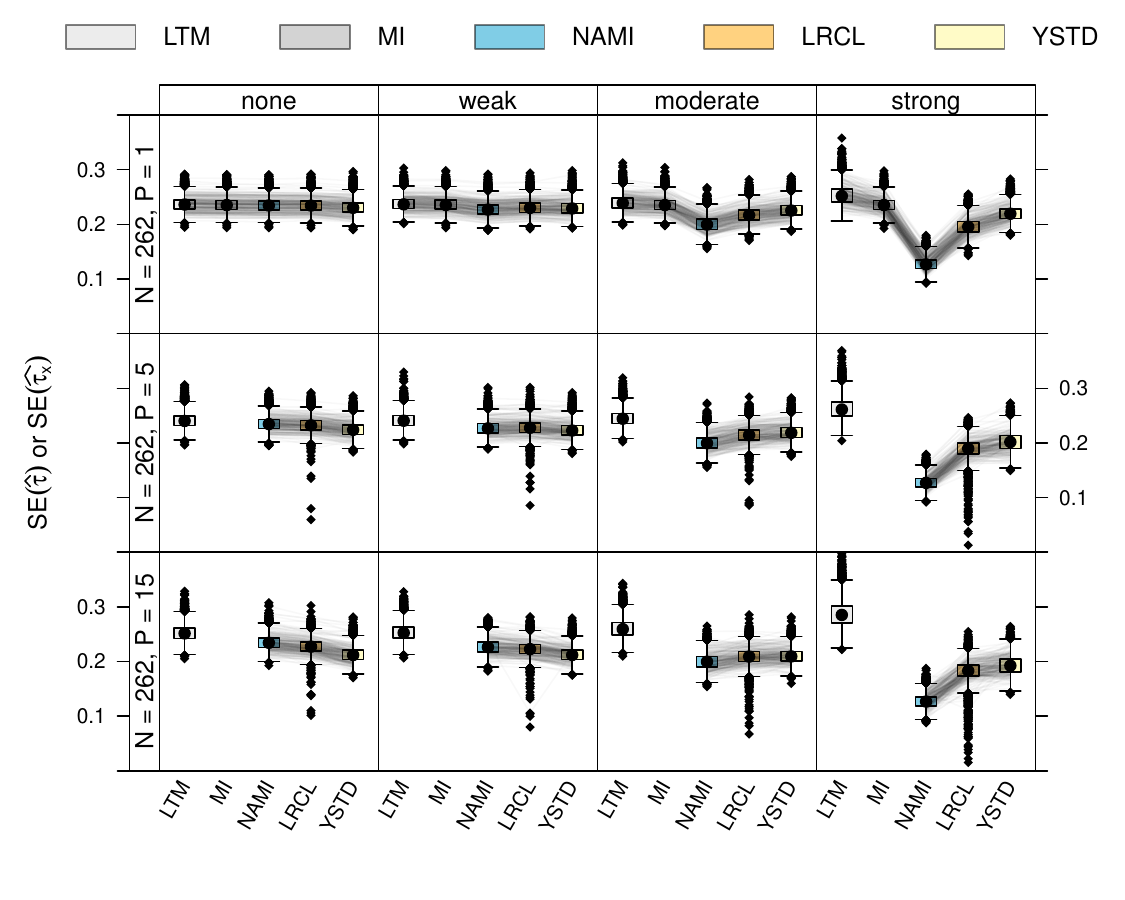} 

  \caption{Empirical experiments for survival outcome under $\tau = 0.5$: Distribution of standard errors of
log-hazard ratio treatment effect estimates $\widehat{\tau}$ obtained from unadjusted marginal inference (MI), \NAMI (NAMI), 
or the standardization approaches of \cite{Ye_Shao_Yi_2024} (LRCL) and \cite{Lu_Tsiatis_2008} (YSTD), and
effect estimates of $\widehat{\tau}_\rx$ by linear transformation models (LTM)  
under varying prognostic strengths of covariate $X_1$ (in columns) and increasing number of noise covariates
($P$, in rows). Standard
errors were computed by inverting the numerically determined negative
Hessian.
    \label{fig:sesurv03_05}}
\end{figure}

\begin{figure}
\centering

\includegraphics[width=.9\textwidth]{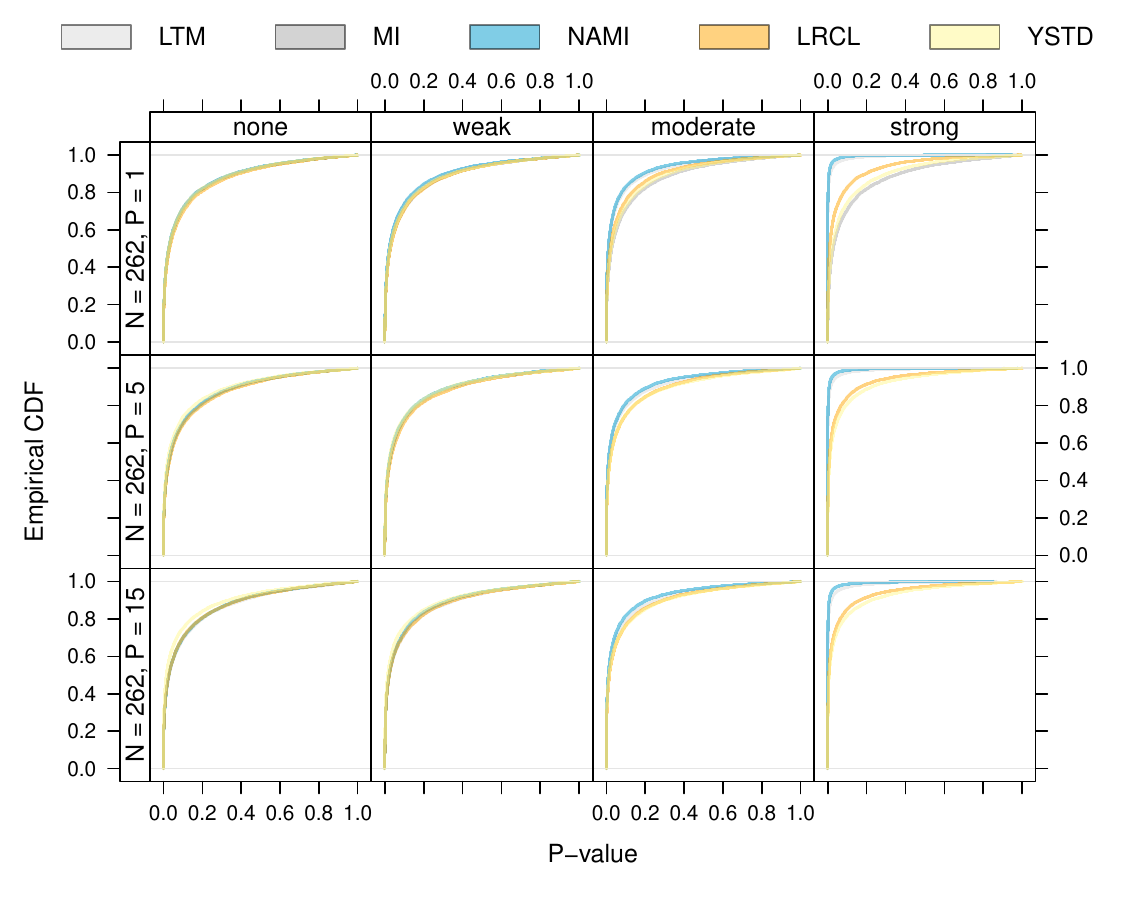} 

\caption{Empirical experiments for survival outcome under \textit{heavy} censoring and $\tau = 0.5$: 
P-value distribution for test against null hypothesis $H_0: \tau = \tau_\rx = 0.5$ obtained from 
unadjusted marginal inference (MI), \NAMI (NAMI), 
or the standardization approaches of \cite{Ye_Shao_Yi_2024} (LRCL) and \cite{Lu_Tsiatis_2008}
(YSTD), and effect estimates of $\widehat{\tau}_\rx$ by linear transformation models (LTM)
under varying prognostic strengths of covariate $X_1$ (in columns) and increasing number of noise covariates
($P$, in rows). \label{fig:surv03pval05}}
\end{figure}

\clearpage
\newpage

 \begin{table}[th!]
	\caption{Empirical size for survival outcomes under \textit{mild} censoring obtained from linear transformation models (LTM), 
		unadjusted marginal inference (MI), \NAMI (NAMI), \cite{Lu_Tsiatis_2008} (YSTD),
		and \cite{Ye_Shao_Yi_2024} (LRCL)
		under varying prognostic strength of covariate $X_1$ (in columns) and varying number of (noise) covariates ($P$, in rows).
		\label{tab:sizemild}}
	\def\arraystretch{1}
	\centering
	\begin{tabular}{lllrrrr}
		\hline
		&&&\multicolumn{4}{c}{Size}\\
		\cline{4-7}
		DGP&Algorithm&P&\multicolumn{1}{c}{none}&\multicolumn{1}{c}{weak}&\multicolumn{1}{c}{moderate}&\multicolumn{1}{c}{strong}\\
		\hline
		survival&MI&P = 1&0.053&     &     &     \\
		&LTM&P = 1&0.054&0.061&0.072&0.115\\
		&&P = 5&0.060&0.064&0.074&0.122\\
		&&P = 15&0.067&0.069&0.085&0.118\\
		&NAMI&P = 1&0.054&0.055&0.056&0.055\\
		&&P = 5&0.055&0.056&0.053&0.060\\
		&&P = 15&0.063&0.062&0.059&0.065\\
		&LRCL&P = 1&0.051&0.050&0.054&0.051\\
		&&P = 5&0.058&0.056&0.053&0.061\\
		&&P = 15&0.069&0.066&0.070&0.074\\
		&YSTD&P = 1&0.056&0.051&0.049&0.041\\
		&&P = 5&0.062&0.059&0.059&0.051\\
		&&P = 15&0.083&0.080&0.069&0.069\\
		\hline
	\end{tabular}
	
\end{table}

\clearpage
\newpage
 
\clearpage
\newpage

\subsection{Misspecification}

\deleted{
	We compare the correctly specified
	conditional Weibull model (LTM), marginal inference (MI, a marginal Cox model comparing the two
	groups), \NAMI \-
	(NAMI, based on the same Cox model) and 
	the standardization approach of \mbox{\cite{Ye_Shao_Yi_2024}} (LRCL) for $N_1 = N_0 = 131$, $\tau_\rx = 0, 0.5$ (as in the simulation study in Section~\ref{sec:simulation}), 
	and $\eparm_1 = 1, \eparm_2 = 4$.
	For $5000$ simulation runs, we estimate
	the corresponding treatment effect parameters and also perform a Wald test
	against the null hypothesis $H_0: \tau = \tau_\rx = 0$.
}

\begin{figure}[th!]

\includegraphics[width=\maxwidth]{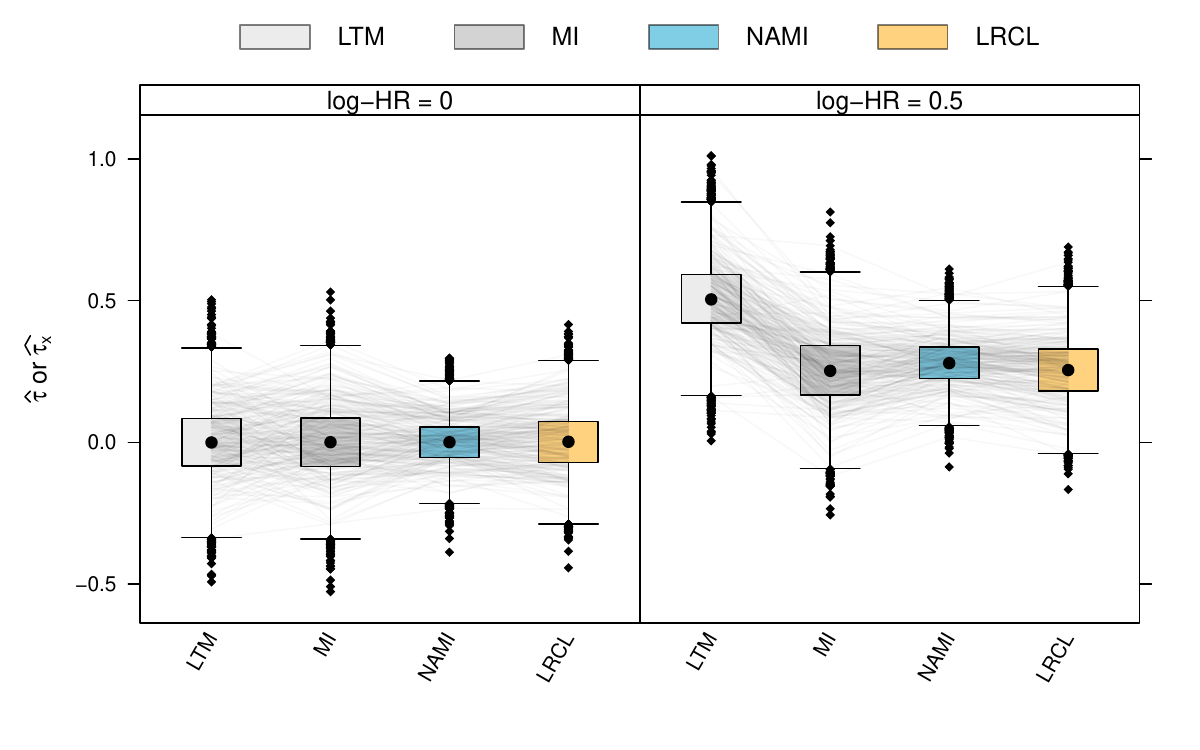} 

\caption{
Empirical experiments for M1 (misspecified marginal model): 
Distribution of log-hazard ratio treatment effect estimates $\widehat{\tau}$
obtained from unadjusted marginal inference (MI), \NAMI (NAMI), 
the standardization approach of \cite{Ye_Shao_Yi_2024} and the
correctly linear transformation models (LTM).
\label{fig:misspec-cf}}
\end{figure}

\deleted{
	The distribution of the parameter
	estimates is given in \Figure~\ref{fig:misspec-cf}. For $\tau_\rx = 0$, all
	three procedures are nearly unbiased, and the parameter variances of NAMI and LRCL are reduced.
	For $\tau_\rx = 0.5$, MI, NAMI and LRCL are biased towards zero, whereas the conditional
	estimate by the Weibull model is right on target.
}

\begin{figure}

\includegraphics[width=\maxwidth]{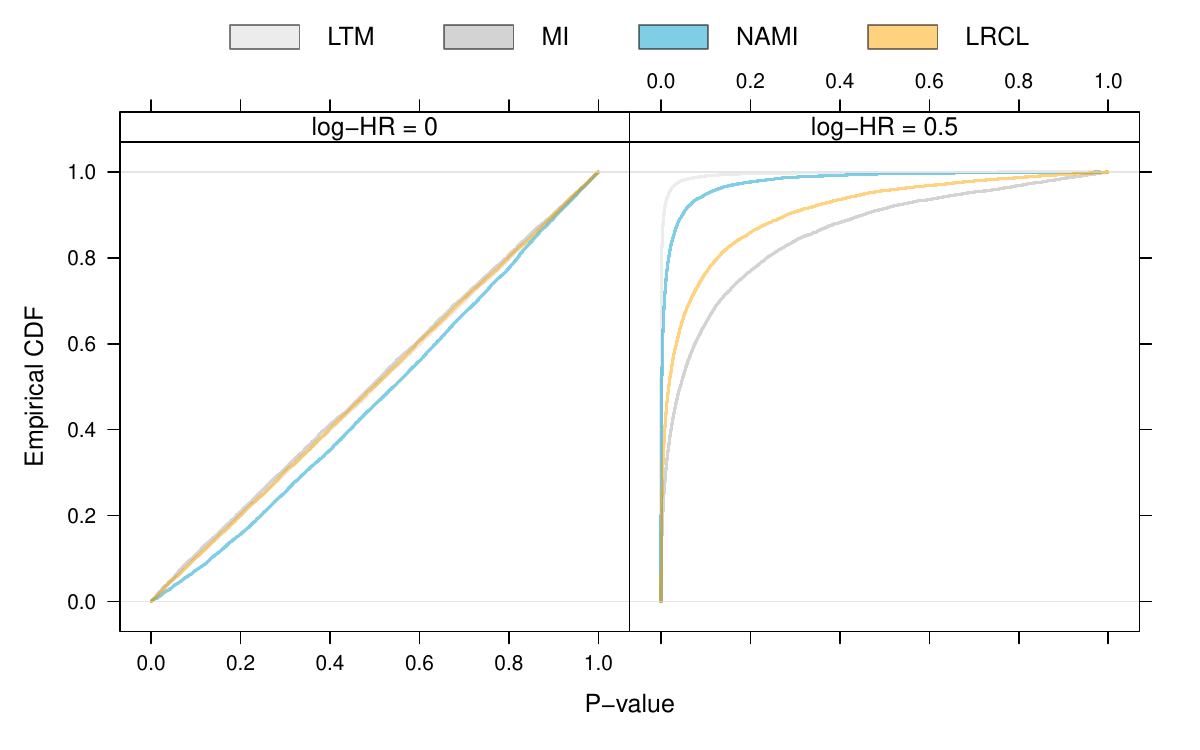} 

\caption{
Empirical experiments for M1 (misspecified marginal model): P-value distribution for test
against null hypothesis $H_0: \tau = \tau_\rx = 0$ obtained 
from unadjusted marginal inference (MI), \NAMI (NAMI), 
the standardization approach of \cite{Ye_Shao_Yi_2024} and the
correctly linear transformation models (LTM).
\label{fig:misspec-pval}}
\end{figure}

\deleted{
	We were in addition interested in the Wald tests' $p$-value distributions
	(\Figure~\ref{fig:misspec-pval}). Under the null hypothesis (left panel), the two
	correctly specified models (MI and LTM) correspond to nearly perfectly
	uniform $p$-value distributions. 
	The same could be observed for LRCL reflecting that the method does not rely on any distributional assumptions
	and is therefore less prone to model misspecifications \mbox{\citep{Ye_Shao_Yi_2024}}.
	The distribution of the misspecified NAMI is stochastically too large, 
	leading to a conservative test. Under the alternative (right panel), 
	the correctly specified Weibull model is most
	powerful. MI leads to much fewer rejections
	at all levels. Covariate adjustment in misspecified NAMI, however, is more
	powerful than the analysis by MI and LRCL, but less powerful than the
	conditional test.
}

\begin{figure}

\includegraphics[width=\maxwidth]{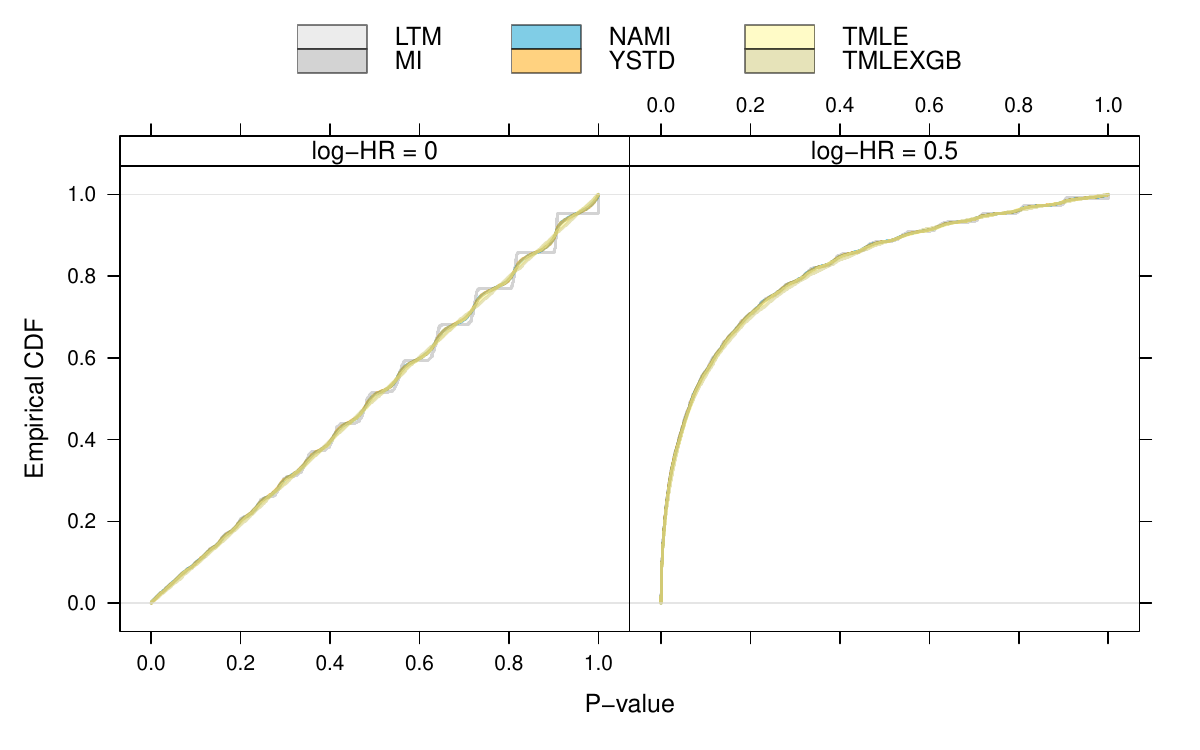} 

\caption{
Empirical experiments for M2 (misspecified Copula structure): P-value distribution for test
against null hypothesis $H_0: \tau = \tau_\rx = 0$ obtained 
from a linear transformation model (LTM), 
unadjusted marginal inference (MI), \NAMI (NAMI), 
the standardization approach of \cite{Ye_Shao_Yi_2024} (YSTD), a
linear transformation model (LTM) and targetted maximum likelihood 
estimation \citep{vanderlaan_tmle_2006} with a (misspecified) logistic regression model (TMLE) 
and a more flexible gradient-boosting-based model (TMLEXGB) to model the outcome.
\label{fig:misspecbin-pval}}
\end{figure}

\newpage
 
\end{document}